\documentclass[moor,nonblindrev]{informs1} 
\pdfoutput=1

\usepackage{bm}
\usepackage{enumerate}
\usepackage{algorithm}
\usepackage{algpseudocode}
\usepackage{mathrsfs}
\usepackage{ulem}
\usepackage{thmtools, thm-restate}
\usepackage{amsmath}
\usepackage{amsfonts}
\usepackage{dsfont}
\usepackage{multirow}
\usepackage{subfigure}
\usepackage{xcolor}

\usepackage{natbib}
 \NatBibNumeric
 \def\bibfont{\small}%
 \bibpunct[, ]{[}{]}{,}{n}{}{,}%

\usepackage[colorlinks=true,breaklinks=true,bookmarks=true,urlcolor=blue,
     citecolor=blue,linkcolor=blue,bookmarksopen=false,draft=false]{hyperref}

\def\EMAIL#1{\href{mailto:#1}{#1}}

\TheoremsNumberedThrough     

\EquationsNumberedThrough    

\begin{document}


\RUNAUTHOR{Shi, Peng, and Zhang}

\RUNTITLE{Top-Two Thompson Sampling for Contextual Selection Problems}

\TITLE{Top-Two Thompson Sampling for Contextual Top-$m_{c}$ Selection Problems}

\ARTICLEAUTHORS{%
\AUTHOR{Xinbo Shi}
\AFF{Department of Management Science and Information Systems, Guanghua School of Management, Peking University, Beijing,
100871, China, \EMAIL{xshi@stu.pku.edu.cn}}
\AUTHOR{Yijie Peng}
\AFF{Department of Management Science and Information Systems, Guanghua School of Management, Peking University, Beijing,
100871, China, \EMAIL{pengyijie@pku.edu.cn}}
\AUTHOR{Gongbo Zhang}
\AFF{Department of Management Science and Information Systems, Guanghua School of Management, Peking University, Beijing,
100871, China, \EMAIL{gongbozhang@pku.edu.cn}}
} 

\ABSTRACT{%
We aim to efficiently allocate a fixed simulation budget to identify the top-$m_{c}$ designs for each context among a finite number of contexts. The performance of each design under a context is measured by an identifiable statistical characteristic, possibly with the existence of nuisance parameters. Under a Bayesian framework, we extend the top-two Thompson sampling method designed for selecting the best design in a single context to the contextual top-$m_{c}$ selection problems, leading to an efficient sampling policy that simultaneously allocates simulation samples to both contexts and designs. To demonstrate the asymptotic optimality of the proposed sampling policy, we characterize the exponential convergence rate of the posterior distribution for a wide range of identifiable sampling distribution families. The proposed sampling policy is proved to be consistent, and asymptotically satisfies a necessary condition for optimality. In particular, when selecting contextual best designs (i.e., $m_{c} = 1$), the proposed sampling policy is proved to be asymptotically optimal. Numerical experiments demonstrate the good finite sample performance of the proposed sampling policy. 
}%


\MSCCLASS{Primary: 62F07; secondary: 62C10, 62L10}
\ORMSCLASS{Primary: Simulation: Efficiency; secondary: Statistics: Sampling}
\HISTORY{}

\maketitle

%


\section{Introduction.}\label{sec:intro}

Simulation optimization (SO) is a powerful technique used to analyze and optimize complex stochastic systems. One well-known problem in SO is ranking and selection (R\&S), which involves comparing a finite number of designs to identify the best one. The performance of each design is measured by a statistical characteristic, such as the mean, which is unknown but can be estimated using simulation samples. In our work, we focus on a contextual top-$m_{c}$ selection problem that requires allocating a fixed number of simulation replications to identify the top-$m_{c}$ designs for each context $c\in\mathcal{C}$, where contexts are known prior to decision-making. The true performance of each design is context-dependent, and the top-$m_{c}$ designs also vary with the context. Contexts can include user profiles and known environmental features, thereby enabling personalized decision-making. Taking product design as an example, a manufacturer aims to identify designs of highest quality within each class of production designs, with each class meeting distinct standards. The production standards are considered as contexts. Other application scenarios include recommendation systems \citep{woerndl2007hybrid}, cancer prevention treatment \citep{li2022efficient}, and assortment optimization \citep{miao2022online}.

We focus on the finite-dimensional context space and does not assume any prior knowledge of the dependence relationship between designs and contexts. Several existing studies assume a dependence structure of designs and contexts by embedding them into a finite-dimensional space \citep{cakmak2021contextual,shen2021ranking}. However, even feasible designs under different contexts may be different due to practical constraints. Therefore, the assumed dependence structure across contexts and designs would be unrealistic in practice. In our work, every context-design pair can be simulated, and the evaluation of the performance of each context-design pair relies solely on simulation samples allocated to that specific pair. A central issue in R\&S problems is how to intelligently allocate the simulation budget to improve sampling efficiency. Metrics like the probability of correct selection (PCS) and the expected opportunity cost (EOC) are commonly used to measure sampling efficiency. An efficient sampling policy must strike a balance between exploiting promising designs and exploring under-sampled designs. In the contextual selection problem, simulation samples are allocated not only to designs but also to contexts. Thus the trade-off between contexts should also be considered. Herein, PCS refers to the probability of correctly selecting the contextual top-$m_{c}$ designs simultaneously, and the false identification of any design under any context is considered as a decision failure.

Under a Bayesian framework, we extend the top-two Thompson Sampling (TTTS) \citep{russo2020simple} algorithm to the contextual top-$m_{c}$ selection problem, resulting in an efficient sampling policy that successively updates the posterior belief on the sampling distributions and sequentially allocates a simulation sample to a context-design pair based on observed sample information. We introduce a tunable hyperparameter to control the allocation of the simulation budget to the promising top-$m_{c}$ designs under each context. The proposed sampling policy is proved to be consistent, meaning that as the number of simulation replicatinos goes to infinity, the posterior distribution definitely identifies the top-$m_{c}$ designs for each context. Beyond the widely studied mean performance in the R\&S literature, with a few exceptions such as \citet{pasupathy2010selecting} and \citet{peng2021efficient}, which study a quantile performance, we consider an identifiable statistical characteristic to measure the performance of each design under a context. Specifically, the family of sampling distributions can be any identifiable family that satisfies the assumption of Glivenko-Cantelli (GC) classes and allows for nuisance parameters. This includes commonly used families of sampling distributions, such as the one-parameter exponential family of distributions considered in \citet{russo2020simple}. This broader range of performance measures allows for greater flexibility in applying our proposed sampling policy. 

To demonstrate the asymptotic optimality of the proposed sampling policy, we rigorously characterize the optimal large deviations rate of the posterior probability of the parameter set that leads to an incorrect selection of the top-$m_{c}$ designs for each context. Technical challenges arise due to the generalizations of the performance metric of context-design pairs and the family of sampling distribution. For example, the strict concavity and monotinicity of the posterior large deviations rate functions may no longer hold, leading to non-uniqueness in the solution of the static optimization problem that maximizes the posterior large deviations rate. The proposed sampling policy is proved to be $\gamma$-optimal when selecting contextual best designs (i.e., $m_{c} = 1$). If $\gamma$ is optimally set or adaptively adjusted towards the optimal value, the sampling policy can be proved to achieve the optimal posterior large deviations rate. In line with the findings in \citet{zhang2021asymptotically}, for $m_{c} \ne 1$, the asymptotic optimality conditions, which consist of several optimal posterior large deviations rate functions, serve only as a necessary condition for asymptotic optimality. Therefore, for the contextual top-$m_{c}$ selection problem (in particular, $m_{c} \ne 1$), even with an optimal or adaptively adjusted $\gamma$, the proposed sampling policy can only be proved to satisfy a necessary condition for asymptotic optimality. In numerical experiments, we consider three metrics to measure the sampling efficiency of each compared sampling policy: the PCS, worst-case PCS (PCSW), and expected PCS (PCSE). The numerical results demonstrate the superior performance of the proposed sampling policy under all three metrics. To summarize, this paper contributes to the existing contextual R\&S literature in three aspects. First, we propose an efficient TTTS policy that simultaneously allocates simulation samples to contexts and designs. The proposed sampling policy allows for a broader range of performance metrics for each context-design pair and families of sampling distributions compared to the existing R\&S literature. Second, we rigorously characterize the posterior large deviations rate under a wide range of families of sampling distributions. Third, we prove that the proposed sampling policy is consistent and asymptotically optimal, satisfying a necessary condition for asymptotic optimality, particularly when $m_{c} \ne 1$,

\subsection{Related work.} \label{sec:literature}
The existing research on the contextual R\&S problem can be roughly categorized based on whether the context space is finite or infinite. In the problem setting where the context space is infinite, not all contexts can be simulated. Therefore, linear functions \citep{shen2021ranking} and Gaussian process regression \citep{ding2022knowledge,hu2017sequential,pearce2018continuous} are often used to capture the relationship between contexts and the performance of each design. \citet{li2018data} generalize the results of \citet{shen2021ranking} by considering high-dimensional contexts and larger classes of dependencies beyond linearity. \citet{keslin2022classification} model the contextual fixed-precision R\&S problem as a classification problem and propose a classifier for any context. However, the model assumptions in these policies may not hold under certain violations of the assumed relationship between designs and contexts. In our work, we focus on a finite context space where every context-design pair can be simulated. \citet{du2022rate} (and their conference version \cite{gao2019selecting}) derive asymptotic optimality conditions for the contextual fixed-budget R\&S problem using the large deviations technique. They propose a sampling policy that iteratively balances the two sides of each equation of the asymptotic optimality conditions, which has been proved to sequentially achieve the asymptotically optimal sampling ratios. \citet{cakmak2021contextual} model the mean performance of each design with a Gaussian process and derive a similar sampling policy to \citet{du2022rate}, which uses the posterior means of the Gaussian process rather than the sample means. \citet{li2022efficient} utilize a Gaussian mixture model to capture the relationship between designs and contexts, enabling the utilization of clustering information to improve sampling efficiency. They propose a myopic sampling policy that has also been proved to sequentially achieve the asymptotically optimal sampling ratios as defined in \citet{du2022rate}. \citet{zhang2023efficient} extend the results of asymptotic optimality from \citet{du2022rate} to the problem of selecting contextual top-$m_{c}$ designs. They develop a sampling policy that maximizes a value function approximation (VFA) one-step look ahead, which is proved to sequentially achieve the corresponding asymptotically optimal sampling ratios. Similar to the problem settings of \citet{du2022rate} and \citet{zhang2023efficient}, our work does not rely on dependence structure between designs and contexts, and samples from one context-design pair do not provide any information about others.

Existing sampling policies in R\&S can roughly be categorized into fixed-precision (or frequentist) and fixed-budget (or Bayesian) branches. The fixed-precision branch aims to guarantee a predetermined PCS level under the assumption of an indifference zone for unknown design performances. However, in practice, it often requires more simulation budget than necessary to achieve the desired PCS level. On the other hand, the fixed-budget branch aims to maximize a performance metric given a simulation budget constraint. Although it does not provide a guarantee on PCS, it typically achieves better performance than the fixed-precision branch under a fixed number of simulation budget. \citet{hong2021review} provide a recent review on R\&S problem. They view the former branch from a hypothesis testing perspective and offer a dynamic programming formulation for the latter branch. In this work, we focus on the Bayesian branch. Bayesian sampling policies for the classical R\&S problem include optimal computing budget allocation (OCBA) \citep{chen1996lower,chen2000simulation}, expected value of improvement (EVI) \citep{chick2001new}, knowledge gradient (KG) \citep{frazier2008knowledge}, expected improvement (EI) \citep{ryzhov2016convergence}, asymptotically optimal allocation policy (AOAP) \citep{peng2018ranking}, and balancing optimal large deviations (BOLD) \citep{chen2022balancing}. OCBA and BOLD are developed based on certain asymptotic conditions, which are solved from a static optimization problem using several approximations or large deviations technique. Sample estimates are used as plug-in estimators for unknown parameters in the asymptotic conditions. However, recent studies have highlighted limitations in the finite-sample performance of policies derived solely based on asymptotic conditions \citep{peng2015non,peng2017gradient,russo2020simple,zhang2021asymptotically}. KG and EI, on the other hand, are developed to sequentially allocate simulation samples to optimize myopic surrogate criteria under a stochastic dynamic programming framework. \citet{peng2018ranking} further formulate the Bayesian sequential sampling and allocation decisions as a Markov decision process (MDP) and rigorously establish Bellman equations for the MDP. However, solving the Bellman equations to compute the optimal sampling policy is computationally intractable due to the curse of dimensionality \citep{peng2018ranking,russo2020simple}. Various approaches have been taken to circumvent this computational difficulty. For example, AOAP is developed by optimizing a VFA one-step look ahead in the MDP. Beyond the classical R\&S problem, recent advances have been made in Bayesian sampling policies for the problem of selecting top-$m_{c}$ designs. Some of the proposed sampling policies follow a similar scheme to OCBA \citep{chen2008efficient,gao2016new,zhang2015simulation}, while others generalize the AOAP \citep{zhang2021asymptotically}. In our work, we study contextual R\&S problems by referring to a sampling policy proposed for the multi-armed bandit (MAB) problem, which is also a popular model for studying the trade-off between exploitation and exploration in sequential decision problems.

The research in R\&S and MAB diverges in several aspects, with the most notable difference being that the reward of MAB is collected at each round when pulling an arm. A stream of recent literature in MAB focuses on the best arm identification (BAI) or pure-exploration problem, which solely focuses on selecting the highest-quality arm at the end. The classical BAI problem has been studied under fixed-precision \citep{even2006action,garivier2016optimal} and fixed-budget \citep{audibert2010best,bubeck2009pure} settings, where sampling policies are often developed to match certain lower bounds of the sample complexity of the BAI problem. The classical BAI problem has been generalized to the top-$m_{c}$ arms identification problem \citep{bubeck2013multiple,gabillon2012best,kalyanakrishnan2012pac,kaufmann2013information} and the contextual selection problem \citep{kato2022best,reda2021top,soare2014best}. Thompson Sampling (TS) \citep{thompson1933likelihood} is a popular Bayesian algorithm for regret minimization in the MAB problem. It assumes a prior on the unknown parameters of the reward distribution and selects an arm according to its posterior probability of being the best arm. A comprehensive review of TS can be found in \citet{russo2018tutorial}. TS can achieve a certain asymptotically optimal regret bound \citep{agrawal2017near,kaufmann2012thompson} and appears to have better finite-sample performance compared to $\epsilon$-Greedy and upper confidence bound (UCB) algorithms \citep{auer2002finite} in the MAB problem, as it does not include any tuning parameters. However, TS can exhibit notoriously poor asymptotic performance for the BAI problem since it exclusively chooses an arm in almost all rounds once that arm is estimated to be the best with a high probability. To enforce exploration of sub-optimal arms in the BAI problem, \citet{russo2020simple} adapts TS to TTTS by adding a re-sampling step and a tuning parameter $\gamma$. With probability $\gamma$, the arm found in the first sampling step is selected, and with probability $\left(1 - \gamma\right)$, the arm found in the re-sampling step is selected. The robust empirical performance of TTTS with $\gamma = {1 \mathord{\left/ {\vphantom {1 2}} \right. \kern-\nulldelimiterspace} 2}$ is demonstrated in \citet{russo2020simple}. However, a practical drawback of TTTS is that, as the posterior distributions become concentrated, finding an arm in the re-sampling step requires many posterior re-samples. Some methods have been proposed to alleviate this computational burden, including finding an arm with the lowest transportation cost \citep{shang2020fixed} and truncating re-sampling steps with a fixed number of re-samples \citep{zhang2022monte}. Recently, \citet{shang2020fixed} extend the restrictive prior assumption on TTTS and analyze the sample complexity of TTTS for the fixed-precision BAI problem. \citet{jourdan2022top} generalize sample distribution families of TTTS to bounded distributions for the fixed-precision BAI problem. \citet{peng2022thompson} provide a comprehensive comparison of the empirical performance of TTTS and existing Bayesian sampling policies developed for the classical R\&S problem. \citet{you2023informationdirected} generalize TTTS to the fixed-precision top-$m_{c}$ arms identification problem and propose a sampling policy that selects an arm from the top-two candidate arms by an information-directed rule without using the tuning parameter $\gamma$. In our work, we generalize TTTS to the fixed-budget contextual top-$m_{c}$ selection problem, resulting in an efficient Bayesian sampling policy for learning top-$m_{c}$ designs for each context. This extension addresses one of the open questions raised by \citet{russo2020simple}.

Several existing studies on R\&S and BAI have established the asymptotic optimality of their proposed sampling policies. The notions of asymptotic optimality differ between the fixed-precision and fixed-budget branches. The fixed-precision branch focuses on the sample complexity associated with the simulation budget required to achieve a given precision \citep{garivier2016optimal,qin2017improving,you2023informationdirected}. In the fixed-budget branch, \citet{glynn2004large} derive asymptotic optimality conditions satisfied by the static asymptotically optimal sampling ratios for the classical R\&S problem using large deviations theory. By fixing a selection policy for choosing a design with the best empirical performance, the frequentist probability of incorrect selection (PICS) decays exponentially at the fastest possible rate under the asymptotically optimal sampling ratios. Although asymptotically optimal sampling ratios are defined for light-tailed sampling distributions in \citet{glynn2004large}, most sampling policies have only been proved to be asymptotically optimal under Gaussian sampling distributions \citep{avci2021getting,chen2019complete,peng2018ranking}. An exception is the work of \citet{chen2022balancing}, which develops an asymptotically optimal sampling policy under light-tailed sampling distributions. The framework of \citet{glynn2004large} has been generalized to other problem settings, including the feasibility determination problem \citep{szechtman2008new}, complete ranking problem \citep{xiao2013optimal}, top-$m_{c}$ selection problem \citep{zhang2021asymptotically}, and contextual selection problem \citep{du2022rate,zhang2023efficient}. The existing proof of the asymptotic optimality of a sampling policy typically involves demonstrating that its asymptotic sampling ratios satisfy the corresponding asymptotic optimality conditions. However, achieving the asymptotically optimal sampling ratios is contingent upon the asymptotic optimality conditions serving as sufficient and necessary conditions for such ratios. \citet{zhang2022efficient} demonstrate the sufficiency and necessity of the asymptotic optimality conditions defined in \citet{glynn2004large} under Gaussian sampling distributions by showing that these conditions determine a unique sampling ratio for each design. However, this is not the case in the top-$m_{c}$ selection problem, where the asymptotic optimlaity conditions are only necessary conditions for asymptotically optimal sampling ratios \cite{zhang2021asymptotically}, which highlights the inherent differences and challenges in characterizing asymptotic optimality for the top-$m_{c}$ selection problem. The sampling policy proposed by \citet{zhang2021asymptotically} is proved to be asymptotically optimal when the solution to the asymptotic optimality conditions is unique, and its asymptotic optimality can only be checked numerically when the solution to the asymptotic optimality conditions is non-unique. The work of \citet{russo2020simple} offers an alternative type of performance criterion to characterize the asymptotic optimality of a Bayesian sampling policy by analyzing the large deviations rate of the posterior probability of the parameter set that leads to an incorrect selection. A sampling policy is asymptotically optimal in \citet{russo2020simple} if it ensures an exponential decay at the largest possible posterior large deviations rate (pLDR). The asymptotic optimaility conditions defined in \citet{glynn2004large} are necessary, if not sufficient, for asymptotic optimality of the performance criterion developed in \citet{russo2020simple}. TTTS, which is developed for sampling distributions in the exponential family, is shown to be asymptotically $\gamma$-optimal, which is a weaker notion of asymptotic optimality. If $\gamma$ is set optimally or adaptively adjusted towards the optimal value, TTTS can be proved to be asymptotically optimal as defined in \citet{russo2020simple}. However, the derivation of the large deviations rate for posterior distributions in both the top-$m_{c}$ selection problem and the contextual selection problem has seldom been discussed in the literature. Our work bridges this gap.

The rest of the paper is organized as follows. Section \ref{sec:probdef} formulates the contextual top-$m_{c}$ selection problem under a Bayesian framework. Section \ref{sec:alg} introduces the proposed sampling policy with a tunable hyperparameter $\bm{\gamma}$ for selecting contextual best designs. Section \ref{sec:asymptotics} characterizes the optimal posterior large deviations rate and establishes the asymptotic optimality of the proposed sampling policy in Section \ref{sec:alg}. Section \ref{sec:var} extends the sampling policy and asymptotic analyses to the contextual top-$m_{c}$ selection problem and discusses consistent tuning of the hyperparameter. Section \ref{sec:numexp} presents numerical experiments, and Section \ref{sec:con} concludes the paper. The proofs of the theoretical results are relegated to the electronic companion.

\section{Problem definition.} \label{sec:probdef}

In this section, we formally introduce the contextual top-$m_{c}$ selection problem under a Bayesian framework. Let $\mathcal{D}$ be a set of competitive designs. Each design $d\in\mathcal{D}$ is associated with a random variable $Y_{d}$ with density $p(y \vert \theta_{d})$ with respect to a base measure $\nu$, where $\theta_{d} = (\mu_{d}, \eta_{d})$ is a finite-dimensional parameter, $\mu_{d} \in M \subseteq \mathbb{R}$ is a parameter of interest, $\eta_{d} \in H \subseteq \mathbb{R}^{k}$ is a nuisance parameter, and $M$ and $H$ are parameter spaces of $\mu$ and $\eta$, respectively. The quality of each design is measured by the parameter $\mu_{d}$ while both parameters are unknown to the decision maker and can only be estimated using simulation samples of $Y_{d}$. Suppose $\mathcal{D}$ can be represented as the union of pairwise disjoint sets $\mathcal{D}_{c}$, such that $\mathcal{D} = \bigcup_{c\in \mathcal{C}}{\mathcal{D}_{c}}$. The index $c\in\mathcal{C}$ is commonly referred to as a context, where $\mathcal{C}$ is the set of all available contexts. For any $c\in\mathcal{C}$, let $m_{c}$ be a positive integer that satisfies $m_{c}\leq \vert \mathcal{D}_{c} \vert$, where $\vert \cdot\vert$ is the cardinality of a set, and let $\mathcal{D}_{c} = \mathcal{P}_{c} \bigcup \mathcal{U}_{c}$ such that $\mathcal{P}_{c} \bigcap \mathcal{U}_{c} = \emptyset$, $\vert \mathcal{P}_{c}\vert = m_{c}$ and $\mu_{d} > \mu_{d^{\prime}}, ~\forall d\in \mathcal{P}_{c}, ~d^{\prime}\in \mathcal{U}_{c}$. Throughout this paper, we assume that $\mu_{d}\neq \mu_{d^{\prime}}$ for $d\neq d^{\prime}$, ensuring that the set $\mathcal{P}_{c}$ is well-defined. In other words, $\mathcal{P}_{c}$ is a set that contains top-$m_{c}$ designs with the highest qualities as measured by the parameter $\mu_{d}$ under a context $c$.

The goal is to identify the preferable subset $\bigcup_{c\in\mathcal{C}}\mathcal{P}_{c} \subseteq \mathcal{D}$ by sequentially allocating simulation samples. Let $\bm{Y}=\left(Y_{d}\right)_{d\in\mathcal{D}}$ denote the random vector for all simulation samples, and $\bm{\theta}=\left(\theta_{d}\right)_{d\in\mathcal{D}} \in \bm{\Theta}$ be the parametric vector for all designs, where $\bm{\Theta} = \Theta^{\vert \mathcal{D}\vert}$ is the complete parameter space, and $\Theta = M\times H$ is the marginal parameter space associated with each certain context. Suppose that the total number of simulation budget is $T$. Let $\{\bm{Y}^{(t)}\}_{t\leq T}$ be a sequence of independent replications of $\bm{Y}$. At each step $t \le T$, a simulation sample is allocated to a context $C_{t}\in\mathcal{C}$ and a design $D_{t}\in\mathcal{D}_{C_{t}}$, and the realization of the corresponding coordinate $Y_{D_{t}}^{(t)}$ is observed. We focus on all possibly randomized sampling policies $\{\psi_{t}\}_{t\leq T}$, where $\psi_{t}$ is an $\mathcal{E}_{t-1}$-measurable random distribution over $\mathcal{D}$, and $\mathcal{E}_{t-1}$ is a $\sigma$-field generated by the observed sample information $\{C_{1},D_{1},Y_{D_{1}}^{(1)},\dots,C_{t-1},D_{t-1},Y_{D_{t-1}}^{(t-1)}\}$ up to step $\left(t-1\right)$, such that $\psi_{t}(d) := \mathbb{E}[\bm{1}\{D_{t}=d\}\vert \mathcal{E}_{t-1}]$. Moreover, let $\alpha_{t}(c) := \mathbb{E}[\bm{1}\{C_{t}=c\}\vert \mathcal{E}_{t-1}]$ and $\beta_{t}(c, d) := \mathbb{E}[\bm{1}\{D_{t}=d\}\vert \mathcal{C}_{t} = c, \mathcal{E}_{t-1}]$. Then, for any $c\in\mathcal{C}$ and $d\in\mathcal{D}_{c}$, we have $\psi_{t}(d) = \alpha_{t}(c)\beta_{t}(c,d)$. In implementating a randomized sampling policy, an independent source of randomness is utilized to realize the context-design pair $(C_{t}, D_{t})$ according to $\psi_{t}$. These policies are known as \textit{sequential sampling policies}, which include a wide range of non-anticipatory policies that depend only on the observed sample information. Once the simulation budget being exhausted, a subset $\bigcup_{c\in\mathcal{C}}\hat{\mathcal{P}}_{c}$ of $\mathcal{D}$ is selected as an estimate of $\bigcup_{c\in\mathcal{C}}\mathcal{P}_{c}$ based on the observed sample information, where $\hat{\mathcal{P}}_{c} \subseteq \mathcal{D}_{c}$ is an estimate of $\mathcal{P}_{c}$ with $\vert \hat{\mathcal{P}}_{c} \vert = m_{c}$.

Before delving into further discussion, we first introduce two examples to elaborate on the notions of the design set $\mathcal{D}$, the associated random variable $Y_{d}$, and the parameter of interest $\mu_{d}$.

\noindent \textbf{Example 1 Connection to the existing formulation of contextual R\&S problems.} Existing contextual R\&S problems focus on Gaussian sampling distributions and aim to find the design with the maximum mean performance for each context. Specifically, consider $p$ complicated simulation systems with random simulation outputs $Y_{1}(x), \dots, Y_{p}(x)$, which depend on a covariate $x\in \mathcal{X}$. Assume that $\mathcal{X}$ contains a finite number of $q$ available covariates $x_{1},\dots, x_{q}$. Then $\mathcal{D}=\{(i, j): ~i\leq p, ~j\leq q\}$ consists of all index pairs of simulation systems and covariates with $(i, j)$ corresponding to the $i$-th system and the $j$-th covariate $x_{j}$. Naturally, $\mathcal{D}_{x} = \{(i, j) \in \mathcal{D}: ~i\leq p, ~x_{j} = x\}$ corresponds to the index pairs under covariate $x$, and $x\in\mathcal{X}$ is understood as a context. The quantity of interest is the mean performance $\mu_{(i, j)} = \mathbb{E}[Y_{i}(x_{j})]$, and the objective is to find the best-performing systems under each context. The simulation outputs are modeled as independent Gaussian random variables, i.e., $Y_{i}(x_{j})\sim N(\mu_{i}(x_{j}), \sigma_{i}^{2}(x_{j}))$, where $\sigma_{i}^{2}(x_{j})$ is the variance of $Y_{i}(x_{j})$, and then $\mu_{(i, j)} = \mu_{i}(x_{j})$ and $\eta_{(i, j)} = \sigma^{2}_{i}(x_{j})$. This problem formulation is considered in \citet{du2022rate,gao2019selecting}, and \citet{zhang2023efficient}. Although the Gaussian assumption is commonly used in the R\&S literature, both the mean and variance parameters have to be estimated via simulation samples. Many R\&S sampling policies are developed assuming known variance while substituting sample variance for the true variance in implementation. In this work, we focus on the unknown nature of the parameter.

\noindent \textbf{Example 2 Production design for diversified demands.} The following real-world setting exemplifies a problem where the parameter of interest is different from the sample mean of Gaussian variables. Suppose an electronic component manufacturer has recently developed a variety of production processes with different costs or meeting different production standards. They plan to offer differentiated products within several ranges while retaining a subset of the production lines. In this case, $\mathcal{D}$ is the set of all production processes and the price ranges or production standards are understood as contexts. 

\begin{figure}
    \centering
    \includegraphics[width = 0.3\linewidth]{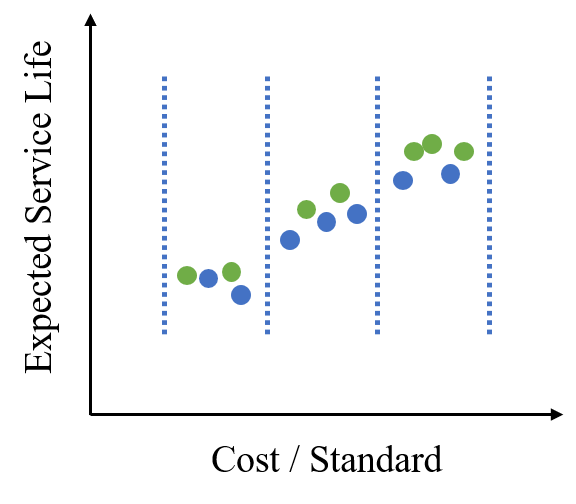}
    \caption{Selecting the set of durable (green) designs in each range.}
    \label{fig:ex2}
\end{figure}

In an attempt to find the best production process that produces the most durable products within each cost or standard range (see Figure \ref{fig:ex2}), the manufacturer conducts accelerated life tests (ALT) to evaluate product durability, where the electronic component is subjected to extreme voltage and thermal conditions, and the time-to-failure is recorded if a failure occurs before the end of the test that lasts for time $\tau$. The lifespan of an electronic component produced by process $d$ is commonly modeled using the two-parameter Weibull distribution $W(\rho_{d}, k_{d})$, which has a shape parameter $k_{d}$ and a scale parameter $\rho_{d}$, as commonly used in survival and reliability analysis \citep{kapur77}. The probability density function of $W(\rho_{d}, k_{d})$ with respect to the Lebesgue measure is given by $f_{W}(y;\rho_{d}, k_{d}) = k_{d}/\rho_{d} \cdot (y/\rho_{d})^{k_{d}-1}\cdot \exp\{-(y/\rho_{d})^{k_{d}}\}$, where $y$ represents the time-to-failure. The manufacturer is interested in the mean time-to-failure, denoted by $\mu_{d} = \rho_{d} \cdot \Gamma(1 + \frac{1}{k_{d}})$, where $\Gamma(z) = \int_{0}^{\infty}{t^{z - 1} e^{-t} dt}$ is the canonical Gamma function. Each $d\in \mathcal{D}$ is associated with the recorded time-to-failure $Y_{d}$, which is a right-censored Weibull random variable at time $\tau$ with density:
\begin{equation*}
    p(y \mid \rho_{d}, k_{d})=f_W(y ; \rho_{d}, k_{d}) \times {\bm{1}\{y<\tau\}} + S_W(\tau ; \rho_{d}, k_{d}) \times {\bm{1}\{y=\tau\}}
\end{equation*}
with respect to measure $(\omega + \delta_{\tau})$, where $S_W(\tau ; \rho_{d}, k_{d}) = \int_{\tau}^{\infty}{f_{W}(y; \rho_{d}, k_{d})dy} = \exp\{-(\tau/\rho_{d})^{k_{d}}\}$ is the survival function, $\omega$ is the Lebesgue measure and $\delta_{\tau}$ is the Dirac measure concentrated at $\tau$. Although $\mu_{d}$ does not explicitly appear in the density, one can perform a reparameterization by setting $\mu_{d} = \rho_{d}\cdot \Gamma(1 + 1/k_{d})$ and $\eta_{d} = k_{d}$. After the reparameterization, the density becomes:
\begin{align*}
    p(y \mid \mu_{d}, \eta_{d})=&f_W(y ; \mu_{d} / \Gamma(1 + 1/\eta_{d}), \eta_{d}) \times {\bm{1}\{y<\tau\}} + S_W(\tau ; \mu_{d} / \Gamma(1 + 1/\eta_{d}), \eta_{d}) \times {\bm{1}\{y=\tau\}}.
\end{align*}

Note that $\mu_{d}$ does not coincide with the mathematical expectation of $Y_{d}$. Therefore, classical R\&S procedures developed based on the mean performance do not directly apply in this example. Similarly, both $\mu_{d}$ and $\eta_{d}$ are estimated using simulation samples.

\subsection{Assumptions.}
We assume that simulation samples $Y_{d}$ are independent across alternative designs. Then the joint density of $\bm{Y}$ belongs to a parametric model $\mathcal{F}=\{p(\bm{Y}\vert \bm{\theta}) = \prod_{c\in\mathcal{C}}\prod_{d\in \mathcal{D}_{c}}{p(Y_{d}\vert \theta_{d})}:\bm{\theta}\in\bm{\Theta}\}$. Moreover, we assume that this model is correctly specified. The true distribution of $\bm{Y}$ passes through $\mathcal{F}$ at the ground truth $\bm{\theta}^{*}$ and the parameter space $\bm{\Theta}$ includes an open neighborhood containing $\bm{\theta}^{*}$. It's important to note that our model considers the true distribution as fixed. Although we will conduct parameter inferences and develop sampling algorithms within a Bayesian framework, as defined in Section \ref{sec:bayes}, the analyses are performed assuming a fixed ground truth.

Furthermore, we make some regularity assumptions on the parametric family as follows.

\begin{assumption}\label{ass:1} The parametric family of sampling distributions $\{p(\cdot\vert\theta): \theta\in \Theta\}$ satisfies:
\begin{enumerate}[(1)]
    \item\label{ass:comp} The parameter space $\Theta$ is compact and its boundary has Lebesgue measure zero;
    \item\label{ass:cont} The parametric family $\{p(\cdot\vert\theta): \theta\in \Theta\}$ is pointwise continuous in $\theta$;
    \item\label{ass:indentifiable} The parametric family $\{p(\cdot\vert\theta): \theta\in \Theta\}$ is identifiable, i.e., for any pair of distinct parameter values $\theta\neq \theta^{\prime}$, the set $\{y\in\mathbb{R}: p(y\vert \theta)\neq p(y\vert \theta^{\prime})\}$ has positive Lebesgue measure;
    \item\label{ass:GC} The set of log-likelihood ratio tests $\mathcal{L}=\{\ln\{p(y\vert \theta^{\prime})/p(y\vert \theta^{\prime\prime})\}: \theta^{\prime},\theta^{\prime\prime}\in \Theta\}$ constitutes a universal Glivenko-Cantelli (GC) class, i.e., for any $\theta\in \Theta$, the following convergence holds almost surely:
    \begin{equation}\label{eq:1}
        \sup_{f\in\mathcal{L}}\left\vert \frac{1}{T}\sum_{t=1}^{T}f(Y^{(t)}) - \mathbb{E}[f(Y)] \right\vert \rightarrow 0, \quad a.s.,
    \end{equation} where $Y\sim p(\cdot\vert \theta)$ and $Y^{(1)},\dots,Y^{(T)}$ are independent and identically distributed samples of $Y$.
\end{enumerate}
\end{assumption}

Assumptions \ref{ass:1}.(\ref{ass:comp}) and \ref{ass:1}.(\ref{ass:cont}) are commonly used regular conditions in Bayesian analysis. Notice that the compactness of $\Theta$ may exclude some conjugate prior distributions of interest, such as Gaussian priors, which do not satisfy the compactness requirement. However, practitioners can easily truncate the extreme values of the distributional parameters while preserving the simplicity of the posterior distribution within our framework. For instance, one may well utilize the Normal-Gamma conjugate prior with unbounded support for Gaussian distributions but can restrict the prior distribution to the compact set $\Theta$, which only contributes to the posterior distribution through a normalizing constant and an indicator function $\bm{1}\{\theta\in\Theta\}$. Assumption \ref{ass:1}.(\ref{ass:indentifiable}) prevents redundant parameterization, where multiple distributions with different parameters in $\mathcal{F}$ are identical and indistinguishable based on simulation samples. This assumption is crucial for establishing the consistency of our proposed sampling policy and is a commonly employed concept in statistics. Assumption \ref{ass:1}.(\ref{ass:GC}) plays a key role in establishing the posterior convergence by stating that log-likelihood ratio tests satisfy the uniform strong law of large numbers. For our asymptotic analysis, it suffices to assume that (\ref{eq:1}) holds for each $\theta^{*}_{d}$. Although the universal GC class assumption imposes a stronger condition, it is necessary due to the unknown nature of $\bm{\theta}^{*}$. The universality of this assumption is a global property of the distribution family that can be verified beforehand, providing a practical and verifiable basis for our asymptotic analysis.

\subsection{Bayesian framework.}\label{sec:bayes}
Suppose that $\Pi_{0}(\bm{\theta})$ is a prior distribution of $\bm{\theta}$. In our framework, this prior does not necessarily represent the true generating mechanism of the model parameter because we assume the true underlying parameter $\bm{\theta}^{*}$ to be fixed. The information on the model parameter $\bm{\theta}$ is accumulated through the recursively updated posterior distribution given by:
\begin{equation*}
\Pi_{t}(\bm{\theta}) = \frac{\Pi_{t-1}(\bm{\theta})L_{t}(\bm{\theta})}{\int_{\bm{\Theta}}{\Pi_{t-1}(\bm{\theta}^{\prime})L_{t}(\bm{\theta}^{\prime})d\bm{\theta}^{\prime}}},
\end{equation*}
where $L_{t}(\bm{\theta}) = \mathbb{E}\left[p\left(\bm{Y}\middle|\bm{\theta}\right)\middle| Y_{D_{t}}^{(t)}\right] = p\left(Y_{D_{t}}^{(t)}\middle|\theta_{D_{t}}\right)$ is the expected likelihood function conditioned on the partial observation $Y_{D_{t}}^{(t)}$. We use the notation $\Pi_{t}(\cdot)$ to refer to both the density and the probability measure of the posterior distribution, with a slight abuse of notation.
Throughout this paper, we use the uninformative prior $\Pi_{0}(\bm{\theta}) = 1$ for the sake of simplicity. However, our results can be easily extended to any prior distribution with density bounded away from zero and infinity. This extension is possible because our focus is on an exponential convergence rate of the posterior distribution, which remains invariant when multiplied by a constant. 

The likelihood function and posterior belief are then given by
\begin{equation*}
    L_{T}(\bm{\theta}) = L_{T}(\bm{\theta}^{*})\exp{\left\{-\ln{\frac{L_{T}(\bm{\theta}^{*})}{L_{T}(\bm{\theta})}}\right\}} = L_{T}(\bm{\theta}^{*})\exp{\left\{-TW_{T}(\bm{\theta})\right\}},
\end{equation*}
and
\begin{equation*}
    \Pi_{T}(\bm{\theta}) = \frac{{L_{T}(\bm{\theta})}}{\int_{\bm{\Theta}}{{L_{T}(\bm{\theta}^{\prime})}d\bm{\theta}^{\prime}}} 
    = \frac{ \exp{\left\{-TW_{T}(\bm{\theta})\right\}} }{\int_{\bm{\Theta}}{ \exp{\left\{-TW_{T}(\bm{\theta}^{\prime})\right\}} d\bm{\theta}^{\prime} }}~,
\end{equation*}
respectively, where 
\begin{equation*}
    W_{T}(\bm{\theta}) = \frac{1}{T}\ln{\frac{L_{T}(\bm{\theta}^{*})}{L_{T}(\bm{\theta})}} = \sum_{d\in \mathcal{D}}\frac{1}{T}\sum_{t=1}^{T}\xi_{t}(d)\lambda\left(\theta^{*}_{d},\theta_{d};Y_{d}^{(t)}\right),
\end{equation*}
is the average exponential rate at which $\Pi_{T}(\bm{\theta})$ decays, $\xi_{t}(d) = \bm{1}\{D_{t} = d\}$, and $\lambda(\theta^{*},\theta;y) = \ln\{p\left(y \middle | \theta^{*}\right)/p\left(y \middle | \theta\right)\}$ is the log-likelihood ratio, which quantifies the amount of information that $y$ provides in favor of $\theta^{*}$ over $\theta$.

Our goal is to maximize the probability of simultaneous correct selection in all contexts. The final selection $\bigcup_{c\in\mathcal{C}}\hat{\mathcal{P}}_{c}$ can be modeled as a Bayesian decision based on the posterior belief $\Pi_{T}$. We define the loss function $\ell(\bigcup_{c\in\mathcal{C}}\hat{\mathcal{P}}_{c}, \bigcup_{c\in\mathcal{C}}\mathcal{P}_{c}) = \max_{c\in\mathcal{C}}\bm{1}\{\hat{\mathcal{P}}_{c} \neq \mathcal{P}_{c}\}$, which equals to one as long as the top-$m_{c}$ designs are incorrectly selected under any context. To make a Bayesian selection decision that minimizes the expected loss, we consider minimizing the posterior risk $\mathbb{E}[\ell(\bigcup_{c\in\mathcal{C}}\hat{\mathcal{P}}_{c}, \bigcup_{c\in\mathcal{C}}\mathcal{P}_{c}) \vert \mathcal{E}_{T}]$. Explicitly, the Bayesian selection decision can be derived as:
\begin{equation}\label{eq:2}
    \bigcup_{c\in\mathcal{C}}\hat{\mathcal{P}}_{c}^{B} := \argmax_{\bigcup_{c\in\mathcal{C}}\hat{\mathcal{P}}_{c}: \vert \hat{\mathcal{P}}_{c} \vert = m_{c}}\Pi_{T}\left(\bigcap_{c\in\mathcal{C}}\bm{\Theta}^{(c,\hat{\mathcal{P}}_{c})}\right),
\end{equation}
where $\bm{\Theta}^{(c,\hat{\mathcal{P}}_{c})} = \{\bm{\theta}\in\bm{\Theta}: \mu_{d} \geq \mu_{d^{\prime}},\;\forall d\in\hat{\mathcal{P}}_{c}, \;d^{\prime}\in \mathcal{D}_{c}\backslash \hat{\mathcal{P}}_{c}\}$ denotes the set of parameters in favor of designs in $\hat{\mathcal{P}}_{c}$ under context $c$, and the superscript $B$ denotes the Bayesian decision. 

While it would be desirable to maximize, if possible, the objective in (\ref{eq:2}) for the correct selection $\mathcal{P}_{c}$, i.e., $\Pi_{T}(\bigcap\nolimits_{c\in\mathcal{C}}\bm{\Theta}^{(c,\mathcal{P}_{c})})$, to achieve a Bayesian decision that aligns with the correct selection, this is impracticable due to the dependence of the posterior belief on random samples. Therefore, we propose to maximize an asymptotic proxy, namely the posterior large deviations rate, which we will discuss in Section \ref{sec:asymptotics}.

\section{Top-two Thompson sampling.}\label{sec:alg}
In this section, we propose the top-two Thompson sampling algorithm for contextual selection problems (TTTS-C). The algorithm effectively allocates the simulation budget, similar to TS, while also controls the effort dedicated to the top-$m_{c}$ designs a posteriori. This control allows for thorough exploration and ensures the algorithm performs well. To provide a clear understanding of how TTTS-C allocates simulation samples to each context and design, we focus on the case with $m_{c} = 1, ~\forall c\in\mathcal{C}$ until the end of Section \ref{sec:asymptotics}. We discuss the more general case in Section \ref{sec:topm}. Specifically, in this case, the preferable set of each context consists of a single design characterized by $\mathcal{P}_{c} = \{d^{*}(c)\}$ and the remaining designs are represented by $\mathcal{U}_{c} = \mathcal{D}_{c}\backslash\{d^{*}(c)\}$. Here, $d^{*}(c):= \arg\max_{d\in\mathcal{D}_{c}}{\mu_{d}}$ denotes the design in $\mathcal{D}_{c}$ that maximizes $\mu_{d}$, and consequently, the selection policy can be expressed as a mapping $\hat{d}: \mathcal{C} \rightarrow \mathcal{D}$, where $\hat{d}(c)\in\mathcal{D}_{c}$ and $\hat{\mathcal{P}}_{c} = \{\hat{d}(c)\}$.

\begin{algorithm}[t]
\caption{Top-Two Thompson Sampling for Selecting Contextual Best Designs} \label{alg:1}
\begin{algorithmic}
\State \textbf{Input:} $\bm{\gamma}=(\gamma(c))\in (0,1)^{\mathcal{C}}$ and posterior distribution $\Pi_{t}$
\State Sample $\hat{\bm{\theta}}^{(t)}_{1}$ from $\Pi_{t}$
\State $\hat{d}^{(t)}_{1}(c) \gets \arg\max_{d\in\mathcal{D}_{c}}(\hat{\mu}_{1}^{(t)})_{d}$, $\forall~c\in\mathcal{C}$ \Comment{Simulate the most possible decisions}
\Repeat 
\State Sample $\hat{\bm{\theta}}^{(t)}_{2}$ from $\Pi_{t}$
\State $\hat{d}^{(t)}_{2}(c) \gets \arg\max_{d\in\mathcal{D}_{c}}(\hat{\mu}_{2}^{(t)})_{d}$, $\forall~c\in\mathcal{C}$ \Comment{Simulate the second most possible decisions}
\State $\Delta_{t} \gets \{c\in\mathcal{C}: \hat{d}^{(t)}_{1}(c)\neq \hat{d}^{(t)}_{2}(c)\}$
\Until{$\Delta_{t}$ is not empty}
\State Randomly sample $C_{t}$ uniformly from $\Delta_{t}$ \Comment{Choose $C_{t}$ from the difference set}
\State Sample $B\sim \text{Bernoulli}(\gamma(C_{t}))$
\If{$B=1$}
\State $D_{t} \gets \hat{d}_{1}^{(t)}(C_{t})$ \Comment{Exploit the most probable best decision}
\Else
\State $D_{t} \gets \hat{d}_{2}^{(t)}(C_{t})$ \Comment{Explore the second most probable best decision}
\EndIf
\end{algorithmic}
\end{algorithm}

We begin by introducing a strategy parameterized by vector $\bm{\gamma} = (\gamma(c))$, where $\gamma(c)\in(0, 1)$, $\forall c\in\mathcal{C}$. At each step $t$, we sample $\hat{\bm{\theta}}^{(t)}_{1}$ from the posterior distribution $\Pi_{t-1}(\cdot)$ and determine the corresponding best design $\hat{d}^{(t)}_{1}(c) := \arg\max_{d\in\mathcal{D}_{c}}\hat{\mu}^{(t)}_{1,d}$ for each context $c\in\mathcal{C}$. Subsequently, we successively re-sample $\hat{\bm{\theta}}^{(t)}_{2}$ from $\Pi_{t-1}(\cdot)$ and find the best design $\hat{d}^{(t)}_{2}(c) := \arg\max_{d\in\mathcal{D}_{c}}\hat{\mu}^{(t)}_{2,d}$ for each context $c\in\mathcal{C}$, until $\hat{d}^{(t)}_{2}(\cdot)$ and $\hat{d}^{(t)}_{1}(\cdot)$ become unequal as a function of $c$. Let $\Delta_{t}$ denote the set of contexts where $\hat{d}^{(t)}_{1}(c)$ differs from $\hat{d}^{(t)}_{2}(c)$, formally defined as $\Delta_{t} := \{c\in\mathcal{C}: \hat{d}^{(t)}_{1}(c) \neq \hat{d}^{(t)}_{2}(c)\}$. Subsequently, the observed context $C_{t}$ is uniformly chosen from $\Delta_{t}$. The observed design $D_{t}$ is equal to $\hat{d}^{(t)}_{1}(C_{t})$ with probability $\gamma(C_{t})$, and with complementary probability, it is equal to $\hat{d}^{(t)}_{2}(C_{t})$. Hence, in the TTTS-C algorithm, $\psi_{t}(d) = \mathbb{E}[\zeta^{(t)}(d)\vert \mathcal{E}_{t-1}]$, where
$$\zeta^{(t)}(d) = \mathbb{E}[\bm{1}\{D_{t} = d\}\vert \mathcal{E}_{t-1}, \hat{d}^{(t)}_{1}(\cdot), \hat{d}^{(t)}_{2}(\cdot)] = \left\{\begin{array}{lcl}{\gamma(c)/\vert \Delta_{t}\vert} , &  & {d = \hat{d}^{(t)}_{1}(c)},\; \exists~c \in \Delta_{t}, \\ {(1 - \gamma(c))/\vert \Delta_{t}\vert} , & & {d = \hat{d}^{(t)}_{2}(c)}, \; \exists~c \in \Delta_{t}, \\ {0} , & & {\text{otherwise}.}\end{array}\right.$$

TTTS-C is a stationary randomized sequential sampling policy, meaning that the sampling policy $\psi_{t}$ is dependent on time $t$ only through the posterior belief. Additionally, the sampling decision is subject to an independent source of randomness introduced to ensure sufficient exploration for all contexts and designs. Through marginalizing $\hat{d}^{(t)}_{1}(\cdot)$ and $\hat{d}^{(t)}_{2}(\cdot)$, we obtain the sampling effort put into each design and context conditioned on $\mathcal{E}_{t-1}$ in Lemma \ref{lem:1}.

\begin{restatable}{lemma}{contextbestratio}
\label{lem:1}%
For any fixed $c\in \mathcal{C}$ and $d\in \mathcal{D}_{c}$, the sampling policy of TTTS-C at time $t$ is
\begin{equation}\label{eq:3}
\begin{aligned}
    &\psi_{t}(d) = \sum_{d_{1}(\cdot): d_{1}(c) = d} \prod_{c^{\prime}\in\mathcal{C}}{\left(\pi_{(c^{\prime},d_{1}(c^{\prime}))}^{(t-1)}\right)} \sum_{\color{black} c \in \Delta} \frac{\prod_{c^{\prime}\in\Delta}{\left(1 - \pi_{(c^{\prime},d_{1}(c^{\prime}))}^{(t-1)}\right)}\prod_{c^{\prime}\notin\Delta}{\left(\pi_{(c^{\prime},d_{1}(c^{\prime}))}^{(t-1)}\right)}}{1 - \prod_{c^{\prime}\in\mathcal{C}}{\left(\pi_{(c^{\prime},d_{1}(c^{\prime}))}^{(t-1)}\right)}} \frac{\gamma(c)}{\vert \Delta\vert} \\
    &+ \sum_{d_{1}(\cdot): d_{1}(c) \neq d} \prod_{c^{\prime}\in\mathcal{C}}{\left(\pi_{(c^{\prime},d_{1}(c^{\prime}))}^{(t-1)}\right)}\sum_{\color{black} c \in \Delta}\frac{\prod_{c^{\prime}\in\Delta}{\left(1 - \pi_{(c^{\prime},d_{1}(c^{\prime}))}^{(t-1)}\right)}\prod_{c^{\prime}\notin\Delta}{\left(\pi_{(c^{\prime},d_{1}(c^{\prime}))}^{(t-1)}\right)}}{1 - \prod_{c^{\prime}\in\mathcal{C}}{\left(\pi_{(c^{\prime},d_{1}(c^{\prime}))}^{(t-1)}\right)}}\cdot\frac{\pi^{(t-1)}_{(c, d)}}{\left(1 - \pi^{(t-1)}_{(c, d_{1}(c))}\right)} \frac{1 - \gamma(c)}{\vert\Delta\vert} ,
\end{aligned}
\end{equation}
\begin{equation}\label{eq:4}
    \alpha_{t}(c) = \sum_{d_{1}(\cdot)}\frac{\prod_{c^{\prime}\in\mathcal{C}}{\left(\pi_{(c^{\prime},d_{1}(c^{\prime}))}^{(t-1)}\right)}} {1 - \prod_{c^{\prime}\in\mathcal{C}}{\left(\pi_{(c^{\prime},d_{1}(c^{\prime}))}^{(t-1)}\right)}} \sum_{\color{black} c \in \Delta}\prod_{c^{\prime}\in\Delta}{\left(1 - \pi_{(c^{\prime},d_{1}(c^{\prime}))}^{(t-1)}\right)}\prod_{c^{\prime}\notin\Delta}{\left(\pi_{(c^{\prime},d_{1}(c^{\prime}))}^{(t-1)}\right)} \frac{1}{\vert \Delta\vert} ,
\end{equation}
\begin{equation}\label{eq:5}
    \beta_{t}(c, d) = \psi_{t}(d) / \alpha_{t}(c),
\end{equation}
where $\pi_{(c,d)}^{(t-1)}:=\Pi_{t-1}(\bm{\Theta}^{(c,\{d\})})$, $\forall c\in\mathcal{C}$, $d\in\mathcal{D}_{c}$. Therein, the summation $\sum_{c \in \Delta}$ is taken over $\Delta\in \mathscr{P}(\mathcal{C})$ that satisfies $c\in\Delta$, and $\mathscr{P}(\mathcal{C})$ is the power set of $\mathcal{C}$ consisting of all subsets of $\mathcal{C}$.
\end{restatable}

When $\vert \mathcal{C} \vert = 1$, TTTS-C reduces to TTTS for selecting the best design within a single context \citep{russo2020simple}. However, when $\vert \mathcal{C}\vert > 1$, the conditional sampling ratio employed by TTTS-C for different designs under a given context generally differ from the sampling ratios obtained by applying TTTS separately for each context, even if the posterior distributions are equitably set for both methods. This difference can be attributed to that the sampling decisions for contexts and designs are not independent. For example, consider a context $c$ and two designs $d$ and $d^{\prime}$, having conditional posterior probabilities of being the best design equal to 0.9 and 0.1, respectively. Moreover, suppose $\gamma(c) = 1$. In TTTS, the probability of sampling design $d$ would simply be 0.9. However, in Algorithm \ref{alg:1}, since, when comparing the scenario where $\hat{d}_{1}^{(t)} = d$, the context $c$ is more likely to be chosen when $\hat{d}_{1}^{(t)} = d^{\prime}$ than when $\hat{d}_{1}^{(t)} = d$, the conditional probability of $d$ being sampled under TTTS-C would be strictly less than 0.9. Therefore, the analysis for the sampling ratios of TTTS-C is more complicated than that of TTTS.

To shed some light on the (conditional) sampling ratios of TTTS-C under each context, we focus on a fixed context $c$ and a design $d\in\mathcal{D}_{c}$. We can decompose the summations in Equation (\ref{eq:3}) and (\ref{eq:4}) over functions $d_{1}:\mathcal{C} \rightarrow \mathcal{D}$ as $\sum_{d_{1}(\cdot)}{(\cdot)} = \sum_{d_{1}(-c)}\sum_{d_{1}(c)\in\mathcal{D}_{c}}{(\cdot)}$, where $d_{1}(-c)$ denotes that the first summation is taken over all possible values of the function $d_{1}$ evaluated at all contexts except for $c$. Given the value of $d_{1}(-c)$, the corresponding summands in (\ref{eq:3}) and (\ref{eq:4}) can be rewritten as 
\begin{equation}\label{eq:6}
     J^{(t-1)}_{d_{1}(-c)} = A^{(t-1)}_{d_{1}(-c)}\left( \gamma(c) \frac{\pi^{(t-1)}_{(c,d)}\left(1 - \pi^{(t-1)}_{(c,d)}\right)}{1 - B^{(t-1)}_{d_{1}(-c)}\cdot \pi^{(t-1)}_{(c,d)}} +  (1 - \gamma(c))\sum_{d_{1}(c)\neq d}\frac{\pi^{(t-1)}_{(c,d)}\pi^{(t-1)}_{(c,d_{1}(c))}}{1 - B^{(t-1)}_{d_{1}(-c)}\cdot \pi^{(t-1)}_{(c,d_{1}(c))}}\right)
\end{equation}
and
\begin{equation}\label{eq:7}
    K^{(t-1)}_{d_{1}(-c)} = A^{(t-1)}_{d_{1}(-c)}\left(\sum_{d_{1}(c)\in\mathcal{D}_{c}}\frac{\pi^{(t-1)}_{(c,d_{1}(c))}\left(1 - \pi^{(t-1)}_{(c,d_{1}(c))}\right)}{1 - B^{(t-1)}_{d_{1}(-c)}\cdot \pi^{(t-1)}_{(c,d_{1}(c))}} \right),
\end{equation}
respectively. Then $\psi_{t}(d) = \sum_{d_{1}(-c)}J^{(t-1)}_{d_{1}(-c)}$ and $\alpha_{t}(c) = \sum_{d_{1}(-c)}K^{(t-1)}_{d_{1}(-c)}$, where
\begin{equation*}
    A^{(t-1)}_{d_{1}(-c)} = \sum_{\Delta\in \mathscr{P}(\mathcal{C}\backslash\{c\}): c\in\Delta}B^{(t-1)}_{d_{1}(-c)}\cdot\prod_{c^{\prime}\in\Delta}{\left(1 - \pi_{(c^{\prime},d_{1}(c^{\prime}))}^{(t-1)}\right)}\prod_{c^{\prime}\notin\Delta}{\left(\pi_{(c^{\prime},d_{1}(c^{\prime}))}^{(t-1)}\right)}\frac{1}{\vert \Delta\vert},
\end{equation*}
and
\begin{equation*}
    B^{(t-1)}_{d_{1}(-c)} = \prod_{c^{\prime}\in\mathcal{C}\backslash\{c\}} {\left(\pi_{(c^{\prime},d_{1}(c^{\prime}))}^{(t-1)}\right)}.
\end{equation*}

Notice that $A^{(t-1)}_{d_{1}(-c)}$ and $B^{(t-1)}_{d_{1}(-c)}$ are irrelevant to the posterior distribution of parameters associated with context $c$. For $d\neq d^{*}(c)$, simple calculation yields that the ratio of (\ref{eq:6}) over (\ref{eq:7}) is asymptotically equivalent to $\pi^{(t-1)}_{(c, d)} / (1-\pi^{(t-1)}_{(c, d^{*}(c))})$. Thus, a sub-optimal design under each context os allocated a minimum conditional sampling effort proportional to its posterior probability of being the best design. On the other hand, the ratio of (\ref{eq:6}) over (\ref{eq:7}) can be lower and upper bounded by $\gamma(c)/(1+\delta_{t})$ and $\gamma(c)\cdot(1+\delta_{t})$, respectively, where $\delta_{t} = \mathcal{O}(1 / \pi^{(t-1)}_{(c,d)} - B^{(t-1)}_{d_{1}(-c)})$. In particular, for $d = d^{*}(c)$, if $B^{(t-1)}_{d_{1}(-c)}\rightarrow 1$ and $\pi^{(t-1)}_{(c,d^{*}(c))}\rightarrow 1$, then the ratio will approach $\gamma(c)$, hence so will $\beta_{t}(c, d^{*}(c))$. This suggests that the conditional sampling effort allocated to contextual best designs is approximately determined by the hyperparameter $\bm{\gamma}$. However, the above discussions are only a sketch of the main ideas and do not constitute a rigorous proof, as $\beta_{t}(c, d^{*}(c))$ is the ratio of summations of (\ref{eq:6}) over (\ref{eq:7}), and the weight $B^{(t-1)}_{d_{1}(-c)}$ does not tend to 1 for each term. The following lemma presents the formal result discussed above.

\begin{lemma}\label{lem:2}
    For any $c\in \mathcal{C}$ and $d\in \mathcal{D}_{c}$, the conditional sampling ratios of TTTS-C with hyperparameter $\bm{\gamma}$ satisfies:
    \begin{enumerate}
        \item for $d = d^{*}(c)$, we have
        \begin{equation*}
            \lim_{t\rightarrow \infty}\beta_{t}(c, d^{*}(c)) = \gamma(c),
        \end{equation*}
        \item for $d\neq d^{*}(c)$, if $\pi^{(t-1)}_{(c, d^{*}(c))} \geq 1/2$ for some $t\geq 1$, then
        \begin{equation*}
            \frac{1-\gamma(c)}{\vert\mathcal{D}_{c}\vert} \frac{\pi^{(t-1)}_{(c, d)}}{1-\pi^{(t-1)}_{(c, d^{*}(c))}} \leq \beta_{t}(c, d) \leq 2 \frac{\pi^{(t-1)}_{(c, d)}}{1-\pi^{(t-1)}_{(c, d^{*}(c))}}.
        \end{equation*}
    \end{enumerate}
\end{lemma}

Lemma 2 provides insights into the sampling policy of TTTS-C, which can be compared to that of TTTS proposed by \cite{russo2020simple}. In both approaches, samples are allocated in a given context based on the posterior probability of each design being optimal. However, TTTS-C stands out due to its unique capability to allocate simulation samples to both contexts and designs simultaneously. Similarly, by leveraging the results from Lemma \ref{lem:1}, we can also characterize the marginalized posterior probability associated with sampling each context.

\begin{lemma}\label{lem:3}
    For any $c\in \mathcal{C}$, under TTTS-C with hyperparameter $\bm{\gamma}>0$, with probability 1, for sufficiently large $t$, the following inequalities hold true:
    \begin{equation}\label{eq47}
        \frac{1}{2}\frac{1 - \pi^{(t-1)}_{(c, d^{*}(c))}}{1 - \prod_{c^{\prime}\in\mathcal{C}}\left(\pi^{(t-1)}_{(c^{\prime}, d^{*}(c^{\prime}))}\right)} \leq \alpha_{t}(c) \leq 2\frac{1 - \pi^{(t-1)}_{(c, d^{*}(c))}}{1 - \prod_{c^{\prime}\in\mathcal{C}}\left(\pi^{(t-1)}_{(c^{\prime}, d^{*}(c^{\prime}))}\right)}.
    \end{equation}
    Furthermore, there exists a random variable $\tilde{C}$ independent of time $T$, such that with probability 1,
    \begin{equation}\label{eq48}
        T\bar{\alpha}_{T}(c) \geq \frac{1}{2\Gamma^{*}}\ln{\left(\tilde{C} + \Gamma^{*}\cdot (T-1)\right)}.
    \end{equation}
\end{lemma}

Lemma \ref{lem:3} states that the sampling effort for each context, $\alpha_{t}(c)$, is asymptotically in proportion to the posterior probability of the corresponding contextual best design being non-optimal. This ensures that under TTTS-C, the sampling effort is focused on vague contexts wherein it is difficult to identify the best design based on the sample information. 

Following Lemma \ref{lem:3}, each context $c\in\mathcal{C}$ receives an increasing amount of cumulative sampling effort $T\bar{\alpha}_{T}(c)$, at least of logarithmic order, regardless of the conditional sampling efforts put into designs. There are two interesting implications of this result. For one thing, Lemma \ref{lem:3} indicates a way in which TTTS-C forces exploration by randomizing over contexts. Though a substantial amount of sampling effort is made for the vague contexts, the cumulative effort for each context is lower bounded. According to Levy's extension of the Borel-Cantelli lemma \citep{will91}, each context receives an infinite number of simulation samples as the simulation budget goes to infinity with probability one. For another, notice that the bounds in (\ref{eq47}) do not explicitly depend on the hyperparameter $\bm{\gamma}$, as (\ref{eq:4}) does not, and the bound in (\ref{eq48}) is independent of $\bm{\gamma}$. Actually, Lemma \ref{lem:3} applies to any sequential sampling policies that determine $C_{t}$ in the same way as TTTS-C and probably differ from TTTS-C in the tradeoff between exploitation (choosing $\hat{d}_{1}^{(t)}(C_{t})$) and exploration (choosing $\hat{d}_{2}^{(t)}(C_{t})$). (\ref{eq48}) provides a robust bound invariant to how one chooses between the top-two candidates to be sampled. This allows for a flexible and possibly time-inhomogenous choice of the hyperparameter $\bm{\gamma}$, for example, consistent tuning of $\bm{\gamma}$, which we will introduce in Section \ref{sec:var}.

Although TTTS-C involves sampling from a posterior distribution, we argue that its computational complexity grows sub-linearly in the number of possible contexts $\vert \mathcal{C} \vert$, resulting in negligible additional computation asymptotically compared to conducting TTTS separately for each context. Specifically, suppose that $\Pi_{t}(\bm{\Theta}_{(c,d^{*}(c))}) \leq 1-\delta$ for each context, where $\delta>0$ is a small positive number. Additionaly, assume that $t$ is sufficiently large such that with high probability, the first sample $\hat{\bm{\theta}}^{(t)}_{1}$ leads to $\hat{d}^{(t)}_{1}(\cdot) = d^{*}(\cdot)$. Then, the probability of stopping re-sampling from the posterior distribution in each attempt is approximately $1 - (1-\delta)^{\vert \mathcal{C}\vert}$, resulting in an expected number of repetitions of $1/\left(1-(1-\delta)^{\vert \mathcal{C}\vert}\right)\approx 1+(1-\delta)^{\vert \mathcal{C}\vert}$. As a consequence, the asymptotic computational complexity is of order $\mathcal{O}(\vert\mathcal{C}\vert + \vert\mathcal{C}\vert(1-\delta)^{\vert \mathcal{C}\vert})$. In comparison, conducting TTTS separately for each context and then choosing the context to be sampled require a computational complexity of order $\mathcal{O}(\vert\mathcal{C}\vert)$. Therefore, we can conclude that the additional computational effort made for choosing contexts in TTTS-C diminishes as the number of contexts grows.

\section{Asymptotics.}\label{sec:asymptotics}
Recall that the posterior probability $\Pi_{T}(\tilde{\bm{\Theta}})$ of a parameter set $\tilde{\bm{\Theta}}$ can be expressed as an integral of the likelihood function $L_{T}(\theta) = L_{T}(\theta^{*})\exp\{-TW_{T}(\theta)\}$, where $W_{T}(\theta)$ is a weighted summation of log-likelihood ratios $\lambda(\theta^{*}_{d}, \theta_{d}; Y_{d}^{(t)})$. It is intuitive that $W_{T}$ converges to $\mathbb{E}[\lambda(\theta^{*}_{d}, \theta_{d}; Y_{d}^{(t)})]$ as $T\rightarrow\infty$ under certain regularity conditions. As a result, $\Pi_{T}(\tilde{\bm{\Theta}})$ can be shown to decay exponentially. In this section, we formalize this idea and establish posterior large deviation rates (pLDR) for general parametric families. It provides an asymptotic upper bound for the objective in (\ref{eq:2}), which characterizes an optimal convergence rate that can be achieved by sequential sampling policies. Then, TTTS-C for selecting contextual best designs is shown to be asymptotically optimal, i.e., achieving this upper bound asymptotically, if the hyperparameter $\bm{\gamma}$ is correctly specified or consistently adjusted.

\subsection{Posterior large deviations rates.}\label{sec:ldr}
Denote $D(\theta^{*}\Vert \theta):= \mathbb{E}_{Y\sim p(\cdot\vert \theta^{*})}[\lambda(\theta^{*},\theta; Y)]$ as the Kullback–Leibler (KL) divergence between two probability measures in $\{p(\cdot\vert \theta): \theta\in \Theta\}$ with parameters $\theta^{*}$ and $\theta$. The overall KL divergence, denoted by $D_{{\bar{\psi}_{T}}}(\bm{\theta}^{*}\Vert\bm{\theta})$, is defined as the sum of the KL divergences weighted by $\bar{\psi}_{T}(d) = \sum_{t=1}^{T}{\psi_{t}(d)}/T$, i.e., $D_{{\bar{\psi}_{T}}}(\bm{\theta}^{*}\Vert\bm{\theta}) := \sum_{c\in\mathcal{C},d\in\mathcal{D}_{c}}\bar{\psi}_{T}(d)D(\theta^{*}_{d}\Vert \theta_{d})$. Additionally, we define $\bar{\alpha}_{T}(c) = \sum_{d^{\prime}\in \mathcal{D}_{c}}\bar{\psi}_{T}(d^{\prime})$ and $\bar{\beta}_{T}(c, d) = \bar{\psi}_{T}(d) / \bar{\alpha}_{T}(c)$ for $c\in\mathcal{C}$ and $d\in\mathcal{D}_{c}$. The partial KL divergence for $c\in\mathcal{C}$, denoted as $D_{\bar{\beta}_{T}(c,\cdot)}(\bm{\theta}^{*}\Vert\bm{\theta})$, is defined as the sum of KL divergences for each $d\in\mathcal{D}_{c}$ weighted by $\bar{\beta}_{T}(c, d)$, i.e., $D_{\bar{\beta}_{T}(c,\cdot)}(\bm{\theta}^{*}\Vert\bm{\theta}):= \sum_{d\in\mathcal{D}_{c}}\bar{\beta}_{T}(c, d)D(\theta^{*}_{d}\Vert \theta_{d})$. We make a regularity assumption on the KL divergence.

\begin{assumption}\label{ass:2}
The KL divergence $D(\theta^{*}\Vert \theta)$ is well defined and continuously differentiable with respect to $\theta$.
\end{assumption}

We then establish the pLDR for general parametric families.

\begin{theorem}\label{thm:1} Under Assumptions \ref{ass:1} and \ref{ass:2}, for any sequential sampling policy $\{\psi_{t}\}_{t\leq T}$ and open set $\tilde{\bm{\Theta}}\subseteq\bm{\Theta}$, we have 
\begin{equation*}
    \lim_{T\rightarrow \infty}{ -\frac{1}{T}\ln{\Pi_{T}(\Tilde{\bm{\Theta}})} - \inf_{\bm{\theta}\in\tilde{\bm{\Theta}}}{D_{\bar{\psi}_{T}}(\bm{\theta}^{*}\Vert\bm{\theta})} } = 0, \quad a.s..
\end{equation*}
Furthermore, suppose $\tilde{\bm{\Theta}}$ is defined by coordinates corresponding to context $c$ for some $c\in\mathcal{C}$, i.e., $\tilde{\bm{\Theta}} = \prod_{c^{\prime}\in\mathcal{C}\backslash\{c\}}(\Theta)^{\vert D_{c^{\prime}} \vert} \times \tilde{\Theta}_{c}$, where $\tilde{\Theta}_{c}$ is an open set in $(\Theta)^{\vert D_{c}\vert}$. Then, with probability 1, if $T\bar{\alpha}_{T}(c) \rightarrow \infty$, we have
\begin{equation*}
    \lim_{T\rightarrow \infty}{ -\frac{1}{T\bar{\alpha}_{T}(c)}\ln{\Pi_{T}(\Tilde{\bm{\Theta}})} - \inf_{\bm{\theta}\in\tilde{\bm{\Theta}}}{D_{\bar{\beta}_{T}(c,\cdot)}(\bm{\theta}^{*}\Vert\bm{\theta})} } = 0.
\end{equation*}
\end{theorem}

Theorem \ref{thm:1} implies that the posterior belief on any open set $\tilde{\bm{\Theta}}$, bounded away from $\bm{\theta}^{*}$, decays exponentially with a rate depending on the distance between $\bm{\theta}^{*}$ and the closest point to it in $\tilde{\bm{\Theta}}$. When $\vert \mathcal{C}\vert = 1$, this result coincides with that of \citet{russo2020simple} for the one-parameter exponential family of sampling distributions. The proof of Theorem \ref{thm:1} first establishes the uniform convergence of the logarithm of the posterior density to the KL divergence by leveraging the GC class assumption. The uniformity of convergence is natural but not trivial, as the posterior density marginalizes all unobserved simulation samples and is equal to the product of expected likelihood functions of observed simulation samples. Notice that the GC class assumption does not directly imply the uniform convergence of the partial average of any subsequence of likelihood ratio tests, and we formalize this idea using independent copies. Based on this, we establish the exponential convergence of the posterior probability, which is an integral of the posterior density, using a Laplace approximation.

The second statement in Theorem \ref{thm:1} extends the main result by analyzing the posterior belief of the designs within a single context. The convergence rate of this marginalized posterior belief associated with context $c$ only depends asymptotically on the empirical conditional sampling ratios $\bar{\beta}_{T}(c, \cdot)$, which plays a key role in justifying the asymptotic optimality of TTTS-C. Notice that neither of the two statements can be directly deduced from the other unless $\vert \mathcal{C}\vert = 1$. The first statement does not impose any requirements on the sampling ratios put into each context, while the second statement holds only for contexts that ultimately receive infinite sampling effort. Additionally, the second statement focuses on a margin of the posterior distribution associated with a context, and the KL divergence is defined in terms of the conditional sampling ratios. When $\liminf_{T\rightarrow \infty}\bar{\alpha}_{T}(c) > 0$, the second statement follows immediately from the first statement. However, it takes more careful treatment solely under the assumption that $T\bar{\alpha}_{T}(c) \rightarrow \infty$.

\subsection{Characterizing the convergence rate.}
Theorem \ref{thm:1} provides an asymptotic approximation to (\ref{eq:2}), which is impossible to maximize due to the dependence of the posterior belief on random samples. Recall that our focus remains on selecting the contextual best designs wherein $\mathcal{P}_{c} = \{d^{*}(c)\}$. We note that $\left(1 - \Pi_{T}\left(\bigcap_{c\in\mathcal{C}}\bm{\Theta}^{(c,\{d^{*}(c)\})}\right)\right)$ is equal to $\Pi_{T}\left(\bigcup_{c\in \mathcal{C}}\bigcup_{d\in\mathcal{U}_{c}}\bm{\Theta}_{(c,d)}\right)$, where $\bm{\Theta}_{(c,d)} = \{\bm{\theta}\in\bm{\Theta}: \mu_{d} \geq \mu_{d^{*}(c)}\}$, $\forall~c\in\mathcal{C}$, $d\in \mathcal{U}_{c}$. The following result characterizes the rate at which $\left(1 - \Pi_{T}\left(\bigcap_{c\in\mathcal{C}}\bm{\Theta}^{(c,\{d^{*}(c)\})}\right)\right)$ converges.

\begin{corollary}\label{corol:1} Under Assumptions \ref{ass:1} and \ref{ass:2}, for any sequential sampling policy $\{\psi_{t}\}_{t\leq T}$, the posterior risk of selection decision $d^{*}(\cdot)$ satisfies:
\begin{equation}\label{eq:8}
    \lim_{T\rightarrow \infty}-\frac{1}{T}\ln\left(1 - \Pi_{T}\left(\bigcap_{c\in \mathcal{C}}\bm{\Theta}^{(c,\{d^{*}(c)\})}\right)\right) - \min_{c\in\mathcal{C}}\min_{d\in\mathcal{D}_{c}\backslash \{d^{*}(c)\}}G_{d}(\bar{\psi}_{T}(d^{*}(c)), \bar{\psi}_{T}(d)) = 0, \quad a.s.,
\end{equation}
where
\begin{align*}
    G_{d}(\psi(d^{*}(c)), \psi(d)) &:=\inf_{\bm{\theta}\in \bm{\Theta}_{(c,d)}}D_{\psi}(\bm{\theta}^{*}\Vert \bm{\theta}) = \inf_{\mu_{d} \geq \mu_{d^{*}(c)}}\sum_{c^{\prime}\in\mathcal{C}}\sum_{d^{\prime}\in\mathcal{D}_{c^{\prime}}}\psi(d^{\prime}) D(\theta_{d^{\prime}}^{*}\Vert \theta_{d^{\prime}}) \\
    &= \inf_{\mu_{d} \geq \mu_{d^{*}(c)}} \psi(d) D(\theta_{d}^{*}\Vert \theta_{d}) + \psi(d^{*}(c)) D(\theta_{d^{*}(c)}^{*}\Vert \theta_{d^{*}(c)}).
\end{align*}
\end{corollary}
\proof{Proof.}
It follows immediately from Theorem \ref{thm:1} and the two-sided bound of the posterior belief
$$\max_{c\in \mathcal{C}}\max_{d\in \mathcal{D}_{c}\backslash \{d^{*}(c)\}}\Pi_{T}\left(\bm{\Theta}_{(c,d)}\right) \leq \Pi_{T}\left(\bigcup\nolimits_{c\in \mathcal{C}}\bigcup\nolimits_{d\in\mathcal{U}_{c}}\bm{\Theta}_{(c,d)}\right) \leq \sum_{c\in\mathcal{C}}\left\vert \mathcal{D}_{c}\right\vert \cdot \max_{c\in \mathcal{C}}\max_{d\in \mathcal{D}_{c}\backslash \{d^{*}(c)\}}\Pi_{T}\left(\bm{\Theta}_{(c,d)}\right).$$ \halmos
\endproof

Following Corollary \ref{corol:1}, the quantity $\left(1 - \Pi_{T}\left(\bigcap_{c\in\mathcal{C}}\bm{\Theta}^{(c,\{d^{*}(c)\})}\right)\right)$ decays at an exponential rate that depends on the sampling policies only through the sampling ratios $\bar{\psi}_{T}$ asymptotically. The bivariate function $G_{d}(\cdot, \cdot)$, referred to as the posterior large deviations rate function, captures the complexity of differentiating design $d$ with $d^{*}(c)$ under context $c$. Equation (\ref{eq:8}) implies that the decay rate is determined by the minimum of rate functions, corresponding to the most challenging design to differentiate. Notice that the quantity $\left(1 - \Pi_{T}\left(\bigcap_{c\in\mathcal{C}}\bm{\Theta}^{(c,\{d^{*}(c)\})}\right)\right)$ differs from the frequentist PICS for the contextual R\&S problems \citep{du2022rate,zhang2023efficient} in two ways. First, the former posterior probability does not rely on any selection policy and represents the posterior risk by selecting the truly contextual best designs based on observed samples. Second, $\left(1 - \Pi_{T}\left(\bigcap_{c\in \mathcal{C}}\bm{\Theta}^{(c,\{d^{*}(c)\})}\right)\right)$ is defined a posteriori, reflecting the updated information after observing the simulation samples. In contrast, the PICS is defined a priori, implying an expected level of performance of sequential sampling policies. Despite these differences, Equation (\ref{eq:8}) implies that $\left(1 - \Pi_{T}\left(\bigcap_{c\in\mathcal{C}}\bm{\Theta}^{(c,\{d^{*}(c)\})}\right)\right)$ decreases at a similar rate as PICS.

For a general parametric family of identifiable distributions, the explicit form and properties of $G_{d}(\cdot, \cdot)$ remain unknown. However, in several widely used parametric families such as Gaussian, Exponential, and Bernoulli distributions, the rate functions can be derived explicitly. Besides, in the frequentist framework, the large deviations rate function for sample means can be calculated using cumulant generating functions, as discussed in \citet{dembo92} on large deviations theory. Unlike some existing R\&S procedures that are derived based on an analytical form of the rate functions \citep{chen2022balancing,du2022rate}, implementing TTTS-C does not require knowledge of the rate functions. For TTTS-C, the rate function is used solely to assist with the asymptotic analysis. To characterize the generic properties of the rate functions $G_{d}(\cdot, \cdot)$, we state the following proposition:
\begin{proposition}\label{prop:1}
For any $c\in\mathcal{C}$ and $d\in\mathcal{D}_{c}$, the large deviations rate function $G_{d}(\cdot, \cdot)$ possesses the following properties:
\begin{enumerate}[(a)]
    \item It is monotonically non-decreasing in both arguments.
    \item It is concave as a bivariate function, and, therefore,  uniformly continuous on any compact subset of $\left(\mathbb{R}^{+}\right)^{2}$.
    \item It exhibits positive homogeneity of degree 1, i.e., $G_{d}(hx, hy) = hG_{d}(x,y), ~\forall h, x, y\geq 0$.
\end{enumerate}
\end{proposition}

\begin{remark}
    Notice that $G_{d}(\cdot, \cdot)$ does not necessarily possess strict monotonicity and strict concavity due to our distribution assumptions, although these strong properties can be justified for certain models. \citet{russo2020simple} identifies the rate function associated with one-parameter exponential families whose sufficient statistic is monotonically increasing with respect to the sample values and proves its strict monotonicity. However, for an arbitrary parameterized family of distributions satisfying Assumptions \ref{ass:1} and \ref{ass:2}, strict monotonicity can be violated. For the literature regarding the PICS, \citet{glynn2004large} show that the large deviations rate associated with the vector of sample means strictly increases in sampling ratios, and \citet{zhang2021asymptotically} prove the strict monotonicity and strict concavity of their large deviations rate function.
\end{remark}

Having discussed the properties of the rate functions in Proposition \ref{prop:1}, we now delve into the exploration of the optimal rate achievable by any sequential sampling strategy. To do so, we consider a static optimization problem that captures the rate functions as follows:
\begin{align}
    \Gamma^{*}:= \max_{\psi\geq 0}&\quad \min_{c\in\mathcal{C}}\min_{d\in\mathcal{U}_{c}}\inf_{\bm{\theta}\in \bm{\Theta}_{(c,d)}}{D_{\psi}(\bm{\theta}^{*}\Vert \bm{\theta})} \label{opt:1}\\
    s.t.& \quad \sum\nolimits_{c\in\mathcal{C}}\sum\nolimits_{d\in \mathcal{U}_{c}}\psi(d) = 1. \notag
\end{align}

According to (\ref{eq:8}), $\Gamma^{*}$ is an asymptotic upper bound on the exponential rate of convergence of the posterior belief under any sequential sampling policy. Consequently, a sequential sampling policy is called asymptotically optimal if its sampling ratios satisfy satisfies the following limit:
\begin{equation*}
    \lim_{T\rightarrow \infty}\min_{c\in\mathcal{C}}\min_{d\in\mathcal{U}_{c}}G_{d}(\bar{\psi}_{T}(d^{*}(c)), \bar{\psi}_{T}(d)) = \Gamma^{*}
\end{equation*}

\begin{remark}
    The failure of strong properties of the rate functions complicates our analysis. In \citep{russo2020simple}, the strict monotonicity of large deviations rate function plays an intermediate role in proving the asymptotic optimality of TTTS. However, in our work, without the strict monotonicity, the uniqueness of the optimal solution to (\ref{opt:1}) cannot be justified. In order to show the asymptotic optimality of TTTS-C, we avoid the need for strict monotonicity of the large deviations rate function (or the uniqueness of the optimal solution to (\ref{opt:1})) by focusing on showing the optimality conditions rather than tracking the unique optimal solution. 
\end{remark}

Following the homogeneity property of the large deviations rate function, the static optimization problem (\ref{opt:1}) can be rewritten as:
\begin{align}
    \Gamma^{*} = \max_{\alpha,\beta\geq 0}&\quad \min_{c\in\mathcal{C}}\alpha(c)\min_{d\in\mathcal{U}_{c}}G_{d}\left(\beta(c,d^{*}(c)), \beta(c, d)\right) \label{opt:2}\\
    s.t.& \quad \sum_{c\in\mathcal{C}}{\alpha(c)} = 1, \quad  \sum_{d\in \mathcal{D}_{c}}{\beta(c,d)} = 1,\quad \forall c\in\mathcal{C}. \notag
\end{align}

With the objective being the minimum of several concave functions, the optimization problem (\ref{opt:2}) is a concave problem. Notably, the concave nature of this optimization problem satisfies Slater's condition. Therefore, the Karush-Kuhn-Tucker (KKT) conditions provide necessary and sufficient conditions for the optimal solutions to problem (\ref{opt:2}). We summarize these conditions in the following proposition:
\begin{proposition}\label{prop:2}
    A feasible solution $\bm{\alpha},\bm{\beta} \geq 0$ to (\ref{opt:2}) is optimal if and only if the following conditions hold:
    \begin{gather}
        \sum_{d\in\mathcal{U}_{c}}{\frac{\partial G_{d}(\beta(c,d^{*}(c)), \beta(c,d))/\partial x_{1}}{\partial G_{d}(\beta(c,d^{*}(c)), \beta(c,d))/\partial x_{2}}} = 1,\quad \forall c\in\mathcal{C}, \label{eq:9}\\
        \alpha(c)G_{d}(\beta(c,d^{*}(c)), \beta(c,d)) = z, \quad \forall c\in\mathcal{C}, d\in\mathcal{U}_{c}, ~\exists z\in\mathbb{R}^{+},\label{eq:10}
    \end{gather}
    where, for all $c\in\mathcal{C}$ and $d\in\mathcal{U}_{c}$, the vector
    \begin{equation*}
        \left(-\frac{\partial}{\partial x_{1}}G_{d}(\beta(c, d^{*}(c)), \beta(c, d)), -\frac{\partial}{\partial x_{2}}G_{d}(\beta(c, d^{*}(c)), \beta(c, d))\right)
    \end{equation*}
    represents a subgradient of $-G_{d}(\cdot, \cdot)$ evaluated at $(\beta(c, d^{*}(c)), \beta(c, d))$.
\end{proposition}

Equation (\ref{eq:9}), derived from the first-order condition of the KKT conditions, plays a crucial role in balancing the trade-off between exploiting the best designs and exploring other designs. It ensures that the allocation of the simulation budget is carefully distributed to achieve an optimal trade-off. On the other hand, Equation (\ref{eq:10}) establishes a balance condition between the allocation of the simulation budget for sub-optimal designs within a single context and those from different contexts. Rather than allocating an equal amount of effort to each sub-optimal design, the optimal allocation guarantees that they are equally differentiated from the best designs, considering the large deviations rate. Notably, at the optimal solution, the value $z$ in Equation (\ref{eq:10}) corresponds to the optimal value $\Gamma^{*}$ obtained from (\ref{opt:2}). This indicates that the balance conditions are achieved, and the asymptotically optimal performance is attained. Note that (\ref{eq:9}) and (\ref{eq:10}) coincide with the asymptotic optimality conditions derived in \citet{du2022rate} by eliminating $z$, despite the undesired fact that $G_{d}(\cdot, \cdot)$'s are not necessarily strictly increasing or strictly concave under our distribution assumptions. 

Although (\ref{eq:9}) and (\ref{eq:10}) are sufficient and necessary conditions for the optimal solution to the optimization problem (\ref{opt:2}), they do not guarantee the uniqueness of the solution to (\ref{opt:2}). In Section \ref{sec:unique} of the online appendix, we provide a straightforward counterexample to demonstrate this lack of uniqueness. Actually, we can show that the strict concavity of rate functions is a crucial condition for the uniqueness of optimal solution, which holds true even in the case where $m_c > 1$ and is discussed in Section \ref{sec:topm}. However, we emphasize that the non-uniqueness of optimal solutions does not hinder our asymptotic analysis of TTTS-C.

According to Lemma \ref{lem:2}, TTTS-C allocates a fraction $\gamma(c)$ of the simulation budget to measure $d^{*}(c)$ under context $c$ in the long run. Therefore, its empirical conditional sampling ratios $\bar{\alpha}_{T}$ and $\bar{\beta}_{T}$ of TTTS-C may not exactly achieve the optimal solutions if $\gamma(c) \neq \beta^{*}(c, d^{*}(c))$, where $(\alpha^{*}, \beta^{*})$ represents an optimal solution to (\ref{opt:2}) that satisfies (\ref{eq:9}) and (\ref{eq:10}). In view of this, we decompose the unconstrained problem (\ref{opt:2}) into two subproblems where the first-layer problem is $\Gamma^{*} = \max_{\bm{\gamma}\geq 0}\Gamma^{*}_{\bm{\gamma}}$ and the second-layer problem is a constrained problem:
\begin{align}
    \Gamma_{\bm{\gamma}}^{*} = \max_{\alpha, \beta \geq 0}& \quad \min_{c\in\mathcal{C}}\alpha(c)\min_{d\in\mathcal{U}_{c}}{ G_{d}(\gamma(c), \beta{(c,d)})} \label{opt:3}\\
    s.t.& \quad \sum_{c\in\mathcal{C}}{\alpha(c)} = 1, \quad  \sum_{d\in \mathcal{U}_{c}}{\beta(c,d)} = 1 - \gamma(c),\quad \forall c\in\mathcal{C}. \notag
\end{align}

By imposing constraints on the conditional sampling ratios of contextual best designs, the second-layer problem (\ref{opt:3}) can be easily solved by balancing the large deviations rate functions. The following proposition shows that the optimal conditions of (\ref{opt:3}) reduce to a balance condition similar to (\ref{eq:10}).

\begin{proposition}\label{prop:3}
   A feasible solution $\bm{\alpha},\bm{\beta} \geq 0$ to (\ref{opt:3}) is optimal if 
    \begin{gather}
        \alpha(c)G_{d}(\gamma(c), \beta(c,d)) = z, \quad \forall c\in\mathcal{C}, d\in\mathcal{U}_{c}, ~\exists z\in\mathbb{R}^{+}.\label{eq:11}
    \end{gather}
    And, (\ref{eq:11}) can be attained at some optimal solution to (\ref{opt:3}). Conversely, if $\gamma(c) = \beta^{*}(c, d^{*}(c))$ for some optimal solution $\bm{\alpha}^{*}, \bm{\beta^{*}}$ to (\ref{opt:2}), then (\ref{eq:11}) is also a necessary condition for optimality.
\end{proposition}

Proposition \ref{prop:3} encompasses three statements that provide insights from different perspectives. In contrast to Proposition \ref{prop:2}, condition (\ref{eq:11}) alone is sufficient to yield optimality of (\ref{opt:3}). This could be attributed to the additional constraints in (\ref{opt:3}) that restrict the sampling ratios of contextual best designs. Furthermore, there always exists an optimal solution to (\ref{opt:3}) that satisfies the balance condition (\ref{eq:11}), even in the presence of non-uniqueness in the optimal solution. This supports the validity of TTTS-C, which asymptotically achieves (\ref{eq:11}). Surprisingly, we emphasize that condition (\ref{eq:11}) may not be a necessity for optimality except when $\gamma(c) = \beta^{*}(c, d^{*}(c))$. For instance, let's consider a context-free case with three alternatives $\mathcal{D} = \{1, 2, 3\}$ and $\mathcal{P} = \{1\}$. Suppose the rate functions are given by $G_{2}(\beta_{1}, \beta_{2}) = 2/(1/\beta_{1} + 1/\beta_{2})$ and $G_{3}(\beta_{1}, \beta_{3}) = \min\{\beta_{1}, \beta_{3}\}$, respectively. If we fix $\beta_{1} = \gamma = 0.1$, the optimal solution to (\ref{opt:3}) can be any feasible solution with $\beta_{2}\geq 0.1$ and $\beta_{3}\geq 0.1$. Specifically, the optimal condition (\ref{eq:11}) is attained when $\beta_{2} = 0.1$ and $\beta_{3} = 0.8$, and it is violated if $\beta_{2} > 0.1$. This example illustrates the possibility of satisfying the optimality condition without it being a necessary requirement.

The second-layer problem (\ref{opt:3}) can be further simplified by exploiting its separability property with respect to $\beta(c, d)$ for different contexts. Specifically, for any $c\in\mathcal{C}$, we define a subproblem as follows:
\begin{align}
    \Gamma_{\gamma(c)}^{*} = \max_{\beta(c, \cdot) \geq 0}& \quad \min_{d\in\mathcal{U}_{c}}{ G_{d}(\gamma(c), \beta{(c,d)})} \label{opt:4}\\
    s.t.& \quad \sum_{d\in \mathcal{U}_{c}}{\beta(c,d)} = 1 - \gamma(c). \notag
\end{align}
By solving each subproblem independently, we obtain $\Gamma_{\gamma(c)}^{*}$, and and $\Gamma_{\bm{\gamma}}^{*}$ can be expressed as an optimization problem $\Gamma_{\bm{\gamma}}^{*} = \max_{\alpha\geq 0}\min_{c\in\mathcal{C}}\alpha(c)\Gamma^{*}_{\gamma(c)}\; s.t.\; \sum_{c\in\mathcal{C}}\alpha(c) = 1$, whose optimal solution can be derived as follows:
\begin{equation}\label{eq63}
    \alpha^{*}_{\bm{\gamma}}(c) := \frac{1 / \Gamma^{*}_{\gamma(c)}}{\sum_{c^{\prime}\in\mathcal{C}}{1/\Gamma^{*}_{\gamma(c^{\prime})}}},
\end{equation}
where the denominator in (\ref{eq63}) represents a normalization term{, and t}he optimal value is given by
$$\Gamma^{*}_{\bm{\gamma}} = \frac{1}{\sum_{c^{\prime}\in\mathcal{C}}{1/\Gamma^{*}_{\gamma(c^{\prime})}}},$$
which is monotonically increasing with respect to each $\Gamma^{*}_{\gamma(c^{\prime})}$. That is, the optimal rate for correctly identifying all of the contextual best designs is positively related to the optimal rate for correctly identifying the best design merely in context $c$, for any $c\in\mathcal{C}$. Finally, the sample allocation problem (\ref{opt:1}) can be solved by optimizing $\beta(c, d)$ for problem (\ref{opt:3}), plugging in the values of $\Gamma^{*}_{\gamma(c)}$ obtained from (\ref{eq63}) to determine $\alpha(c)$, and then separately optimizing $\Gamma^{*}_{\bm{\gamma}}$ for each coordinate $\gamma(c)$.

\subsection{Asymptotic optimality of TTTS-C.}
To close this section, we introduce the asymptotic optimality of TTTS-C for selecting contextual best designs.
\begin{restatable}{theorem}{rateoptone}
\label{thm:3}%
    Under TTTS-C with hyperparameter $\bm{\gamma} > 0$ in Algorithm \ref{alg:1}, the balance condition (\ref{eq:10}) is asymptotically satisfied, i.e., there exists $z\in\mathbb{R}^{+}$, such that for all $c\in\mathcal{C}$, $d\in\mathcal{U}_{c}$, we have
    \begin{equation*}
        \lim_{T\rightarrow \infty}\bar{\alpha}_{T}(c)G_{d}(\gamma(c), \bar{\beta}_{T}(c,d)) = z.
    \end{equation*}
\end{restatable}
 
This establishes that the optimality condition is asymptotically satisfied by empirical sampling ratios, namely $\bar{\alpha}_{T}$ and $\bar{\beta}_{T}$, for the optimization problem (\ref{opt:3}). However, it should be noted that Theorem \ref{thm:3} does not necessarily imply convergence of the empirical sampling ratios to an optimal solution of (\ref{opt:3}), as the optimal solution for $\beta$ in (\ref{opt:3}) may not be unique. Nevertheless, we can demonstrate that the posterior large deviations rate achieves its optimal value as follows.

\begin{restatable}[Asymptotic Quasi-Optimality]{corollary}{rateopt}
\label{thm:euclid}%
Under TTTS-C with hyperparameter $\bm{\gamma} > 0$ in Algorithm \ref{alg:1}, 
\begin{equation*}
    \lim_{T\rightarrow \infty} -\frac{1}{T}\ln \Pi_{T}\left( \bigcup_{c\in\mathcal{C}}\bigcup_{d\in\mathcal{D}\backslash\{d^{*}(c)\}}\bm{\Theta}_{(c, d)} \right) = \Gamma_{\bm{\gamma}}^{*}.
\end{equation*}
Under any sequential sampling policy with $\lim_{T\rightarrow \infty}\bar{\beta}_{T}(c, d^{*}(c)) = \gamma(c)$,
\begin{equation*}
    \limsup_{T\rightarrow \infty} -\frac{1}{T}\ln \Pi_{T}\left( \bigcup_{c\in\mathcal{C}}\bigcup_{d\in\mathcal{D}\backslash\{d^{*}(c)\}}\bm{\Theta}_{(c, d)} \right) \leq \Gamma_{\bm{\gamma}}^{*}.
\end{equation*}
\end{restatable}

By asymptotic quasi-optimality, we mean that TTTS-C performs at least as well as any sequential sampling policy in an asymptotic sense, provided that these sequential sampling policies allocate a fraction of samples, $\gamma(c)$, to the true best design $d^{*}(c)$ in each context $c$, as measured by the posterior large deviations rate. Specially, if hyperparameter $\bm{\gamma}$ is appropriately chosen to maximize $\Gamma^{*}_{\bm{\gamma}}$, the large deviations rate can attain its maximum value $\Gamma^{*}$. In this way, the proposed TTTS-C achieves true asymptotic optimality. This result extends the results of Theorem 1 of \citet{russo2020simple} in two ways. First, TTTS-C is capable of simultaneously allocating samples for both contexts and designs. Corollary \ref{corol:2} implies that not only do the conditional sampling ratios for each design approach their optimal allocations, but also the aggregate sampling ratios for each context tend towards their optimal values. Second, the proposed sampling policy is distinguished from existing ones by enabling its applicability to a wider range of sampling distribution families. This flexibility is a notable advantage as it allows TTTS-C to be used in various scenarios where the strict monotonicity assumption may not hold. 

In the following, we provide a high-level proof outline for Theorem \ref{thm:3}. The proof consists of three steps. The first step establishes the consistency of TTTS-C, which is summarized in Theorem \ref{thm:5} below. A direct consequence is that under TTTS-C, the Bayesian decision $\hat{d}_{B}(\cdot)$ is a consistent estimation of $d^{*}(\cdot)$ if $\vert\mu_{d} - \mu_{d^{\prime}}\vert > 0$ for $d\neq d^{\prime}$. Moreover, it is demonstrated that for each $c\in\mathcal{C}$ and $d\in\mathcal{U}_{c}$, we have $\pi^{(T)}_{(c, d^{*}(c))} \rightarrow 1$ and $\pi^{(T)}_{(c, d)} \rightarrow 0$, which lead to the asymptotic bounds of sampling effort for $\alpha_{t}, \beta_{t}$ as shown in Lemmas \ref{lem:2} and \ref{lem:3}. 

\begin{restatable}[Consistency]{theorem}{consistent}\label{thm:5}
    For any hyperparameter $\bm{\gamma} > 0$, the proposed TTTS-C is consistent, i.e., ${\sum_{t=1}^{T}\xi_{t}(d)}\rightarrow \infty$, $a.s.$, as $T\rightarrow \infty$, $\forall~c\in\mathcal{C}$, $\forall~d\in\mathcal{D}_{c}$.
\end{restatable}

Theorem \ref{thm:5} relies on the identifiability of the parameter model, without the need to specify the distribution family of the simulation samples.

The second step is to show the balance condition for a fixed context $c\in\mathcal{C}$, which states that $\lim_{T\rightarrow \infty}G_{d}(\gamma(c), \bar{\beta}_{T}(c, d)) = z_{c}, \forall d\in\mathcal{U}_{c}$ for some $z_{c}\in\mathbb{R}$. Actually, we can show that $z_{c} = \Gamma^{*}_{\gamma(c)}$, indicating that the empirical conditional sampling effort $\bar{\beta}_{T}(c, \cdot)$ asymptotically maximizes (\ref{opt:4}). To prove this, we argue by contradiction, taking into account the approximation of the sampling policy provided by Lemma \ref{lem:2} and Theorem \ref{thm:1}:
\begin{equation}\label{expoequ}
\begin{aligned}
    \beta_{T}(c, d) & \leq 2 \frac{\pi_{(c, d)}^{(T-1)}}{1 - \pi_{(c, d^{*}(c)}^{(T-1)}} \\
    & \approx 2\exp\bigg\{-(T-1)\Bar{\alpha}_{T-1}(c)\bigg(G_{d}(\gamma(c), \bar{\beta}_{T-1}(c, d)) - \min_{d^{\prime} \in \mathcal{U}_{c}}G_{d^{\prime}}(\gamma(c), \bar{\beta}_{T-1}(c, d^{\prime}))\bigg)\bigg\}.
\end{aligned}
\end{equation}

When a design receives more samples than optimal solution, the parenthesized term in the exponent of (\ref{expoequ}) will be positive, i.e., $G_{d} - \min\nolimits_{d^{\prime} \in \mathcal{U}_{c}}G_{d^{\prime}} > 0$, since $G_{d}(\cdot, \cdot)$ is increasing in both arguments. This leads to a negligible sampling effort since $(T-1)\bar{\alpha}_{T-1}(c)$ diverges. Assuming, for the sake of contradiction, that the parenthesized term converges to a positive number $\delta>0$, we have $\beta_{T}(c, d)\leq 2\exp\{-(T-1)\bar{\alpha}_{T-1}(c)\delta/2\}$ for $T$ sufficiently large. However, when combined with the Abel transformation, this implies $\bar{\beta}_{T}(c, d) = \sum_{t=1}^{T}\alpha_{t}(c)\beta_{t}(c, d) / \sum_{t=1}^{T}\alpha_{t}(c) \rightarrow 0$, contradicting the assumption, as $G_{d}(\cdot, \cdot)$ becomes zero when any of its arguments equals zero. Our proof rigorously addresses the convergence of the parenthesized term in the electronic companion.

The last step completes the proof by showing that the sampling ratio $\bar{\alpha}_{T}$ for each context balances the optimality condition (\ref{eq:10}). This can be justified by repeating the argument used in the second step with Lemma (\ref{lem:3}). Specifically, we can show that $\lim_{T\rightarrow \infty}\bar{\alpha}_{T}(c) = \alpha^{*}_{\bm{\gamma}}(c)$. Therefore, we have $$\lim_{T\rightarrow \infty}\bar{\alpha}_{T}(c)G_{d}(\bar{\beta}_{T-1}(c, d^{*}(c)), \bar{\beta}_{T-1}(c, d)) = \alpha^{*}_{\bm{\gamma}} \Gamma^{*}_{\gamma(c)} = \Gamma^{*}_{\bm{\gamma}}.$$ 
This completes the proof for the optimality condition.

\section{Extensions of TTTS-C.}\label{sec:var}
In this section, we discuss two important extensions of TTTS-C. First, we modify TTTS-C for the general contextual top-$m_{c}$ selection problem, where $m_{c}\geq 1$. We can justify a similar balance condition, although it may not be sufficient to yield asymptotic optimality. Thereafter, we consider tuning the hyperparameter in the proposed sampling policies to further enhance the asymptotic efficiency. With the hyperparameter adjusted towards its optimal value, the posterior large deviations rate of the proposed sampling policies approaches the optimal value $\Gamma^{*}$ asymptotically.

\subsection{Selecting contextual top-$m_{c}$ designs.}\label{sec:topm}
For general contextual top-$m_{c}$ selection problems defined in Section \ref{sec:probdef} where $m_{c} \ge 1$, the preferable set $\mathcal{P}_{c}$ is not necessarily a singleton for each $c\in\mathcal{C}$. Analogous to Corollary \ref{corol:1}, the following result characterizes the posterior convergence for identifying the top-$m_{c}$ designs under each context.
\begin{corollary}\label{corol:2} Under Assumptions \ref{ass:1} and \ref{ass:2}, for any sequential sampling policy $\{\psi_{t}\}_{t\leq T}$, the posterior risk of selection decision $\bigcup_{c\in\mathcal{C}}\mathcal{P}_{c}$ satisfies
\begin{equation}\label{eq:13}
    \lim_{T\rightarrow \infty}-\frac{1}{T}\ln\left(1 - \Pi_{T}\left(\bigcap_{c\in \mathcal{C}}\bm{\Theta}^{(c,\mathcal{P}_{c})}\right)\right) - \min_{c\in\mathcal{C}}\min_{d\in\mathcal{P}_{c}}\min_{d^{\prime}\in\mathcal{U}_{c}}G_{d,d^{\prime}}(\bar{\psi}_{T}(d^{*}(c)), \bar{\psi}_{T}(d)) = 0, \quad a.s.,
\end{equation}
where $G_{d, d^{\prime}}(\psi(d), \psi(d^{\prime})) := \inf_{\mu_{d^{\prime}} \geq \mu_{d}} \psi(d) D(\theta^{*}_{d}\Vert \theta_{d}) + \psi(d^{\prime}) D(\theta_{d^{\prime}}^{*}\Vert \theta_{d^{\prime}})$ is the rate function.
\end{corollary}

The proof is a simple tautology of Corollary \ref{corol:1}. Then, the rate optimization problem associated with the sample allocation decisions can be formulated as the following static concave problem:
\begin{align}
    \Gamma^{*} = \max_{\alpha,\beta\geq 0}&\quad \min_{c\in\mathcal{C}}\alpha(c)\min_{d\in\mathcal{P}_{c}}\min_{d^{\prime}\in\mathcal{U}_{c}}G_{d,d^{\prime}}\left(\beta(c, d), \beta(c, d^{\prime})\right) \label{opt:5}\\
    s.t.& \quad \sum_{c\in\mathcal{C}}{\alpha(c)} = 1, \quad  \sum_{d\in \mathcal{D}_{c}}{\beta(c,d)} = 1,\quad \forall c\in\mathcal{C}. \notag
\end{align}
We repeatedly use the symbol $\Gamma^{*}$ for the optimal value, as the optimization problem reduces to (\ref{opt:2}) when $\mathcal{P}_{c}=\{d^{*}(c)\}$ is a singleton.

Regarding this problem, the KKT conditions can be derived similarly. However, sufficient optimality conditions are generally intractable for an arbitrarily parameterized sampling distribution without introducing dual variables. While a set of balance conditions can be derived for Gaussian sampling distributions, the number of candidates for optimal solutions typically grows combinatorially as $\min\{\vert\mathcal{P}{c}\vert, \vert\mathcal{U}{c}\vert\}$ increases. An example of additional optimality conditions under Gaussian assumptions can be found in Section \ref{sec:gausscondition} of the electronic companion. Nevertheless, the following proposition establishes the uniqueness of optimal solutions by assuming the strict concavity of the rate function. This assumption is to some extent sharp since the violation of this assumption may lead to multiple optimal solutions, as illustrated by an example in Section \ref{sec:unique} of the electronic companion.

\begin{proposition}\label{prop:4}
Assuming that the rate function $G_{d,d^{\prime}}(\cdot, \cdot)$ is strictly concave as a bivariate function, for all $c\in\mathcal{C}$, $d\in\mathcal{P}_{c}$ and $d^{\prime}\in\mathcal{U}_{c}$, the solution to the problem (\ref{opt:5}) is unique.
\end{proposition}

Proposition \ref{prop:4} solves an open question in the work of \citep{zhang2021asymptotically}, which study an optimization problem that optimizes the frequentist large deviations rate function under Gaussian sampling distributions and leave the uniqueness of their optimization problem as an open question. In the following, we focus on a necessary condition for optimality of (\ref{opt:5}).

\begin{proposition}\label{prop:5}
    A feasible solution $\alpha,\beta\geq 0$ to (\ref{opt:5}) is optimal only if there exists $z\in\mathbb{R}^{+}$ such that for all $c\in\mathcal{C}$, $d\in\mathcal{P}_{c}$, and $d^{\prime}\in\mathcal{U}_{c}$,
    \begin{equation}\label{eq:12}
        \min_{\tilde{d}\in\mathcal{P}_{c}}\alpha(c) G_{\tilde{d}, d^{\prime}}(\beta(c, \tilde{d}), \beta(c, d^{\prime})) = \min_{\tilde{d}^{\prime}\in\mathcal{U}_{c}} \alpha(c) G_{d, \tilde{d}^{\prime}}(\beta(c, d), \beta(c, \tilde{d}^{\prime})) = z.
    \end{equation}
\end{proposition}

Unlike selecting the contextual best designs, the balance condition does not necessarily require that all of the bivariate rate functions take the same value, unless $\vert \mathcal{P}_{c}\vert = 1$ or $\vert\mathcal{U}_{c} \vert = 1$. Instead, for each $d\in\mathcal{P}_{c}$, the rate at which design $d$ is distinguished from the set $\mathcal{U}_{c}$ is approximately the same, i.e.,
\begin{equation*}
    -\frac{1}{T}\ln \Pi_{T}\left(\bigcup_{\tilde{d}^{\prime}\in\mathcal{U}_{c}}\{\bm{\theta}\in\bm{\Theta}: \mu_{d} < \mu_{\tilde{d}^{\prime}}\}\right) \approx \min_{\tilde{d}^{\prime}\in\mathcal{U}_{c}}\alpha(c)G_{d, \tilde{d}^{\prime}}(\beta(c, d), \beta(c, \tilde{d}^{\prime})) = z.
\end{equation*}

We can naturally extend TTTS-C proposed in Algorithm \ref{alg:1} to asymptotically achieve this necessary optimality condition with a hyperparameter $\bm{\gamma}$, where $\gamma(c)$ now restricts the sum of conditional sampling ratios put into the preferable set $\mathcal{P}_{c}$ of designs instead of being solely allocated to the best design $\{d^{*}(c)\}$. This extension is described in Algorithms \ref{alg:2} and \ref{alg:3}. The following theorem implies that the sampling ratios induced by TTTS-C asymptotically solve the balance condition.

\begin{restatable}{theorem}{rateopttwo}
    \label{thm:4}
    Under TTTS-C with hyperparameter $\bm{\gamma} > 0$ in Algorithms \ref{alg:2} and \ref{alg:3}, the balance condition (\ref{eq:12}) is asymptotically satisfied, i.e., there exists a sequence $\{z_{T}\}_{T\geq 1}$ with $z_{T}\in\mathbb{R}^{+}$, such that for all $c\in\mathcal{C}$, $d\in\mathcal{P}_{c}$, $d^{\prime}\in\mathcal{U}_{c}$,
    \begin{equation*}
        \lim_{T\rightarrow \infty}\min_{\tilde{d}\in\mathcal{P}_{c}}\bar{\alpha}_{T}(c) G_{\tilde{d}, d^{\prime}}(\bar{\beta}_{T}(c, \tilde{d}), \bar{\beta}_{T}(c, d^{\prime})) - z_{T} = \lim_{T\rightarrow \infty}\min_{\tilde{d}^{\prime}\in\mathcal{U}_{c}} \bar{\alpha}_{T}(c) G_{d, \tilde{d}^{\prime}}(\bar{\beta}_{T}(c, d), \bar{\beta}_{T}(c, \tilde{d}^{\prime})) - z_{T} = 0.
    \end{equation*}
\end{restatable}

We provide some discussions regarding the non-uniqueness of the solution to Equation (\ref{eq:12}). Typically, there exists a continuum of solutions, especially when $\vert \mathcal{P}{c}\vert \gg 1$ and $\vert \mathcal{U}{c}\vert \gg 1$, as there are significantly more variables than equations. Consequently, $z$ does not necessarily coincide with $\Gamma^{*}$, and the asymptotic sampling ratios of TTTS-C may not be asymptotically optimal as in Corollary \ref{corol:2}. Nevertheless, TTTS-C remains an ideal algorithm for top-$m_{c}$ selection under a contextual setting. Although further necessary optimality conditions could be formalized for certain sampling distributions, they are mathematically intractable for most parametric families of distributions. As discussed in Section \ref{sec:gausscondition} of the electronic companion, these additional conditions do not yield sufficiency for optimality, and there still remains a combinatorial number of candidates for the optimal allocation. Moreover, we note that TTTS-C, by tracking the balance condition in Proposition \ref{prop:3}, which admits multiple solutions, empirically performs well compared to existing sampling policies. This could be attributed to the sufficient exploration of contexts under TTTS-C.

\begin{algorithm}
\caption{Extension of TTTS-C for Selecting Contextual Top-$m_{c}$ Designs.}
\label{alg:2}
\begin{algorithmic}
\State \textbf{Input:} $m_{c}\geq 1$ and posterior distribution $\Pi_{t}$
\State Sample $\hat{\bm{\theta}}^{(t)}_{1}$ from $\Pi_{t}$
\State $\hat{\mathcal{P}}^{(t)}_{c,1} \gets \arg\max^{m_{c}}_{d\in\mathcal{D}_{c}}(\hat{\mu}_{1}^{(t)})_{d}$, $\forall~c\in\mathcal{C}$ \Comment{$\arg\max^{m_{c}}$ returns the largest $m_{c}$ candidates}
\Repeat 
\State Sample $\hat{\bm{\theta}}^{(t)}_{2}$ from $\Pi_{t}$
\State $\hat{\mathcal{P}}^{(t)}_{c,2} \gets \arg\max^{m_{c}}_{d\in\mathcal{D}_{c}}(\hat{\mu}_{2}^{(t)})_{d}$, $\forall~c\in\mathcal{C}$ 
\State $\Delta_{t} \gets \{c\in\mathcal{C}: \hat{\mathcal{P}}^{(t)}_{c,1}\neq \hat{\mathcal{P}}^{(t)}_{c,2}\}$ \Comment{`$\neq$' denotes set-wise inequality }
\Until{$\Delta_{t}\neq \phi$}
\State Randomly sample $C_{t}$ uniformly from $\Delta_{t}$ 
\State $\tilde{D}_{t} \gets \operatorname{Unif}(\hat{\mathcal{P}}^{(t)}_{c,1} \backslash  \hat{\mathcal{P}}^{(t)}_{c,2})$ \Comment{`$\backslash$' denotes set difference}
\State $\bar{D}_{t} \gets \operatorname{Unif}(\hat{\mathcal{P}}^{(t)}_{c,2} \backslash  \hat{\mathcal{P}}^{(t)}_{c,1})$ \Comment{Uniformly sample $\bar{D}_{t}$ from the difference set}
\State Decide $D_{t}$ from candidates $\tilde{D}_{t}$ and $\bar{D}_{t}$ by Algorithm \ref{alg:3} (or \ref{alg:4})
\end{algorithmic}
\end{algorithm}

\begin{algorithm}
\caption{Sample Allocation for Top-Two Candidates with Hyperparameter $\bm{\gamma}$}
\label{alg:3}
\begin{algorithmic}
\State \textbf{Input:} $\bm{\gamma}=(\gamma(c))\in (0,1)^{\mathcal{C}}$, context $C_{t}$, candidates $\tilde{D}_{t}$ and $\bar{D}_{t}$
\State Sample $B \sim \operatorname{Bernoulli}(\gamma(C_{t}))$
\State $D_{t} \gets B \tilde{D}_{t} + (1 - B) \bar{D}_{t}$
\end{algorithmic}
\end{algorithm}

\subsection{Tuning of hyperparameter.}
TTTS-C requires a smart choice of the hyperparameter $\bm{\gamma}$. The optimal choice of $\bm{\gamma}$, which allows TTTS-C to be asymptotically optimal, depends on the underlying distribution parameters and is unknown a priori. A natural choice for the parameter is a predetermined constant that ensures fair allocation for all instances. When $\left| \mathcal{C} \right| = 1$ and $m_{c} = 1$, \citet{russo2020simple} suggests using $\gamma(c)=1/2$ and shows that it leads to a near-optimal convergence rate, at least half of the optimal rate. Our numerical experiments demonstrate that such a choice also performs robustly well in contextual R\&S problems.

In this work, we propose a modification of TTTS-C by allowing $\bm{\gamma}$ to be learned sequentially as simulation samples are gathered. Algorithm \ref{alg:4} presents the pseudo-code of the TTTS-C algorithm with tuning hyperparameter $\bm{\gamma}$. At each step $t$, the hyperparameter $\bm{\gamma}$ is updated based on the estimated optimal conditional sampling ratios for contextual best designs if a user-specified updating rule $\mathcal{R}$ is satisfied. Specifically, $\mathcal{R}$ is a sequence of finite and increasing stopping times that adapt to $\{\mathcal{E}_{t}\}_{t\geq 1}$ when $\bm{\gamma}$ is tuned. For example, updating $\bm{\gamma}$ at a predetermined sequence of time steps suffices to deliver a consistent estimation of the best hyperparameter $\bm{\gamma}^{*}$, leading to a posterior large deviations rate approaches $\Gamma^{*}$. The asymptotic optimality of TTTS-C with tuning parameter $\bm{\gamma}$ is established in Section \ref{sec:rateoptproof} of the electronic companion.

\begin{algorithm}
\begin{algorithmic}
\caption{Sample Allocation for Top-Two Candidates with tuning hyperparameter}
\label{alg:4}
\State \textbf{Input}: Current hyperparameter $\bm{\gamma}$, updating rule $\mathcal{R}$, context $C_{t}$, candidates $\tilde{D}_{t}$ and $\bar{D}_{t}$
\If{$\mathcal{R}$ is satisfied}
\State Consistently estimate $\hat{\bm{\theta}}$
\State Plug in $\hat{\bm{\theta}}$ to estimate $\hat{G}_{d, d^{\prime}}$ and the preferable (undesired) set $\hat{\mathcal{P}}_{c}$ ($\hat{\mathcal{U}}_{c}$) for $c\in\mathcal{C}$
\State Solve optimal allocation $\hat{\alpha}$ and $\hat{\beta}$ with $G_{d, d^{\prime}}=\hat{G}_{d, d^{\prime}}$, $\mathcal{P}_{c} = \hat{\mathcal{P}}_{c}$ and $\mathcal{U}_{c} = \hat{\mathcal{U}}_{c}$
\State Update $\gamma(c)$ with $\sum_{d\in\hat{\mathcal{P}}_{c}}\hat{\beta}(c, d)$
\EndIf
\State Sample $B \sim \operatorname{Bernoulli}(\gamma(C_{t}))$
\State $D_{t} \gets B \tilde{D}_{t} + (1 - B) \bar{D}_{t}$
\end{algorithmic}
\end{algorithm}

Updating the hyperparameter $\bm{\gamma}$ involves searching for the optimal static allocation $\hat{\alpha}$ and $\hat{\beta}$ to (\ref{opt:5}) for the instance with estimated parameter $\hat{\bm{\theta}}$. Recall that $\hat{\alpha}(c) = \left(\hat{\Gamma}^{*}_{\gamma(c)}\right)^{-1} \Big/ \sum_{c^{\prime}\in \mathcal{C}}{\left( \hat{\Gamma}^{*}_{\gamma(c^{\prime})}\right)^{-1}}$, where $\hat{\Gamma}^{*}_{\gamma(c)}$ is defined analogous to (\ref{opt:3}) as the optimal value of the optimization problem with the objective $\min_{d\in\hat{\mathcal{P}}_{c}}\min_{d^{\prime}\in\hat{\mathcal{U}}_{c}}\hat{G}_{d, d^{\prime}}(\beta(c, d), \beta(c, d^{\prime}))$ and the constraint $\sum_{d^{\prime}\in\hat{\mathcal{U}}_{c}}\beta(c, d^{\prime}) = 1-\gamma(c)$. Here, $\hat{G}_{d, d^{\prime}}(\cdot,\cdot)$, $\hat{\mathcal{P}}_{c}$, and $\hat{\mathcal{U}}_{c}$ are plug-in estimations of $G_{d, d^{\prime}}$, $\mathcal{P}_{c}$, and $\mathcal{U}_{c}$ using $\hat{\bm{\theta}}$. Then it remains to solve this optimization problem numerically for each context $c\in\mathcal{C}$. Specially, for Gaussian sampling distributions with known variance, the optimal allocation problem can be reduced to a one-dimensional convex program utilizing balance condition (\ref{eq:11}) \cite{peng12share}.

\section{Numerical experiments.}\label{sec:numexp}
In this section, we numerically test the sample allocation and finite sample performance of TTTS-C under both Gaussian sampling distributions and non-Gaussian sampling distributions. We consider two synthetic settings corresponding to Example 1 and Example 2 in Section \ref{sec:probdef}, respectively. We test TTTS-C with equal allocation (EA), balancing optimal large deviations for contextual top-$m_{c}$ selection ({BOLDmc}) and asymptotically optimal allocation policy for contextual top-$m_{c}$ selection (AOAmc). 

EA equally allocates the simulation budget to each context and equally to each design within a context. BOLDmc and AOAmc are designed for Gaussian sampling distributions with an unknown mean $\mu_{d}$ and known variance $\sigma_{d}^{2}$ for $d\in\mathcal{D}$. BOLDmc is designed to maximize $\operatorname{PCS} := \mathbb{P} (\bigcup_{c\in\mathcal{C}}\hat{\mathcal{P}}_{c} = \bigcup_{c\in\mathcal{C}}\mathcal{P}_{c} )$, $\operatorname{PCSW} := \min_{c\in\mathcal{C}}\mathbb{P} (\hat{\mathcal{P}}_{c} = \mathcal{P}_{c} )$, and $ \operatorname{PCSE} := \sum_{c\in\mathcal{C}}w_{c}\mathbb{P} (\hat{\mathcal{P}}_{c} = \mathcal{P}_{c} )$, where $w_{c}$ is a weighting coefficient indicating the relative importance of contexts, whereas AOAmc is developed solely to maximize $\operatorname{PCSW}$. BOLDmc is a compared sampling policy proposed in \citet{zhang2023efficient}, which extends \citet{du2022rate} to the contextual top-$m_{c}$ selection problem. At each step, BOLDmc allocates one simulation sample to context $c_{T}\in\mathcal{C}$ and one of the two candidate designs $d_{T}\in\mathcal{P}_{c_{T}}, d_{T}^{\prime}\in\mathcal{U}_{c_{T}}$ according to
\begin{equation}
    (c_{T}, d_{T}, d_{T}^{\prime})  = \argmin_{c\in\mathcal{C}, d\in\mathcal{P}_{c}, d^{\prime}\in\mathcal{U}_{c} } \frac{(\hat{\mu}_{d} - \hat{\mu}_{d^{\prime}})^{2}}{\hat{\sigma}_{d}^{2} / N_{T-1}(d) + \hat{\sigma}_{d^{\prime}}^{2} / N_{T-1}(d^{\prime})}, \label{eq:14}
\end{equation}
where $\hat{\mu}_{d}$ and $\hat{\sigma}_{d}^{2}$ denote the sample mean and variance of design $d$, and $N_{T}(d) = \sum_{t=1}^{T}\bm{1}\{D_{t} = d\}$ represents the number of simulation samples received by $d$ up to time step $T$. Then, $C_{T} = c_{T}$ and $D_{T} = d_{T}$ if $$\sum_{d\in\mathcal{P}_{c_{T}}} \frac{\bar{\psi}^{2}_{T-1}(d)}{\hat{\sigma}_{d}^{2}} < \sum_{d^{\prime}\in\mathcal{U}_{c_{T}}} \frac{\bar{\psi}^{2}_{T-1}(d^{\prime})}{\hat{\sigma}_{d^{\prime}}^{2}};$$ 
otherwise, $D_{T} = d^{\prime}_{T}$. AOAmc is a sampling policy proposed in \citet{zhang2023efficient}, which has a similar form to BOLDmc in determining the candidates $(c_{T}, d_{T}, d_{T}^{\prime})$, but with the replacement of $\hat{\mu}_{d}$ and $\hat{\sigma}_{d}^{2}$ by the posterior expectation of the mean parameter $\mu^{*}_{d}$ of interest and posterior variance, respectively. At each step, AOAmc allocates a simulation sample to $C_{T} = c_{T}$ and $D_{T} = d_{T}$ if
\begin{equation*}
    \min_{d\in\mathcal{P}_{c}, d^{\prime}\in\mathcal{U}_{c}}\frac{(\hat{\mu}_{d} - \hat{\mu}_{d^{\prime}})^{2}}{\hat{\sigma}_{d}^{2} / N_{T}^{d_{T}}(d) + \hat{\sigma}_{d^{\prime}}^{2} / N_{T}^{d_{T}}(d^{\prime})} > \min_{d\in\mathcal{P}_{c}, d^{\prime}\in\mathcal{U}_{c}} \frac{(\hat{\mu}_{d} - \hat{\mu}_{d^{\prime}})^{2}}{\hat{\sigma}_{d}^{2} / N_{T}^{d^{\prime}_{T}}(d) + \hat{\sigma}_{d^{\prime}}^{2} / N_{T}^{d^{\prime}_{T}}(d^{\prime})},
\end{equation*}
where $N_{T}^{\tilde{d}}(d) = N_{T-1}(d) + \bm{1}\{d = \tilde{d}\}$; otherwise, $D_{T} = d^{\prime}_{T}$. Notice that BOLDmc chooses between the two candidate designs $d_{T}\in\mathcal{P}_{c_{T}}$ and $d_{T}^{\prime}\in\mathcal{U}_{c_{T}}$ by tracking the first-order condition of the KKT conditions, whereas AOAmc maximizes the one-step look ahead large deviations rate.

In each numerical example, we use PCS, PCSW, and PCSE as performance metrics for each sampling policy. For all numerical examples, we set $w_{c} = 1 / \vert \mathcal{C} \vert$. Although the three metrics are shown to converge to one at the same exponential rate \citep{du2022rate}, PCS is more conservative than the others, as it only favors the event where the preferable set is estimated correctly as a whole.

\subsection{Gaussian designs.}
\begin{figure}
    \centering
    \includegraphics[width = 0.325\linewidth]{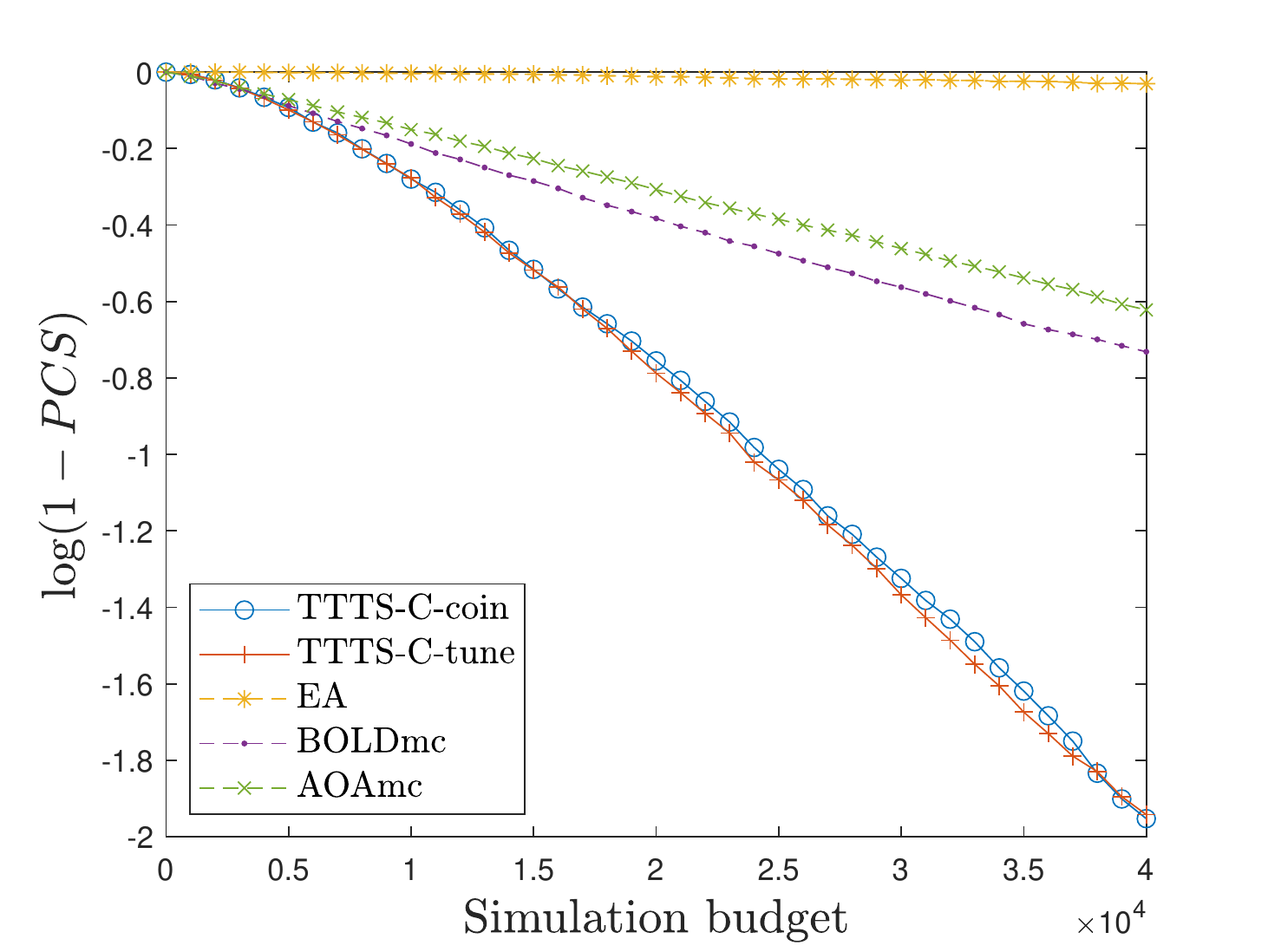}
    \includegraphics[width = 0.325\linewidth]{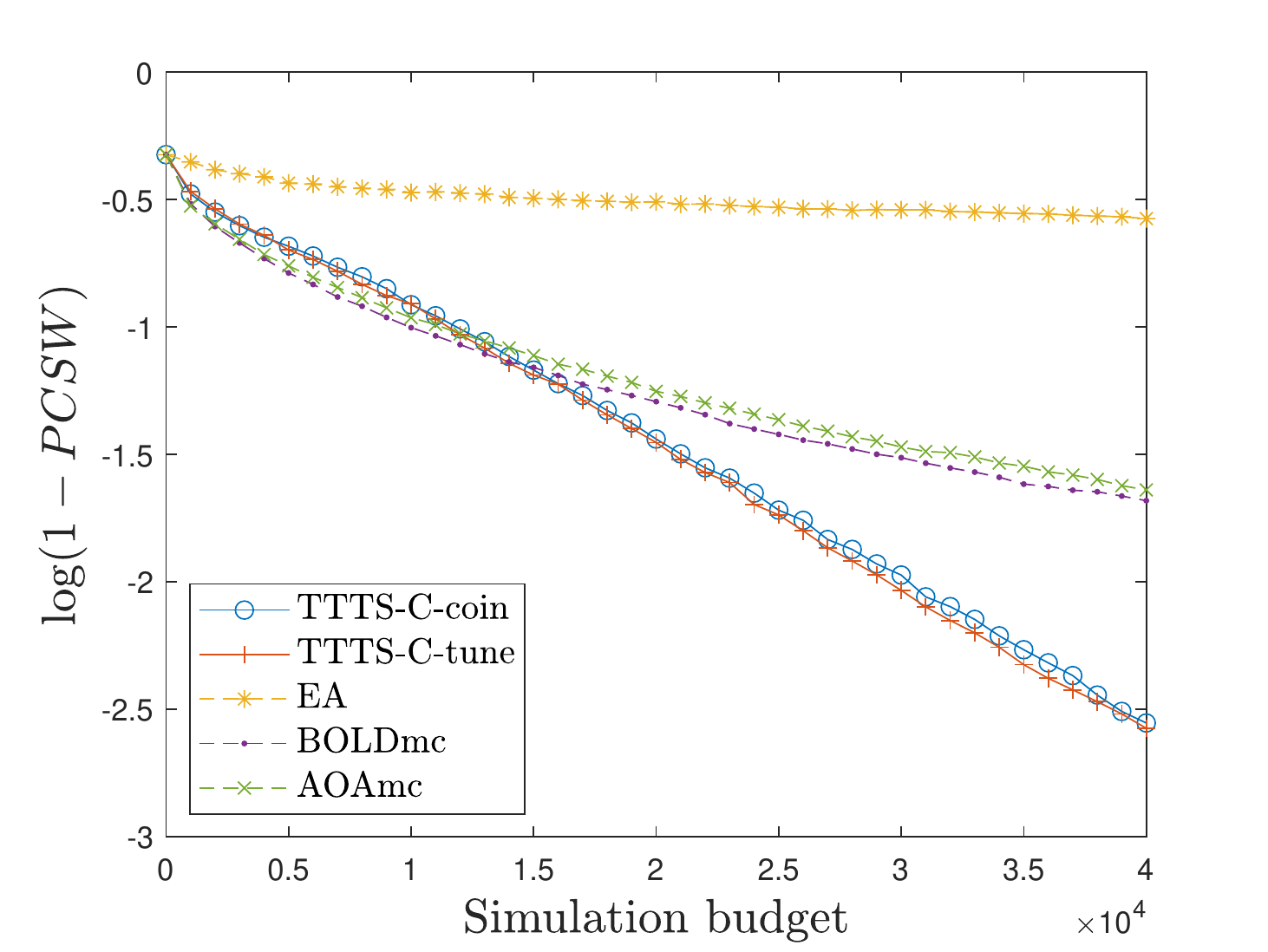}
    \includegraphics[width = 0.325\linewidth]{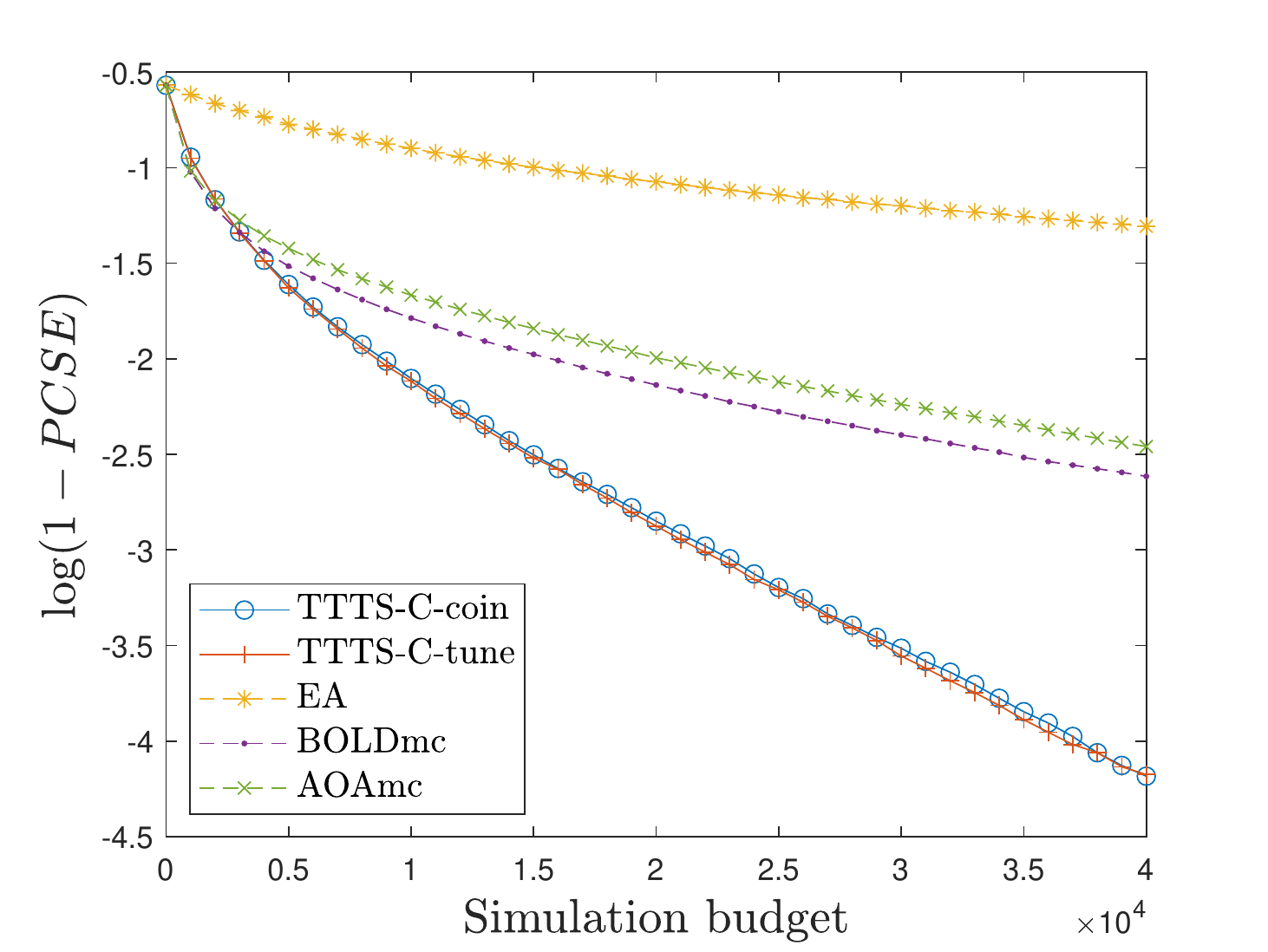}
    \caption{Log probability of incorrect selection for selecting the contextual best designs.}
    \label{fig:num1}
\end{figure}

\begin{figure}
    \centering
    \includegraphics[width = 0.325\linewidth]{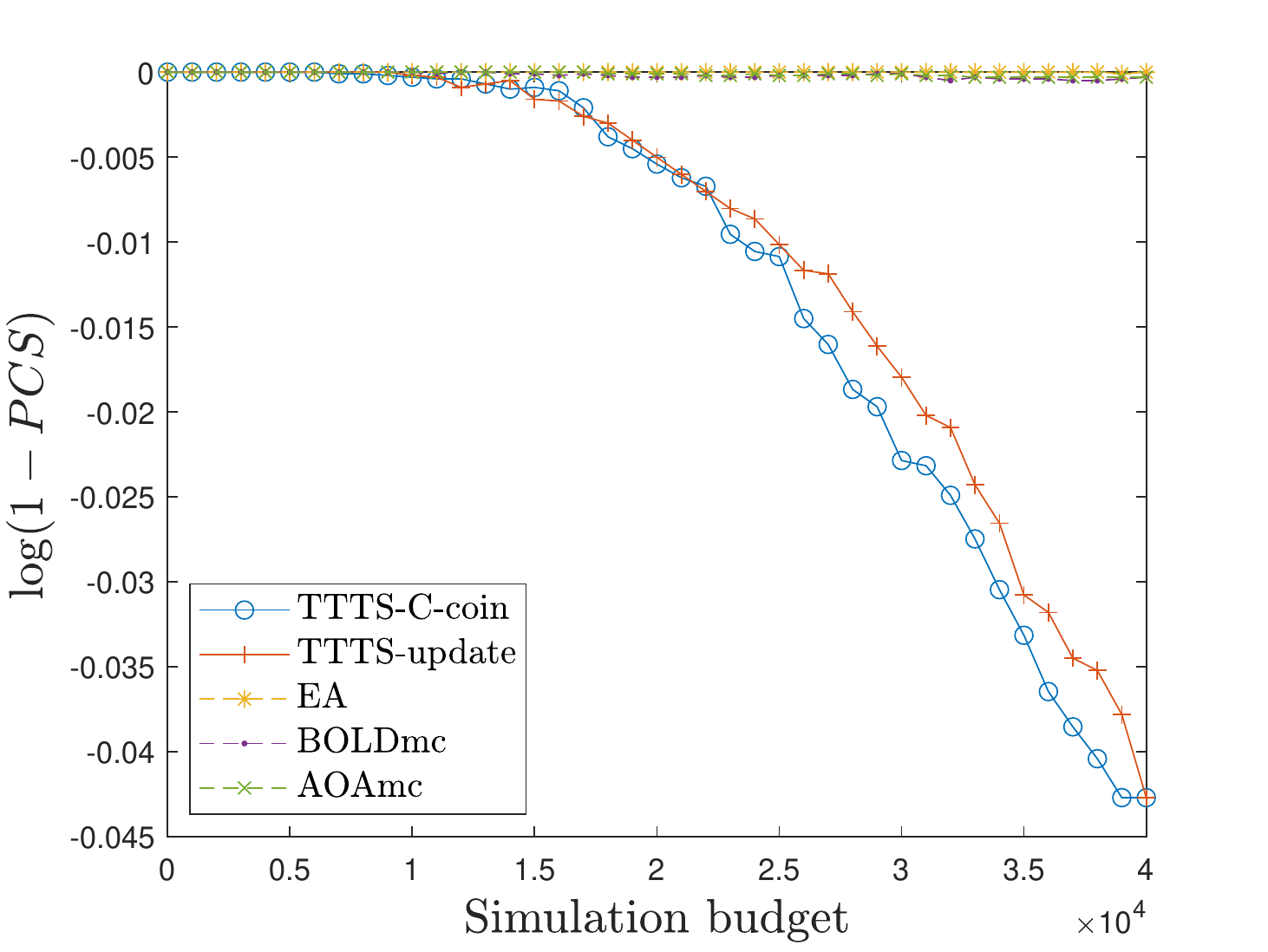}
    \includegraphics[width = 0.325\linewidth]{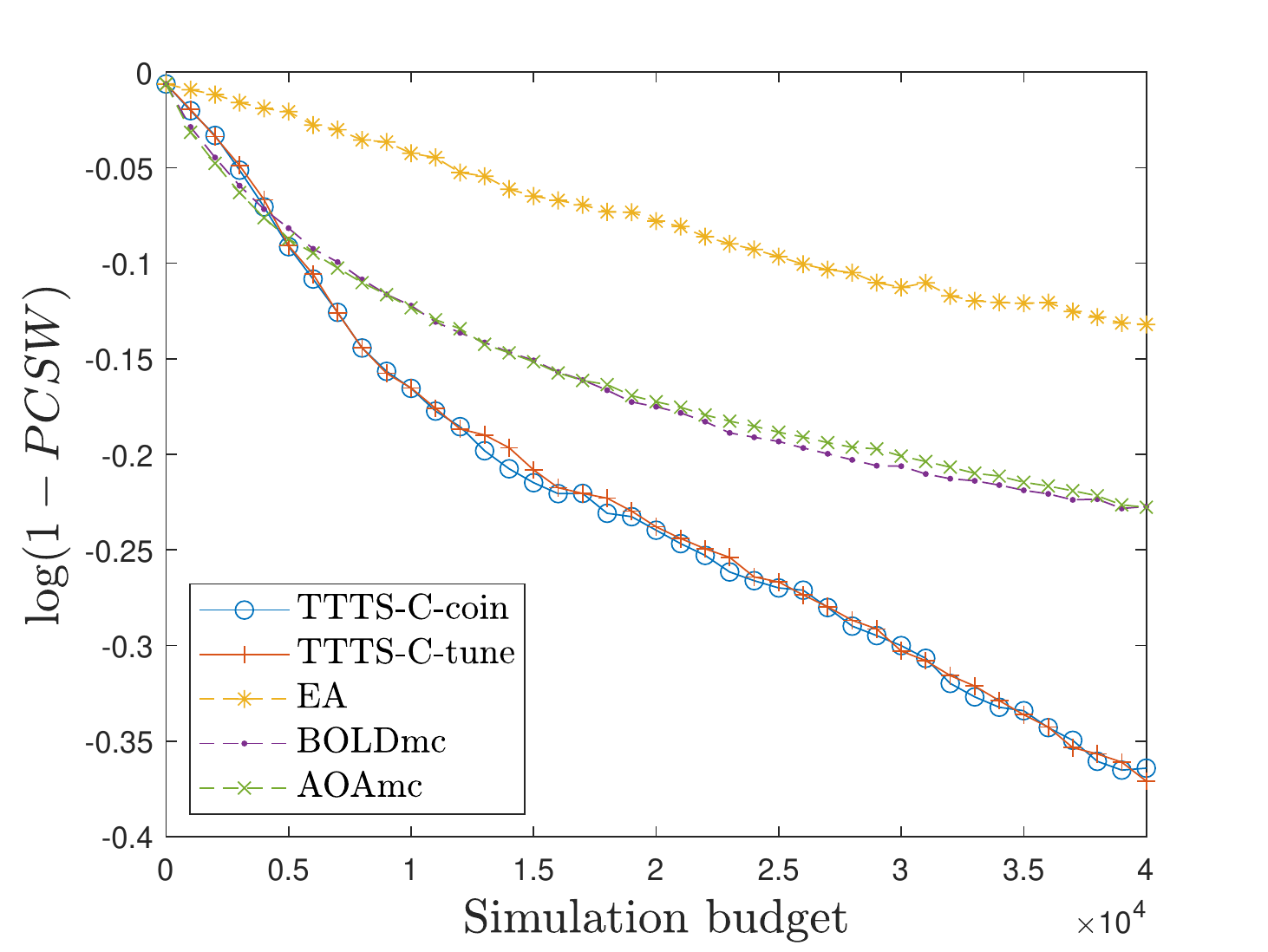}
    \includegraphics[width = 0.325\linewidth]{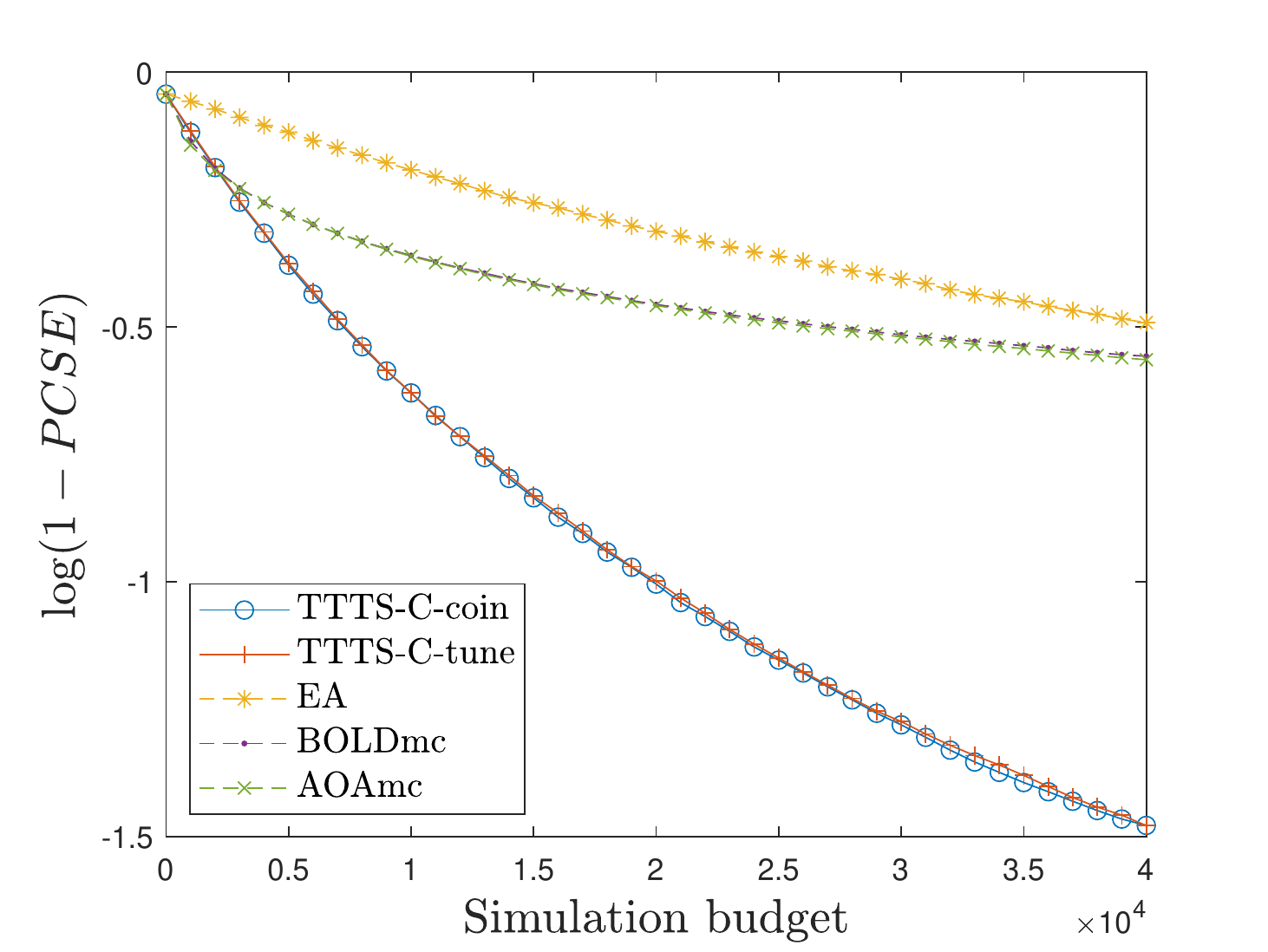}
    \caption{Log probability of incorrect selection for selecting the contextual top-5 designs.}
    \label{fig:num2}
\end{figure}

Consider a synthetic example with $\vert \mathcal{C} \vert = 10$ and $\vert \mathcal{D}_{c} \vert = 50$, $\forall c\in\mathcal{C}$. Each design follows a Gaussian sampling distribution $N(\mu_{d}, \sigma^{2}_{d})$, where $\mu_{d}$ and $\sigma_{d}$ are generated from $N(0, 10)$ and $\operatorname{Unif}([4, 6])$, respectively. Both the mean parameter $\mu_{d}$ and the nuisance parameter $\eta_{d} = \sigma_{d}^{2}$ are assumed unknown throughout. Once an instance is generated, it is fixed for all macro replications when testing the three metrics.

The prior distribution is chosen as the Normal-Gamma distribution with parameters $m_{0, d}, n_{0, d}, a_{0, d}, b_{0, d}$, which is the conjugate prior distribution of the Gaussian distribution with unknown mean and variance. The prior distribution is defined as follows:
\begin{equation*}
    \mu_{d}\vert \sigma_{d}^{2} \sim N\left(m_{0, d}, \frac{\sigma_{d}^{2}}{n_{0, d}}\right),\quad \text{ and } 1/\sigma_{d}^{2}\sim \operatorname{Gamma}(a_{0, d}, b_{0, d}),
\end{equation*}
where $\operatorname{Gamma}(a, b)$ is a Gamma distribution with shape parameter $a$ and scale parameter $b$. The parameters of the posterior distribution can be updated by
\begin{align*}
    &m_{T, d} = \frac{n_{0, d} m_{0, d} + N_{T}(d) \hat{\mu}_{d}}{n_{0, d} + N_{T}(d)}, & & n_{T, d} = n_{0, d} + N_{T}(d), \\
    &a_{T, d} = a_{0, d} + \frac{N_{T}(d)}{2}, & & b_{T, d} = b_{0, d} + \frac{N_{T}(d)}{2}\left(\hat{\sigma}_{d}^{2} + \frac{n_{0, d} (\hat{\mu}_{d} - \mu_{0, d})^{2}}{n_{0, d} + N_{T}(d)}\right).
\end{align*}

We test both cases of selecting the contextual best designs and contextual top-$m_{c}$ designs. In the first experiment, we set $m_{c} = 1$ for $c\in\mathcal{C}$, while in the second experiment, $m_{c} = 5$. Two versions of TTTS-C are implemented, including TTTS-C-coin, which utilizes an unbiased tossing coin, i.e., using $\gamma(c) = 1/2$ for all contexts, and TTTS-C-tune, which updates $\bm{\gamma}$ by solving the static optimal allocation problem (\ref{opt:1}) or (\ref{opt:5}) every time the number of consumed simulation samples reaches powers of $10$, (i.e., updating $\gamma$ when $T = 10, 100,\cdots$). Figures \ref{fig:num1} and \ref{fig:num2} show the logarithm of three PICS terms based on 10,000 macro replications of 40,000 simulation samples, with 10 initial samples allocated to each design at the beginning.

In the first experiment, TTTS-C-coin and TTTS-C-tune significantly outperform the other compared sampling policies in terms of all three metrics. Notably, TTTS-C-coin and TTTS-C-tune are the only two policies that achieve PCS$> 0.8$, PCSW$> 0.9$, and PCSE$> 0. 95$ within 40,000 samples. Specifically, TTTS-C yields a satisfactory level of PCS, implying its capability in simultaneously identifying the best designs for all contexts. TTTS-C-coin, which replies merely on a predetermined hyperparameter, performs comparably to {TTTS-C-tune, which is an} asymptotically optimal sampling policy. This validates the robustness of the choice of the hyperparameter.

BOLDmc and AOAmc have a comparable finite sample performance, which have an edge over EA but are outperformed by TTTS-C. It's worth noting that, in terms of PCSW, BOLDmc and AOAmc slightly surpass TTTS-C when the simulation budget is small. This could be attributed to the different sampling effort made for contexts. BOLDmc and AOAmc, as discussed above in (\ref{eq:14}), feature in choosing the context to be sampled by searching for the context with the global minimum rate functions attained among its designs, i.e., the context whose best design is least distinguished. This effectively elevates the rate function of the most vague context evaluated at the conditional sampling ratios of its designs at the initial steps. However, allocating simulation samples to the most vague context with the smallest rate function could be misleading when simulation samples have been scarcely collected, as the estimated rate function incorporates random noises. Additionally, BOLDmc and AOAmc are tempted to allocate successive samples to the same context, thereby reducing the actual rate at which the integrated PICS decreases. In contrast, TTTS-C forces exploration for each context according to Lemma \ref{lem:3}, resulting in better performance compared to the other sampling policies.

\begin{figure}
    \centering
    \includegraphics[width = 0.48\linewidth]{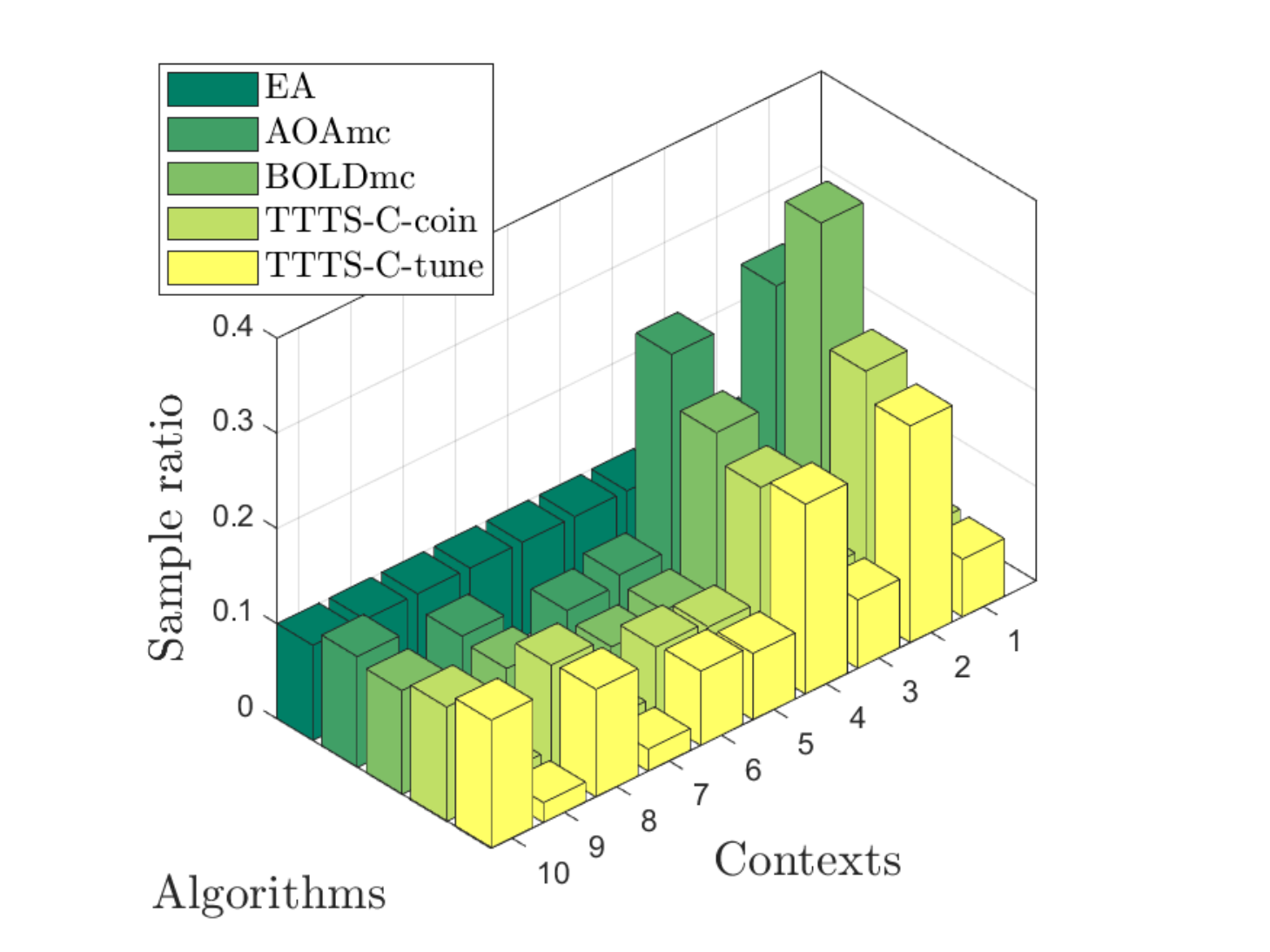}
    \includegraphics[width = 0.48\linewidth]{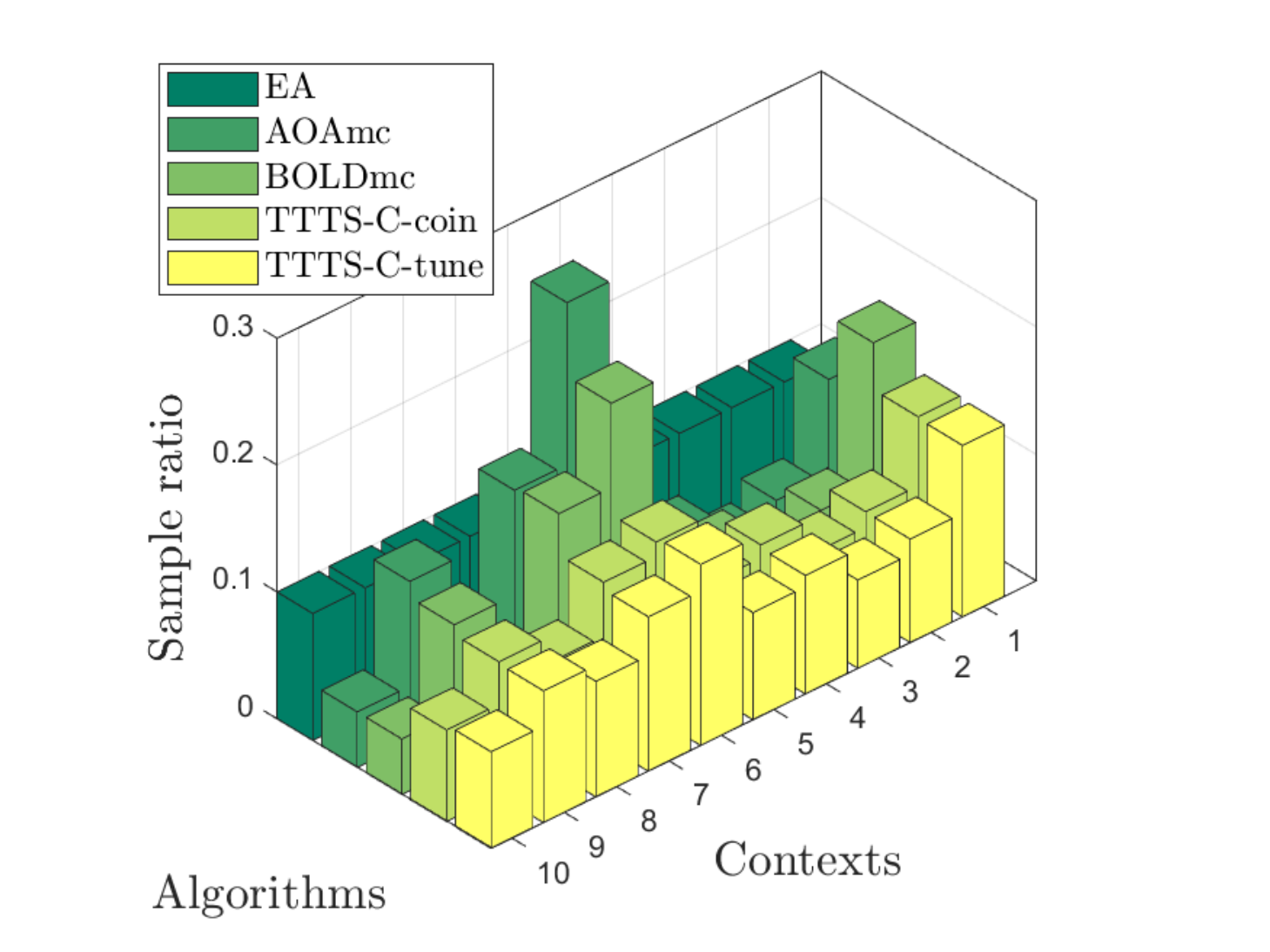}
    \caption{Empirical sample ratio of contexts for compared algorithms. The left (right) panel illustrates the empirical sample ratio for each context in the first (second) experiment.}
    \label{fig:num4}
\end{figure}

The second experiment, which aims to find the top-5 designs for each context, presents an apparently harder task than the first experiment. In terms of PCS, all sampling policies except for TTTS-C fail to provide even a satisfactory level of PCS. Regarding other metrics, TTTS-C-coin and TTTS-C-tune consistently and significantly outperform the other compared sampling policies. TTTS-C-coin, once again, performs no worse than TTTS-C-tune, implying the efficacy of simply tracking the balance condition of optimality. 

Figure \ref{fig:num4} illustrates the sampling ratios of each context under the five sampling policies. In the first experiment, the aggregate sampling ratios of designs under contexts $2$ and $4$ are higher than other contexts for all tested sequential sampling policies except for EA, indicating the fact that these two contexts are harder to distinguish. Furthermore, AOAmc and BOLDmc allocate more simulation samples to contexts $2$ and $4$ than TTTS-C. On the contrary, the latter allocates simulation samples more fairly in comparison, and no context receives an extremely large or small size of samples. The second experiment yields consistent findings.

\subsection{Non-Gaussian designs.}
\begin{figure}
    \centering
    \includegraphics[width = 0.325\linewidth]{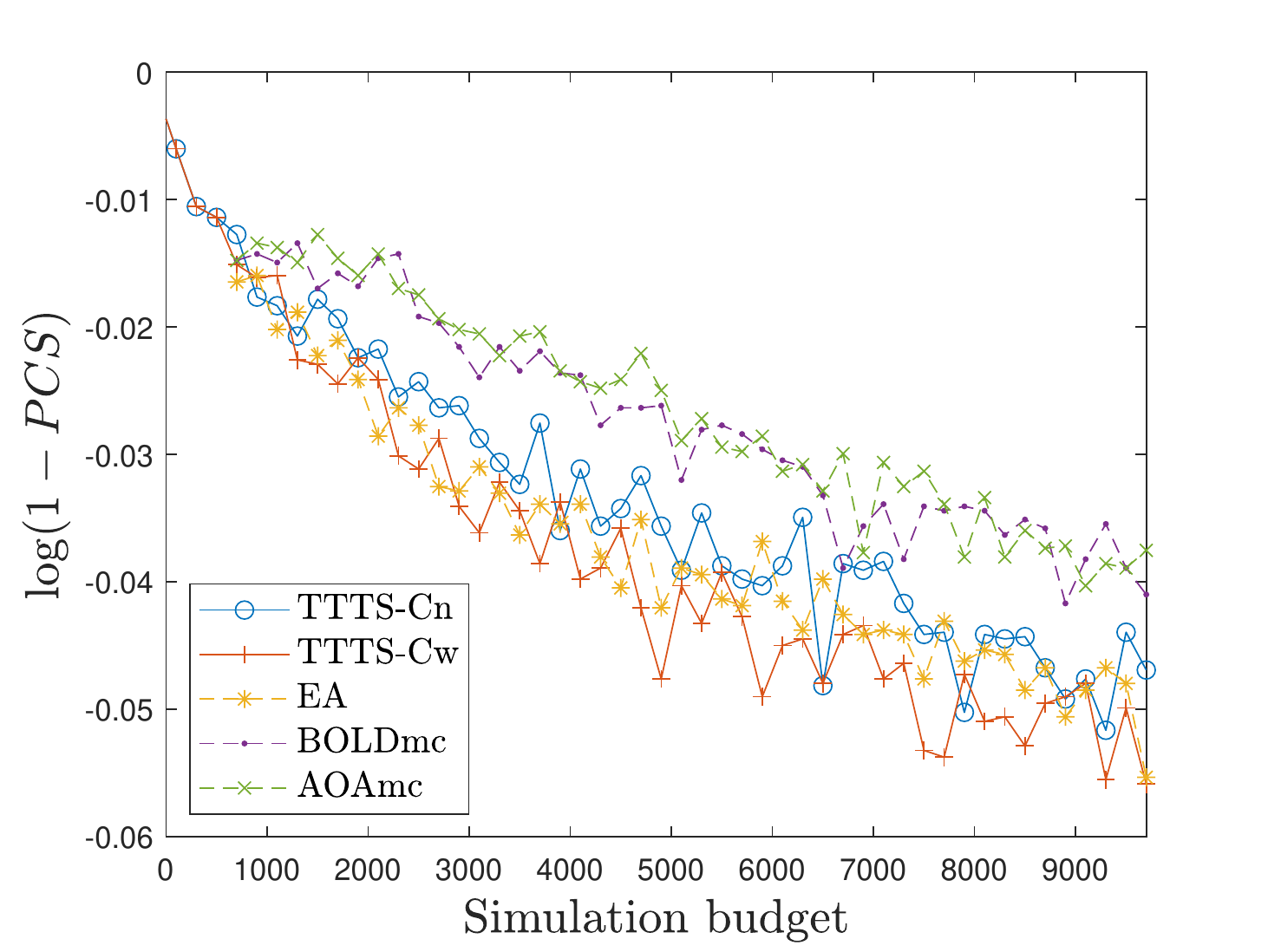}
    \includegraphics[width = 0.325\linewidth]{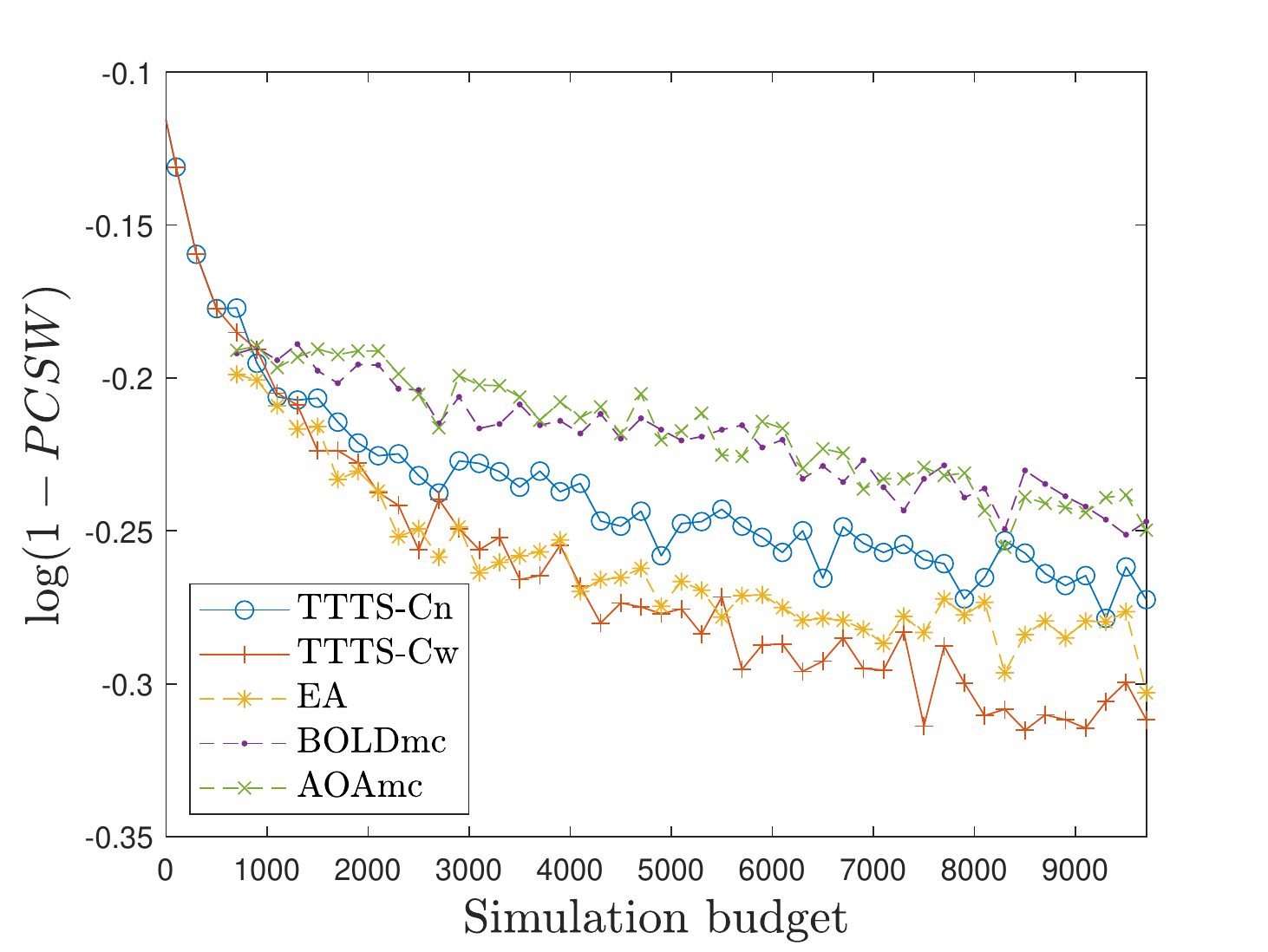}
    \includegraphics[width = 0.325\linewidth]{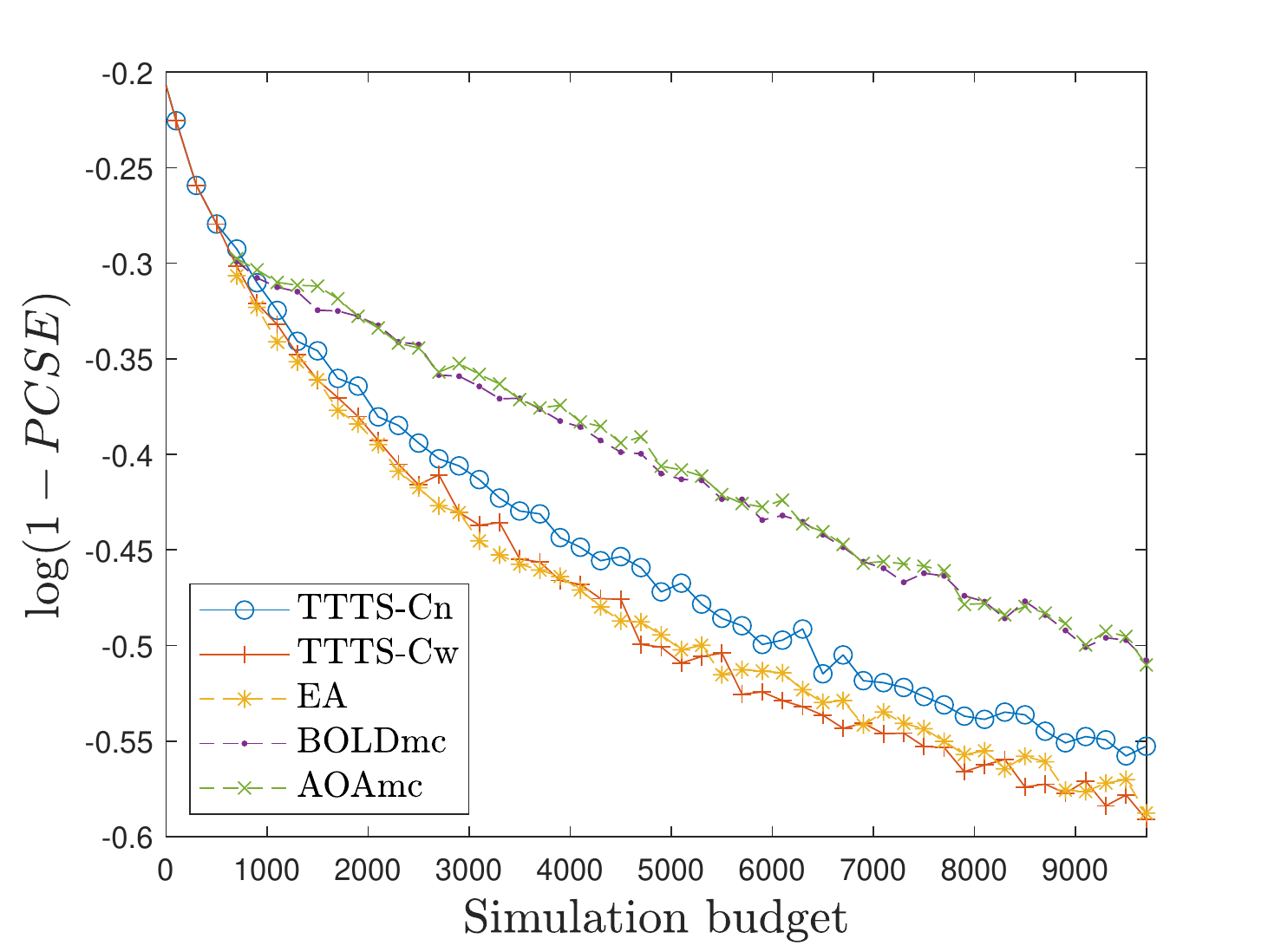}
    \caption{Log probability of incorrect selection for selecting the best design for three contexts and selecting the top-2 designs for the other two contexts.}
    \label{fig:num3}
\end{figure}

We consider a synthetic instance based on Example 2 in Section \ref{sec:probdef}, where there are 5 categories (contexts) of production lines, each with an unequal number of designs. Specifically, let $\mathcal{C} = \{1, 2, 3, 4, 5\}$ denote $5$ increasingly demanding and costly production standards, and $\mathcal{D}_{c}$ contains $5, 5, 7, 6$, and $7$ production lines, respectively, for $c = 1, 2, 3, 4, 5$. For $c = 1, 2, 3$, the corresponding output of products may be less lucrative, and our aim is to select the best design. For $c = 4, 5$, we aim to select the top-2 designs. 

In this instance, the underlying parameters of the Weibull distributions are randomly generated from $\mu_{d} \sim \operatorname{Unif}([90, 110])$ and $\eta_{d} \sim \operatorname{Unif}([2, 4])$. The prior distribution is set as uninformative on the compact set $\Theta = [0, 200] \times [0, 20]$. It's worth noting that we do not have to perform parameter transformation as described in Example 2. Instead, we can maintain a posterior distribution for the natural parameters $\rho_{d}$ and $k_{d}$ of Weibull distributions, and equivalently implement the parameter transformation after the posterior samples of $\rho_{d}$ and $k_{d}$ are simulated in Algorithm \ref{alg:1} or \ref{alg:3}. Recall that BOLDmc and AOAmc are designed for Gaussian designs. To fairly compare TTTS-C with these sampling policies, we also compare them with TTTS-C derived assuming Gaussian distributions, denoted by TTTS-Cn. Meanwhile, TTTS-Cw denotes the variant of TTTS-C derived assuming Weibull distributions. In this experiment, we fix the hyperparameter $\bm{\gamma}$ as $\gamma(c) = 1/2$.

Figure \ref{fig:num3} demonstrates the logarithm of PICS based on 6,000 macro replications of 10,000 simulation samples, with 10 initial samples allocated to each design at the beginning. TTTS-Cw is the only procedure that has an edge over EA within 10,000 samples. It significantly outperforms other compared algorithms in all the metrics. As expected, all three algorithms designed for selecting the design with the largest mean sample performance cannot efficiently identify the best design in terms of other characteristics. This implies that R\&S procedures designed for identifying the best performance may fail, which substantiates the superiority of TTTS-C. 

\section{Conclusion.}\label{sec:con}
In this work, we consider a generic contextual top-$m_{c}$ selection problem under a Bayesian framework where the sample distributions are to be estimated among a parametric family of distributions with potential nuisance parameters. The aim is to identify the top-$m_{c}$ designs with the largest values of the parameter of interest by efficiently allocating the limited simulation budget. Our contribution can be summarized in three aspects. First, we propose an efficient sequential sampling policy based on Thompson sampling that can simultaneously allocate simulation samples to contexts and designs. This policy is easy to implement for the aforementioned distribution families. This allows practitioners to consider various performance metrics of the simulation systems, which can be measured by an arbitrary identifiable distribution parameter apart from the mean performance, and thus enhances the efficient selection of an effective design. Second, we show that the proposed sampling policy leads to an exponential concentration of the posterior distribution around the truth values. We characterize the exponential convergence rate, known as the posterior larger deviations rate, which provides deeper insights into the generic contextual selection problem. We also demonstrate that the proposed sampling policy attains the maximum possible rate asymptotically, despite that fact that the rate function cannot be derived explicitly for any distribution family and may not have strong properties such as strict monotinicity and strict concavity, and thus the optimal allocation may not be unique. Finally, we provide extensive synthetic experiments to demonstrate the finite performance of our proposed method. 

Future research may include a full characterization of the optimal rate for selecting contextual top-$m_{c}$ designs, since sufficient and necessary optimality conditions are in general unavailable given an arbitrary parametric model without introducing dual variables. In addition, leveraging the proposed sampling policy to design asymptotically optimal sequential sampling policies may be of interest for future researchers. Another direction to explore is the consideration of large-scale problems with copious contexts and designs. Although the proposed sampling policy has been shown to be computationally efficient, it still involves repeated simulations from a high-dimensional posterior distribution. Therefore, developing sampling policies capable of parallel computing could be a viable solution for addressing these large-scale problems.

%
%
%

\section*{Acknowledgments.}
This work was supported in part by the National Natural Science Foundation of China (NSFC) under Grants 72250065, 72022001, and 71901003, and by the China Scholarship Council (CSC) under Grant CSC202206010152. A preliminary version of this work without rigorous proofs has been accepted by the 2023 Winter Simulation Conference.


\bibliographystyle{informs2014} 
\bibliography{ref} 

\begin{thebibliography}{67}
\providecommand{\natexlab}[1]{#1}
\providecommand{\url}[1]{\texttt{#1}}
\providecommand{\urlprefix}{URL }

\bibitem[{Agrawal \protect\BIBand{} Goyal(2017)}]{agrawal2017near}
Agrawal S, Goyal N (2017) Near-optimal regret bounds for thompson sampling.
  \emph{Journal of the ACM} 64(5):1--24.

\bibitem[{Audibert et~al.(2010)Audibert, Bubeck, \protect\BIBand{}
  Munos}]{audibert2010best}
Audibert JY, Bubeck S, Munos R (2010) Best arm identification in multi-armed
  bandits. \emph{In Proceedings of the 23rd Conference on Learning Theory
  (COLT)}, 41--53.

\bibitem[{Auer et~al.(2002)Auer, Cesa-Bianchi, \protect\BIBand{}
  Fischer}]{auer2002finite}
Auer P, Cesa-Bianchi N, Fischer P (2002) Finite-time analysis of the multiarmed
  bandit problem. \emph{Machine Learning} 47:235--256.

\bibitem[{Avci et~al.(2021)Avci, Nelson, \protect\BIBand{}
  W{\"a}chter}]{avci2021getting}
Avci H, Nelson L Barry, W{\"a}chter A (2021) Getting to ``rate-optimal" in
  ranking \& selection. \emph{In Proceedings of the 2021 Winter Simulation
  Conference (WSC)}, volume~1, 1--12 (IEEE).

\bibitem[{Bubeck et~al.(2009)Bubeck, Munos, \protect\BIBand{}
  Stoltz}]{bubeck2009pure}
Bubeck S, Munos R, Stoltz G (2009) Pure exploration in multi-armed bandits
  problems. \emph{In Proceedings of the 20th International Conference on
  Algorithmic Learning Theory (ALT)}, 23--37 (Springer).

\bibitem[{Bubeck et~al.(2013)Bubeck, Wang, \protect\BIBand{}
  Viswanathan}]{bubeck2013multiple}
Bubeck S, Wang T, Viswanathan N (2013) Multiple identifications in multi-armed
  bandits. \emph{In Proceedings of the 30th International Conference on Machine
  Learning (ICML)}, 258--265 (PMLR).

\bibitem[{Cakmak et~al.(2021)Cakmak, Zhou, \protect\BIBand{}
  Gao}]{cakmak2021contextual}
Cakmak S, Zhou E, Gao S (2021) Contextual ranking and selection with gaussian
  processes. \emph{In Proceedings of the 2021 Winter Simulation Conference
  (WSC)}, 1--12 (IEEE).

\bibitem[{Chen(1996)}]{chen1996lower}
Chen CH (1996) A lower bound for the correct subset-selection probability and
  its application to discrete-event system simulations. \emph{IEEE Transactions
  on Automatic Control} 41(8):1227--1231.

\bibitem[{Chen et~al.(2008)Chen, He, Fu, \protect\BIBand{}
  Lee}]{chen2008efficient}
Chen CH, He D, Fu M, Lee LH (2008) Efficient simulation budget allocation for
  selecting an optimal subset. \emph{INFORMS Journal on Computing}
  20(4):579--595.

\bibitem[{Chen et~al.(2000)Chen, Lin, Y{\"u}cesan, \protect\BIBand{}
  Chick}]{chen2000simulation}
Chen CH, Lin J, Y{\"u}cesan E, Chick SE (2000) Simulation budget allocation for
  further enhancing the efficiency of ordinal optimization. \emph{Discrete
  Event Dynamic Systems} 10:251--270.

\bibitem[{Chen \protect\BIBand{} Ryzhov(2019)}]{chen2019complete}
Chen Y, Ryzhov IO (2019) Complete expected improvement converges to an optimal
  budget allocation. \emph{Advances in Applied Probability} 51(1):209--235.

\bibitem[{Chen \protect\BIBand{} Ryzhov(2022)}]{chen2022balancing}
Chen Y, Ryzhov IO (2022) Balancing optimal large deviations in sequential
  selection. \emph{Management Science, early access}
  \urlprefix\url{http://dx.doi.org/10.1287/mnsc.2022.4527}.

\bibitem[{Chick \protect\BIBand{} Inoue(2001)}]{chick2001new}
Chick SE, Inoue K (2001) New two-stage and sequential procedures for selecting
  the best simulated system. \emph{Operations Research} 49(5):732--743.

\bibitem[{Chow(1971)}]{chow71}
Chow YS (1971) On the $l_p$-convergence for $n^{-1/p} s_n, 0 <p < 2$. \emph{The
  Annals of Mathematical Statistics} 42(1):393--394,
  \urlprefix\url{http://dx.doi.org/10.1214/aoms/1177693530}.

\bibitem[{Dembo \protect\BIBand{} Zeitouni(1992)}]{dembo92}
Dembo A, Zeitouni O (1992) \emph{Large Deviations Techniques and Applications}
  (Jones and Bartlett).

\bibitem[{Ding et~al.(2022)Ding, Hong, Shen, \protect\BIBand{}
  Zhang}]{ding2022knowledge}
Ding L, Hong LJ, Shen H, Zhang X (2022) Knowledge gradient for selection with
  covariates: Consistency and computation. \emph{Naval Research Logistics
  (NRL)} 69(3):496--507.

\bibitem[{Du et~al.(2022)Du, Gao, \protect\BIBand{} Chen}]{du2022rate}
Du J, Gao S, Chen CH (2022) Rate-optimal contextual ranking and selection.
  \emph{arXiv preprint arXiv:2206.12640} .

\bibitem[{Even-Dar et~al.(2006)Even-Dar, Mannor, Mansour, \protect\BIBand{}
  Mahadevan}]{even2006action}
Even-Dar E, Mannor S, Mansour Y, Mahadevan S (2006) Action elimination and
  stopping conditions for the multi-armed bandit and reinforcement learning
  problems. \emph{Journal of machine learning research} 7(6).

\bibitem[{Frazier et~al.(2008)Frazier, Powell, \protect\BIBand{}
  Dayanik}]{frazier2008knowledge}
Frazier PI, Powell WB, Dayanik S (2008) A knowledge-gradient policy for
  sequential information collection. \emph{SIAM Journal on Control and
  Optimization} 47(5):2410--2439.

\bibitem[{Gabillon et~al.(2012)Gabillon, Ghavamzadeh, \protect\BIBand{}
  Lazaric}]{gabillon2012best}
Gabillon V, Ghavamzadeh M, Lazaric A (2012) Best arm identification: A unified
  approach to fixed budget and fixed confidence. \emph{In Advances in Neural
  Information Processing Systems (NIPS)} 25.

\bibitem[{Gao \protect\BIBand{} Chen(2016)}]{gao2016new}
Gao S, Chen W (2016) A new budget allocation framework for selecting top
  simulated designs. \emph{IIE Transactions} 48(9):855--863.

\bibitem[{Gao et~al.(2019)Gao, Du, \protect\BIBand{} Chen}]{gao2019selecting}
Gao S, Du J, Chen CH (2019) Selecting the optimal system design under
  covariates. \emph{In Proceddings of the 15th International Conference on
  Automation Science and Engineering (CASE)}, 547--552 (IEEE).

\bibitem[{Garivier \protect\BIBand{} Kaufmann(2016)}]{garivier2016optimal}
Garivier A, Kaufmann E (2016) Optimal best arm identification with fixed
  confidence. \emph{In Proceedings of the 29th Conference on Learning Theory
  (COLT)}, 998--1027 (PMLR).

\bibitem[{Glynn \protect\BIBand{} Juneja(2004)}]{glynn2004large}
Glynn P, Juneja S (2004) A large deviations perspective on ordinal
  optimization. \emph{In Proceedings of the 2004 Winter Simulation Conference
  (WSC)}, volume~1 (IEEE).

\bibitem[{Hong et~al.(2021)Hong, Fan, \protect\BIBand{} Luo}]{hong2021review}
Hong LJ, Fan W, Luo J (2021) Review on ranking and selection: A new
  perspective. \emph{Frontiers of Engineering Management} 8(3):321--343.

\bibitem[{Hu \protect\BIBand{} Ludkovski(2017)}]{hu2017sequential}
Hu R, Ludkovski M (2017) Sequential design for ranking response surfaces.
  \emph{SIAM/ASA Journal on Uncertainty Quantification} 5(1):212--239.

\bibitem[{Jourdan et~al.(2022)Jourdan, Degenne, Baudry, de~Heide,
  \protect\BIBand{} Kaufmann}]{jourdan2022top}
Jourdan M, Degenne R, Baudry D, de~Heide R, Kaufmann E (2022) Top two
  algorithms revisited. \emph{In Proceedings of the 36th Annual Conference on
  Neural Information Processing System (NIPS)}.

\bibitem[{Kalyanakrishnan et~al.(2012)Kalyanakrishnan, Tewari, Auer,
  \protect\BIBand{} Stone}]{kalyanakrishnan2012pac}
Kalyanakrishnan S, Tewari A, Auer P, Stone P (2012) Pac subset selection in
  stochastic multi-armed bandits. \emph{In Proceedings of the 29th
  International Conference on Machine Learning (ICML)}, volume~12, 655--662.

\bibitem[{Kapur \protect\BIBand{} Lamberson(1977)}]{kapur77}
Kapur KC, Lamberson LR (1977) Reliability in engineering design. \emph{New
  York} .

\bibitem[{Kato et~al.(2022)Kato, Ariu, Imaizumi, Uehara, Nomura,
  \protect\BIBand{} Qin}]{kato2022best}
Kato M, Ariu K, Imaizumi M, Uehara M, Nomura M, Qin C (2022) Best arm
  identification with contextual information under a small gap. \emph{arXiv
  preprint arXiv:2209.07330} .

\bibitem[{Kaufmann \protect\BIBand{}
  Kalyanakrishnan(2013)}]{kaufmann2013information}
Kaufmann E, Kalyanakrishnan S (2013) Information complexity in bandit subset
  selection. \emph{In Proceedings of the 26th Conference on Learning Theory
  (COLT)}, 228--251 (PMLR).

\bibitem[{Kaufmann et~al.(2012)Kaufmann, Korda, \protect\BIBand{}
  Munos}]{kaufmann2012thompson}
Kaufmann E, Korda N, Munos R (2012) Thompson sampling: An asymptotically
  optimal finite-time analysis. \emph{In Proceddings of the 23rd International
  Conference on Algorithmic Learning Theory (ALT)}, 199--213 (Springer).

\bibitem[{Keslin et~al.(2022)Keslin, Nelson, Plumlee, Pagnoncelli,
  \protect\BIBand{} Rahimian}]{keslin2022classification}
Keslin G, Nelson BL, Plumlee M, Pagnoncelli BK, Rahimian H (2022) A
  classification method for ranking and selection with covariates. \emph{In
  Proceedings of the 2022 Winter Simulation Conference (WSC)}, 1--12 (IEEE).

\bibitem[{Li et~al.(2022)Li, Lam, \protect\BIBand{} Peng}]{li2022efficient}
Li H, Lam H, Peng Y (2022) Efficient learning for clustering and optimizing
  context-dependent designs. \emph{Operations Research}
  \urlprefix\url{http://dx.doi.org/10.1287/opre.2022.2368}.

\bibitem[{Li et~al.(2018)Li, Zhang, \protect\BIBand{} Zheng}]{li2018data}
Li X, Zhang X, Zheng Z (2018) Data-driven ranking and selection:
  High-dimensional covariates and general dependence. \emph{In Proceedings of
  the 2018 Winter Simulation Conference (WSC)}, 1933--1944 (IEEE).

\bibitem[{Miao \protect\BIBand{} Chao(2022)}]{miao2022online}
Miao S, Chao X (2022) Online personalized assortment optimization with
  high-dimensional customer contextual data. \emph{Manufacturing \& Service
  Operations Management} 24(5):2741--2760.

\bibitem[{Milgrom \protect\BIBand{} Segal(2002)}]{milgrom02}
Milgrom PR, Segal I (2002) Envelope theorems for arbitrary choice sets.
  \emph{Econometrica} 70:583--601.

\bibitem[{Munkres(2000)}]{munkres2000topology}
Munkres JR (2000) \emph{Topology} (Prentice Hall), 2nd edition.

\bibitem[{Pasupathy et~al.(2010)Pasupathy, Szechtman, \protect\BIBand{}
  Y{\"u}cesan}]{pasupathy2010selecting}
Pasupathy R, Szechtman R, Y{\"u}cesan E (2010) Selecting small quantiles.
  \emph{In Proceedings of the 2010 Winter Simulation Conference (WSC)},
  2762--2770 (IEEE).

\bibitem[{Pearce \protect\BIBand{} Branke(2018)}]{pearce2018continuous}
Pearce M, Branke J (2018) Continuous multi-task bayesian optimisation with
  correlation. \emph{European Journal of Operational Research}
  270(3):1074--1085.

\bibitem[{Peng et~al.(2013)Peng, Chen, Fu, \protect\BIBand{} Hu}]{peng12share}
Peng Y, Chen CH, Fu MC, Hu JQ (2013) Efficient simulation resource sharing and
  allocation for selecting the best. \emph{IEEE Transactions on Automatic
  Control} 58(4):1017--1023.

\bibitem[{Peng et~al.(2015)Peng, Chen, Fu, \protect\BIBand{} Hu}]{peng2015non}
Peng Y, Chen CH, Fu MC, Hu JQ (2015) Non-monotonicity of probability of correct
  selection. \emph{In Proceedings of the 2015 Winter Simulation Conference
  (WSC)}, 3678--3689 (IEEE).

\bibitem[{Peng et~al.(2017)Peng, Chen, Fu, \protect\BIBand{}
  Hu}]{peng2017gradient}
Peng Y, Chen CH, Fu MC, Hu JQ (2017) Gradient-based myopic allocation policy:
  An efficient sampling procedure in a low-confidence scenario. \emph{IEEE
  Transactions on Automatic Control} 63(9):3091--3097.

\bibitem[{Peng et~al.(2021)Peng, Chen, Fu, Hu, \protect\BIBand{}
  Ryzhov}]{peng2021efficient}
Peng Y, Chen CH, Fu MC, Hu JQ, Ryzhov IO (2021) Efficient sampling allocation
  procedures for optimal quantile selection. \emph{INFORMS Journal on
  Computing} 33(1):230--245.

\bibitem[{Peng et~al.(2018)Peng, Chong, Chen, \protect\BIBand{}
  Fu}]{peng2018ranking}
Peng Y, Chong EK, Chen CH, Fu MC (2018) Ranking and selection as stochastic
  control. \emph{IEEE Transactions on Automatic Control} 63(8):2359--2373.

\bibitem[{Peng \protect\BIBand{} Zhang(2022)}]{peng2022thompson}
Peng Y, Zhang G (2022) Thompson sampling meets ranking and selection. \emph{In
  Proceedings of the 2022 Winter Simulation Conference (WSC)}, 3075--3086
  (IEEE).

\bibitem[{Qin et~al.(2017)Qin, Klabjan, \protect\BIBand{}
  Russo}]{qin2017improving}
Qin C, Klabjan D, Russo D (2017) Improving the expected improvement algorithm.
  \emph{In Proceedings of the 31st Annual Conference on Neural Information
  Processing System (NIPS)} 30.

\bibitem[{R{\'e}da et~al.(2021)R{\'e}da, Kaufmann, \protect\BIBand{}
  Delahaye-Duriez}]{reda2021top}
R{\'e}da C, Kaufmann E, Delahaye-Duriez A (2021) Top-m identification for
  linear bandits. \emph{In Proceedings of the 24th International Conference on
  Artificial Intelligence and Statistics (AISTATS)}, 1108--1116 (PMLR).

\bibitem[{Rockafellar(1970)}]{rockafellar70}
Rockafellar RT (1970) \emph{Convex Analysis}. Princeton Mathematical Series
  (Princeton University Press).

\bibitem[{Russo(2020)}]{russo2020simple}
Russo D (2020) Simple bayesian algorithms for best-arm identification.
  \emph{Operations research} (6):1625--1647.

\bibitem[{Russo et~al.(2018)Russo, Van~Roy, Kazerouni, Osband, Wen
  et~al.}]{russo2018tutorial}
Russo DJ, Van~Roy B, Kazerouni A, Osband I, Wen Z, et~al. (2018) A tutorial on
  thompson sampling. \emph{Foundations and Trends{\textregistered} in Machine
  Learning} 11(1):1--96.

\bibitem[{Ryzhov(2016)}]{ryzhov2016convergence}
Ryzhov IO (2016) On the convergence rates of expected improvement methods.
  \emph{Operations Research} 64(6):1515--1528.

\bibitem[{Shang et~al.(2020)Shang, Heide, Menard, Kaufmann, \protect\BIBand{}
  Valko}]{shang2020fixed}
Shang X, Heide R, Menard P, Kaufmann E, Valko M (2020) Fixed-confidence
  guarantees for bayesian best-arm identification. \emph{In Proceedings of the
  23rd International Conference on Artificial Intelligence and Statistics
  (AISTATS)}, 1823--1832 (PMLR).

\bibitem[{Shen et~al.(2021)Shen, Hong, \protect\BIBand{}
  Zhang}]{shen2021ranking}
Shen H, Hong LJ, Zhang X (2021) Ranking and selection with covariates for
  personalized decision making. \emph{INFORMS Journal on Computing}
  33(4):1500--1519.

\bibitem[{Soare et~al.(2014)Soare, Lazaric, \protect\BIBand{}
  Munos}]{soare2014best}
Soare M, Lazaric A, Munos R (2014) Best-arm identification in linear bandits.
  \emph{Advances in Neural Information Processing Systems} 27:828--836.

\bibitem[{Szechtman \protect\BIBand{} Yucesan(2008)}]{szechtman2008new}
Szechtman R, Yucesan E (2008) A new perspective on feasibility determination.
  \emph{In Proceedings of the 2008 Winter Simulation Conference (WSC)},
  273--280 (IEEE).

\bibitem[{Thompson(1933)}]{thompson1933likelihood}
Thompson WR (1933) On the likelihood that one unknown probability exceeds
  another in view of the evidence of two samples. \emph{Biometrika}
  25(3-4):285--294.

\bibitem[{van~der Vaart(1998)}]{vdv98}
van~der Vaart AW (1998) \emph{Asymptotic Statistics}. Cambridge Series in
  Statistical and Probabilistic Mathematics (Cambridge University Press).

\bibitem[{Williams(1991)}]{will91}
Williams D (1991) \emph{Probability with Martingales} (Cambridge University
  Press).

\bibitem[{Woerndl et~al.(2007)Woerndl, Schueller, \protect\BIBand{}
  Wojtech}]{woerndl2007hybrid}
Woerndl W, Schueller C, Wojtech R (2007) A hybrid recommender system for
  context-aware recommendations of mobile applications. \emph{In Proceedings of
  the 23rd International Conference on Data Engineering (ICDE)}, 871--878
  (IEEE).

\bibitem[{Xiao et~al.(2013)Xiao, Lee, \protect\BIBand{} Ng}]{xiao2013optimal}
Xiao H, Lee LH, Ng KM (2013) Optimal computing budget allocation for complete
  ranking. \emph{IEEE Transactions on Automation Science and Engineering}
  11(2):516--524.

\bibitem[{You et~al.(2022)You, Qin, Wang, \protect\BIBand{}
  Yang}]{you2023informationdirected}
You W, Qin C, Wang Z, Yang S (2022) Information-directed selection for top-two
  algorithms.

\bibitem[{Zhang et~al.(2022{\natexlab{a}})Zhang, Chen, Jia, \protect\BIBand{}
  Peng}]{zhang2022efficient}
Zhang G, Chen B, Jia QS, Peng Y (2022{\natexlab{a}}) Efficient sampling policy
  for selecting a subset with the best. \emph{IEEE Transactions on Automatic
  Control, early access}
  \urlprefix\url{http://dx.doi.org/10.1109/TAC.2022.3207871}.

\bibitem[{Zhang et~al.(2023)Zhang, Chen, Huang, \protect\BIBand{}
  Peng}]{zhang2023efficient}
Zhang G, Chen S, Huang K, Peng Y (2023) Efficient learning for selecting top-m
  context-dependent designs. \emph{arXiv preprint arXiv:2305.04086} .

\bibitem[{Zhang et~al.(2022{\natexlab{b}})Zhang, Peng, \protect\BIBand{}
  Xu}]{zhang2022monte}
Zhang G, Peng Y, Xu Y (2022{\natexlab{b}}) An efficient dynamic sampling policy
  for monte carlo tree search. \emph{In Proceedings of the 2022 Winter
  Simulation Conference (WSC)}, 2760--2771, number 1--12.

\bibitem[{Zhang et~al.(2021)Zhang, Peng, Zhang, \protect\BIBand{}
  Zhou}]{zhang2021asymptotically}
Zhang G, Peng Y, Zhang J, Zhou E (2021) Asymptotically optimal sampling policy
  for selecting top-m alternatives. \emph{arXiv preprint arXiv: 2111.15172} .

\bibitem[{Zhang et~al.(2015)Zhang, Lee, Chew, Xu, \protect\BIBand{}
  Chen}]{zhang2015simulation}
Zhang S, Lee LH, Chew EP, Xu J, Chen CH (2015) A simulation budget allocation
  procedure for enhancing the efficiency of optimal subset selection.
  \emph{IEEE Transactions on Automatic Control} 61(1):62--75.

\end{thebibliography}


\begin{APPENDICES}
\section{Posterior large deviations rates.}
In this section, we prove Theorem \ref{thm:1} which extends the posterior large deviations principle for general identifiable distribution to sequential sampling policies.

\proof{Proof of Theorem \ref{thm:1}.}
    We first establish a uniform approximation to $W_{T}(\cdot)$ by the overall KL divergence, namely,
    \begin{equation}\label{eq1}
        \sup_{\bm{\theta}\in\bm{\Theta}}\left\vert W_{T}(\bm{\theta}) - D_{\bar{\psi}_{T}}(\bm{\theta}^{*}\Vert \bm{\theta})\right\vert \rightarrow 0,\quad a.s..
    \end{equation}
    
    Fix a context $c\in\mathcal{C}$ and a design $d\in\mathcal{D}_{c}$, we define a sequence of strictly increasing stopping times by recursion 
    \begin{equation*}
        \tau_{1} = \min\{t\in\mathbb{N}: \xi_{t}(d) = 1\},\quad and \quad \tau_{k+1} = \min\{\tau_{k}<t\in\mathbb{N}: \xi_{t}(d) = 1\},\quad k=1,2,\dots,
    \end{equation*}
    with the convention $\min\phi=\infty$. It's worth mentioning that these stopping times are well defined on the extended probability space filtered by $\mathcal{E}^{\prime}_{t} = \sigma(\mathcal{E}_{t}\bigcup \sigma(\xi_{1}(d), \dots,\xi_{t}(d)))$ rather than $\mathcal{E}_{t}$. Then, 
    \begin{equation}\label{eq24}
        \frac{1}{T}\sum_{t=1}^{T}\xi_{t}(d)\lambda\left(\theta^{*}_{d}, \theta_{d}; Y^{(t)}_{d}\right) = \frac{1}{T}\sum_{k=1}^{T}\bm{1}\{\tau_{k}\leq T\}\lambda\left(\theta^{*}_{d}, \theta_{d}; Y^{(\tau_{k})}_{d}\right).
    \end{equation}
    
    Let $Z^{(t)}_{d}$ and ${Z^{\prime}}^{(t)}_{d}$ be two independent copies of $Y^{(t)}_{d}$ that are also independent of anything else. We investigate the convergence of the right-hand side of equation (\ref{eq24}) by its modification with $Y^{(\tau_{k})}_{d}$ replaced by $Z^{(k)}_{d}$. Denote $a_{k}=\bm{1}\{\tau_{k}<\infty\}\lambda\left(\theta^{*}_{d}, \theta_{d}; Y^{(\tau_{k})}_{d}\right) + \bm{1}\{\tau_{k}=\infty\}\lambda\left(\theta^{*}_{d}, \theta_{d}; {Z^{\prime}}^{(k)}_{d}\right)$ and $b_{k}=\lambda\left(\theta^{*}_{d}, \theta_{d}; Z^{(k)}_{d}\right)$. Note that $a_{k}$ is well-defined even when $\tau_{k}=\infty$. Then we will show that $\{a_{k}\}$ is a process identically distributed as $\{b_{k}\}$. For any Borel measurable set $A_{1},A_{2},\dots$, the following equality holds for all $k$:
    \begin{align*}
         \mathbb{E}\left[\bm{1}\{a_{k}\in A_{k}\}\middle\vert \mathcal{E}^{\prime}_{\tau_{k}}\right]
        = & P\left(\lambda\left(\theta^{*}_{d},\theta_{d};Y^{(\tau_{k})}_{d}\right)\in A_{k}\right) P(\tau_{k} < \infty\vert \mathcal{E}^{\prime}_{\tau_{k}}) \\
        + P\left(\lambda\left(\theta^{*}_{d},\theta_{d};{Z^{\prime}}^{(k)}_{d}\right)\in A_{T}\right)P(\tau_{k} = \infty\vert \mathcal{E}^{\prime}_{\tau_{k}})\\
        = & P\left(\lambda\left(\theta^{*}_{d},\theta_{d};Y_{d}\right)\in A_{k}\right) \left(P(\tau_{k} < \infty\vert \mathcal{E}^{\prime}_{\tau_{k}}) + P(\tau_{k} = \infty\vert \mathcal{E}^{\prime}_{\tau_{k}})\right) \\
        = & P\left(\lambda\left(\theta^{*}_{d},\theta_{d};Y_{d}\right)\in A_{k}\right).
    \end{align*}
    
    Note that the second equality follows from the fact that $Y^{(\tau_{k})}_{d}$ is independent of the $\sigma$-algebra $\mathcal{E}^{\prime}_{\tau_{k}}$. Therefore, 
    \begin{align*}\allowdisplaybreaks[4]
         &\ \mathbb{P}((a_{1},a_{2},\dots,a_{T})\in A_{1}\times A_{2}\times\cdots\times A_{T}) \\
        =&\  \mathbb{E}\left[\prod_{k=1}^{T}\bm{1}\{a_{k}\in A_{k}\}\right] \\
        =&\  \mathbb{E}\left[\mathbb{E}\left[\prod_{k=1}^{T}\bm{1}\{a_{k}\in A_{k}\}\middle\vert \mathcal{E}^{\prime}_{\tau_{T}}\right]\right] \\
        =&\ \mathbb{E}\left[\prod_{k=1}^{T-1}\bm{1}\{a_{k}\in A_{k}\} \cdot \mathbb{E}\left[\bm{1}\{a_{T}\in A_{T}\}\middle\vert \mathcal{E}^{\prime}_{\tau_{T}}\right]\right] \\
        =&\ \mathbb{E}\left[\prod_{k=1}^{T-1}\bm{1}\{a_{k}\in A_{k}\}\right] \cdot \mathbb{P}\left(\lambda\left(\theta^{*}_{d},\theta_{d};Y_{d}\right)\in A_{T}\right) \\
        =&\  \prod_{k=1}^{T}\mathbb{P}\left(\lambda\left(\theta^{*}_{d},\theta_{d};Y_{d}\right)\in A_{k}\right) \\
        =&\ \mathbb{P}((b_{1},b_{2},\dots,b_{T})\in A_{1}\times A_{2}\times\cdots\times A_{T}).
    \end{align*}
    
    Thus $(a_{1},a_{2},\dots,a_{T})$ and $(b_{1},b_{2},\dots,b_{T})$ have the same probability on any rectangular set in $\mathbb{R}^{T}$ for any $T$. Therefore, $\{a_{k}\}$ and $\{b_{k}\}$ are equally distributed by Kolmogorov's extension theorem. Moreover, let $B$ be the event where the design $d$ under context $c$ is sampled infinitely often, that is, $B:=\{\tau_{k}<\infty, ~\forall k\in\mathbb{N}\}$. Then we see that
    \begin{align}
         &\ \mathbb{P}\left( \left\{\lim\inf_{T\rightarrow \infty}\sup_{\theta_{d}\in \Theta}\bigg\vert \frac{1}{T}\sum_{k=1}^{T}\bm{1}\{\tau_{k}\leq T\}\left(\lambda\left(\theta^{*}_{d}, \theta_{d}; Y^{(\tau_{k})}_{d}\right) - D(\theta^{*}_{d}\Vert \theta_{d})\right) \bigg\vert > 0\right\} \bigcap B\right) \notag\\
        \leq &\ \mathbb{P}\left( \left\{\lim\inf_{T\rightarrow \infty}\sup_{\theta_{d}\in \Theta}\bigg\vert \sum_{k=1}^{T}\frac{\bm{1}\{\tau_{k}\leq T\}}{\sum_{l=1}^{T}\bm{1}\{\tau_{l}\leq T\}}\left(\lambda\left(\theta^{*}_{d}, \theta_{d}; Y^{(\tau_{k})}_{d}\right) - D(\theta^{*}_{d}\Vert \theta_{d})\right) \bigg\vert > 0\right\} \bigcap B\right) \notag\\
        = &\ \mathbb{P}\left( \left\{\lim\inf_{T\rightarrow \infty}\sup_{\theta_{d}\in \Theta}\bigg\vert \frac{1}{T}\sum_{k=1}^{T}\left(a_{k} - D(\theta^{*}_{d}\Vert \theta_{d})\right) \bigg\vert > 0\right\} \bigcap B\right) \notag\\
        \leq &\ \mathbb{P}\left( \left\{\lim\inf_{T\rightarrow \infty}\sup_{\theta_{d}\in \Theta}\bigg\vert \frac{1}{T}\sum_{k=1}^{T}\left(a_{k} - D(\theta^{*}_{d}\Vert \theta_{d})\right) \bigg\vert > 0\right\}\right) \label{eq12}\\
        = &\ \mathbb{P}\left( \left\{\lim\inf_{T\rightarrow \infty}\sup_{\theta_{d}\in \Theta}\bigg\vert \frac{1}{T}\sum_{k=1}^{T}\left(b_{k} - D(\theta^{*}_{d}\Vert \theta_{d})\right) \bigg\vert > 0\right\}\right) = \ 0, \notag
    \end{align}
    where the last equality follows from the GC class assumption in Assumption 1. Moreover, we also have 
    \begin{align}
        \mathbb{P}\left( \left\{\lim\inf_{T\rightarrow \infty}\sup_{\theta_{d}\in \Theta}\bigg\vert\frac{1}{T}\sum_{k=1}^{\infty}\bm{1}\{\tau_{k}\leq T\}\left(\lambda\left(\theta^{*}_{d}, \theta_{d}; Z^{(k)}_{d}\right) - D(\theta^{*}_{d}\Vert \theta_{d})\right) \bigg\vert > 0 \right\} \bigcap B^{C}\right) = 0, \label{eq13}
    \end{align}
    since the log-likelihood ratio function and the KL divergence are bounded due to the compactness and smoothness assumptions in Assumptions 1 and 2. Combining Equations (\ref{eq24})-(\ref{eq13}),
    \begin{equation}\label{eq3}
        \sup_{\theta_{d}\in \Theta}\bigg\vert \frac{1}{T}\sum_{t=1}^{T}\xi_{t}(d)\lambda\left(\theta^{*}_{d}, \theta_{d}; Y^{(t)}_{d}\right) - \frac{1}{T}\sum_{t=1}^{T}\xi_{t}(d)D(\theta^{*}_{d}\Vert \theta_{d}) \bigg\vert \rightarrow 0,\quad a.s..
    \end{equation}
    
    Note that $\sum_{t=1}^{T}(\xi_{t}(d) - \psi_{t}(d))$ is a (bounded) martingale difference sequence with respect to the filtration $\mathcal{E}^{\prime}_{t}$, and the KL divergence $D(\theta^{*}_{d}\Vert \theta_{d})$ is universally bounded. Therefore, the law of large numbers of martingale difference sequences \citep{chow71} implies
    \begin{equation}\label{eqec4}
        \sup_{\theta_{d}\in \Theta}\bigg\vert \frac{1}{T}\sum_{t=1}^{T}\xi_{t}(d)D(\theta^{*}_{d}\Vert \theta_{d}) - \frac{1}{T}\sum_{t=1}^{T}\psi_{t}(d)D(\theta^{*}_{d}\Vert \theta_{d}) \bigg\vert \rightarrow 0, \quad a.s..
    \end{equation}
    
    To partially conclude, Equation (\ref{eq1}) follows from (\ref{eq3}) and (\ref{eqec4}), and the triangle inequality. An immediate consequence of (\ref{eq1}) is
    \begin{equation}\label{eq6}
        \lim_{T\rightarrow \infty}{ -\frac{1}{T}\ln{ \int_{\Tilde{\bm{\Theta}}}{\exp\{-TW_{T}(\bm{\theta})\}d\bm{\theta}} } + \frac{1}{T}\ln{ \int_{\Tilde{\bm{\Theta}}}{\exp\{-TD_{\bar{\psi}_{T}}(\bm{\theta}^{*}\Vert \bm{\theta})\}d\bm{\theta}} } } = 0,\quad a.s..
    \end{equation}
    
    Then using Laplace approximations, we show that for any open set $\Tilde{\bm{\Theta}}\subseteq \bm{\Theta}$,
    \begin{equation}\label{eq5}
        \lim_{T\rightarrow \infty}{ -\frac{1}{T}\ln{ \int_{\Tilde{\bm{\Theta}}}{\exp\{-TD_{\bar{\psi}_{T}}(\bm{\theta}^{*}\Vert \bm{\theta})\}d\bm{\theta}} } - \inf_{\bm{\theta}\in\tilde{\bm{\Theta}}}{D_{\bar{\psi}_{T}}(\bm{\theta}^{*}\Vert\bm{\theta})} } = 0.
    \end{equation}
    
    The left-hand side is apparently non-negative. For the other direction, let $r:=\sup_{\theta\in \Theta} \Vert \partial D(\theta^{*}\Vert \theta)/\partial \theta\Vert$, which is finite since $\Theta$ is compact, and $D(\theta^{*}\Vert \theta)$ is continuously differentiable. For any $r\varepsilon>0$, one can choose a finite $\varepsilon$-cover $O$ of $\Tilde{\bm{\Theta}}$ such that each $o\in O$ intersects $\tilde{\bm{\Theta}}$. Let $\bm{\theta}_{T}\in\bm{\Theta}$ that achieves the infimum $\inf_{\bm{\theta}\in\tilde{\bm{\Theta}}}{D_{\bar{\psi}_{T}}(\bm{\theta}^{*}\Vert\bm{\theta})}$, which possibly does not lie in $\Tilde{\theta}$, and $o_{T}\in O$ be any cover whose closure contains $\bm{\theta}_{T}$. Then,
    \begin{align*}
        -\frac{1}{T}\ln{ \int_{\Tilde{\bm{\Theta}}}{\exp\{-TD_{\bar{\psi}_{T}}(\bm{\theta}^{*}\Vert \bm{\theta})\}d\bm{\theta}} } &\leq -\frac{1}{T}\ln{ \int_{\Tilde{\bm{\Theta}}\bigcap o_{T}}{\exp\{-TD_{\bar{\psi}_{T}}(\bm{\theta}^{*}\Vert \bm{\theta})\}d\bm{\theta}} } \\
        &\leq -\frac{1}{T}\ln{ \int_{\Tilde{\bm{\Theta}}\bigcap o_{T}}{\exp\{-T(D_{\bar{\psi}_{T}}(\bm{\theta}^{*}\Vert \bm{\theta}_{T}) + r\varepsilon)\}d\bm{\theta}} } \\
        &\leq -\frac{1}{T}\ln\int_{\Tilde{\bm{\Theta}}\bigcap o_{T}}{d\bm{\theta}} + \inf_{\bm{\theta}\in\tilde{\bm{\Theta}}}{D_{\bar{\psi}_{T}}(\bm{\theta}^{*}\Vert\bm{\theta})} + r\varepsilon,
    \end{align*}
    where the second inequality follows from $\big\vert D(\bm{\theta}^{*}\Vert \bm{\theta}) - D(\bm{\theta}^{*}\Vert \bm{\theta}_{T})\big\vert \leq r\Vert\bm{\theta} -\bm{\theta}_{T} \Vert \leq r\varepsilon$. Sending $T$ to infinity and choosing $\varepsilon$ to be arbitrarily small yields equation (\ref{eq5}).

    Equation{s} (\ref{eq6}) and (\ref{eq5}) combine to yield
    \begin{equation}\label{eq7}
        \lim_{T\rightarrow \infty}{ -\frac{1}{T}\ln{ \int_{\Tilde{\bm{\Theta}}}{\exp\{-TW_{T}(\bm{\theta})\}d\bm{\theta}} } - \inf_{\bm{\theta}\in\tilde{\bm{\Theta}}}{D_{\bar{\psi}_{T}}(\bm{\theta}^{*}\Vert\bm{\theta})} } = 0,\quad a.s..
    \end{equation}
    
    Note that $\ln\Pi_{T}{(\Tilde{\bm{\Theta}})} = \ln\int_{\tilde{\bm{\Theta}}}{ \exp\{-TW_{T}(\bm{\theta})\} d\bm{\theta} } - \ln\int_{\operatorname{int}(\bm{\Theta})}{ \exp\{-TW_{T}(\bm{\theta})\} d\bm{\theta} }$, where $\operatorname{int}(\cdot)$ indicates the interior part of a set. Then the conclusion follows from (\ref{eq7}) and the fact that $\inf_{\bm{\theta}\in \operatorname{int}(\bm{\Theta})}{D_{\bar{\psi}_{T}}(\bm{\theta}^{*}\Vert \bm{\theta})} = \inf_{\bm{\theta}\in \bm{\Theta}}{D_{\bar{\psi}_{T}}(\bm{\theta}^{*}\Vert \bm{\theta})} = D_{\bar{\psi}_{T}}(\bm{\theta}^{*}\Vert \bm{\theta}^{*}) = 0$.

    Finally, we prove the second statement. Fix a context $c$ and a design $d\in\mathcal{D}_{c}$. A similar argument to (\ref{eq12}) implies with probability 1, if $\sum_{t=1}^{T}\sum_{d^{\prime}\in\mathcal{D}_{c}}{\xi_{t}(d^{\prime})}\rightarrow \infty$, then
    \begin{equation}\label{eq25}
        \sup_{\theta_{d}\in \Theta}\Bigg\vert \frac{\sum_{t=1}^{T}\xi_{t}(d)\lambda\left(\theta^{*}_{d}, \theta_{d}; Y^{(t)}_{d}\right)}{\sum_{t=1}^{T}\sum_{d^{\prime}\in\mathcal{D}_{c}}{\xi_{t}(d^{\prime})}} - \frac{\sum_{t=1}^{T}\xi_{t}(d)D(\theta^{*}_{d}\Vert \theta_{d})}{\sum_{t=1}^{T}\sum_{d^{\prime}\in\mathcal{D}_{c}}{\xi_{t}(d^{\prime})}} \Bigg\vert \rightarrow 0.
    \end{equation}
    
    By Levy's extension of the Borel-Cantelli Lemma \cite{will91}, we see that with probability 1, if $T\bar{\alpha}_{T}(c) \rightarrow \infty$, then 
    \begin{equation}\label{eq26}
        \frac{\sum_{t=1}^{T}\sum_{d^{\prime}\in\mathcal{D}_{c}}{\xi_{t}(d^{\prime})}}{T\bar{\alpha}_{T}(c)} = \frac{\sum_{t=1}^{T}\sum_{d^{\prime}\in\mathcal{D}_{c}}{\xi_{t}(d^{\prime})}}{\sum_{t=1}^{T}\sum_{d^{\prime}\in\mathcal{D}_{c}}{\psi_{t}(d^{\prime})}} \rightarrow 1,
    \end{equation}
    and
    \begin{equation}\label{eq27}
        \sup_{\theta_{d}\in \Theta}\Bigg\vert \frac{\sum_{t=1}^{T}\xi_{t}(d)D(\theta^{*}_{d}\Vert \theta_{d})}{\sum_{t=1}^{T}\sum_{d^{\prime}\in\mathcal{D}_{c}}{\xi_{t}(d^{\prime})}} - \frac{\sum_{t = 1}^{T}{\psi_{t}(d)D(\theta^{*}_{d}\Vert \theta_{d})}}{T\bar{\alpha}_{T}(c)} \Bigg\vert \rightarrow 0.
    \end{equation}
    
    Combining (\ref{eq25})-(\ref{eq27}) and the triangle inequality, we arrive at a counterpart of (\ref{eq1})
    \begin{equation*}
        \sup_{\bm{\theta}\in\bm{\Theta}}\left\vert \sum_{d\in \mathcal{D}_{c}}\frac{1}{T\bar{\alpha}_{T}(c)}\sum_{t=1}^{T}\xi_{t}(d)\lambda\left(\theta^{*}_{d},\theta_{d};Y_{d}^{(t)}\right) - D_{\bar{\beta}_{T}(c,\cdot)}(\bm{\theta}^{*}\Vert \bm{\theta})\right\vert \rightarrow 0,\quad a.s..
    \end{equation*}
    
    Then the second statement follows immediately by applying the Laplace approximation trick again.
    \halmos
\endproof

\section{Consistency proof of TTTS-C.}
Then we establish Lemma \ref{lem:ec3}, which is an intermediate result for the consistency proof. 
\begin{lemma}\label{lem:ec3}
Under Assumptions 1 and 2, with probability 1, if $\Psi_{T}(d) := T\bar{\psi}_{T}(d) \rightarrow \infty$ for some $c\in \mathcal{C}$ and $d\in\mathcal{D}_{c}$, then for all $\varepsilon>0$ such that $\left\{\bm{\theta}\in\bm{\Theta} \middle \vert \Vert \bm{\theta} - \bm{\theta}^{*} \Vert_{2} < \varepsilon \right\}$ has positive Lebesgue measure, we have
\begin{equation}\label{eq11}
    \Pi_{T}\left(\left\{\bm{\theta}\in\bm{\bm{\Theta}} \middle| \big\vert \mu_{d} - \mu_{d}^{*} \big\vert\geq \varepsilon\right\} \right) \rightarrow 0,
\end{equation}
where $\vert \cdot\vert$ is the Euclidean norm; if $\lim_{T\rightarrow \infty}\Psi_{T}(d) < \infty$ for some $c\in \mathcal{C}$ and $d\in\mathcal{D}_{c}$, then for any open set $(\mu^{\prime}, \mu^{\prime\prime}) \subseteq M$, 
\begin{equation}\label{eq14}
    \inf_{T}\ \Pi_{T}\left(\left\{\bm{\theta}\in\bm{\Theta} \middle| \mu_{d}\in(\mu^{\prime}, \mu^{\prime\prime})\right\} \right)> 0.
\end{equation}
\end{lemma}

\proof{Proof of Lemma \ref{lem:ec3}.}
Without loss of generality, assume $\bm{\theta}^{*}$ is an interior point of $\bm{\Theta}$. Now, fix a context $c\in\mathcal{C}$ and a design $d\in\mathcal{D}_{c}$. Recall that we assume an uninformative prior so that the marginal density of the posterior distribution in $\theta_{d}$ is 
\begin{equation}\label{eq10}
    \Pi_{T,d}(\theta) = \frac{L_{T,d}(\theta)}{\int_{\Theta}{L_{T,d}(\theta^{\prime})d\theta^{\prime}}} = \frac{\exp\{-\sum_{t=1}^{T}\xi_{t}(d)\lambda(\theta^{*},\theta;Y^{(t)})\}}{\int_{\Theta}{\exp\{-\sum_{t=1}^{T}\xi_{t}(d)\lambda(\theta^{*},\theta^{\prime};Y^{(t)})\}d\theta^{\prime}}}.
\end{equation}

Note that we leave out any subscript $d$ if no confusion arises. First, we consider the case when $\Psi_{T}(d)\rightarrow \infty$. Recall that equation (\ref{eq12}) implies that with probability 1, if $\Psi_{T}(d) \rightarrow \infty$, then 
\begin{equation*}
    \sup_{\theta\in \Theta} \left(\sum_{t=1}^{T}{\xi_{t}(d)}\right)^{-1}\left\vert \sum_{t=1}^{T}\xi_{t}(d)\lambda(\theta^{*},\theta;Y^{(t)}) - \Psi_{T}(d)D(\theta^{*}\Vert \theta) \right\vert \rightarrow 0.
\end{equation*}

Together with Levy's extension of the Borel-Cantelli Lemma \citep{will91}, which implies that $\sum_{t=1}^{T}{\xi_{t}(d)}/\Psi_{T}(d)\rightarrow 1$, we can conclude that, with probability 1, if $\Psi_{T}(d) \rightarrow \infty$, then 
\begin{equation*}
    \sup_{\theta\in \Theta} \Psi_{T}^{-1}(d)\left\vert \sum_{t=1}^{T}\xi_{t}(d)\lambda(\theta^{*},\theta;Y^{(t)}) - \Psi_{T}(d)D(\theta^{*}\Vert \theta) \right\vert \rightarrow 0.
\end{equation*}

Let $v_{T}$ denote the supremum, and then (\ref{eq10}) can be bounded by
\begin{equation*}
    \Pi_{T,d}(\theta) \leq \frac{ \exp\{-\Psi_{T}(d)(D(\theta^{*}\Vert \theta) - v_{T})\} }{ \int_{\Theta}{ \exp\{-\Psi_{T}(d)(D(\theta^{*}\Vert \theta^{\prime}) + v_{T})\} d\theta^{\prime}} }.
\end{equation*}

To bound the posterior probability in (\ref{eq11}), we investigate the behavior of the KL divergence. Let $F:=\inf_{\theta:\Vert \theta -\theta^{*}\Vert_{2}\geq \varepsilon}D(\theta^{*}\Vert \theta)$. For each $\theta$ such that $\Vert \theta-\theta^{*} \Vert_{2}\geq \varepsilon$, we have $D(\theta^{*}\Vert \theta) > D(\theta^{*}\Vert \theta^{*}) = 0$ by Lemma 5.35 of \cite{vdv98}. Since every continuous function on a compact set attains its infimum, we conclude that $F>0$. Recall that $r=\sup_{\theta\in \Theta} \Vert \partial D(\theta^{*}\Vert \theta)/\partial \theta\Vert\leq \infty$. Then for $\varepsilon^{\prime} = F/2r$, and for any $\theta$ such that $\Vert \theta-\theta^{*}\Vert_{2} \leq \varepsilon^{\prime}$, 
\begin{equation*}
    D(\theta^{*}\Vert \theta) \leq D(\theta^{*}\Vert \theta^{*}) + r\Vert \theta - \theta^{*}\Vert \leq F/2.
\end{equation*}

Consequently, we see that
\begin{align*}
     \Pi_{T,d}\left(\left\{\bm{\theta} \in \bm{\Theta}\middle\vert \big\vert \mu_{d}-\mu_{d}^{*} \big\vert \geq \varepsilon\right\}\right)
    \leq &\frac{\int_{\{\Vert \theta -\theta^{*}\Vert_{2}\geq \varepsilon\}}{ \exp\{-\Psi_{T}(d)(D(\theta^{*}\Vert \theta) - v_{T})\} d\theta}}{\int_{\{\Vert \theta-\theta^{*}\Vert_{2}\leq \varepsilon^{\prime}\}}{ \exp\{-\Psi_{T}(d)(D(\theta^{*}\Vert \theta) + v_{T})\} d\theta}} \\
    \leq &\frac{\operatorname{Diam}(\Theta)^{d+1}}{(\varepsilon^{\prime})^{d+1}}\exp\left\{ -\Psi_{T}(d)\bigg( F - F/2 - 2v_{T} \bigg) \right\},
\end{align*}
where $\operatorname{Diam}(\cdot)$ indicates the diameter of sets. Since $\Psi_{T}\rightarrow\infty$, $v_{T}\rightarrow 0$, and $F > 0$, this bound vanishes almost surely.

Now consider the second part. Again by Levy's extension of the Borel-Cantelli Lemma, with probability 1, if $\lim_{T\rightarrow \infty}\Psi_{T} < \infty$, then $\lim_{T\rightarrow \infty}\sum_{t=1}^{T}\xi_{t}(d) < \infty$. Fix any realization of $\xi_{t}(d)$'s and $Y^{(t)}$'s with $\lim_{T\rightarrow \infty}\sum_{t=1}^{T}\xi_{t}(d) < \infty$, we have $\sup_{\theta\in \Theta}\sup_{T\in\mathbb{N}}{\left\vert \sum_{t=1}^{T}{\xi_{t}(d)\lambda(\theta^{*},\theta;Y^{(t)})} \right\vert} < \infty$ thanks to Assumption 1. By (\ref{eq10}), $\delta<\Pi_{T,d}(\theta)<\delta^{-1}$ is bounded both from below and above for some $\delta>0$, and (\ref{eq14}) follows immediately. 
\halmos
\endproof

Lemma \ref{lem:ec3} shows that the posterior distribution concentrates around the true value as long as the cumulative sampling ratio put into the design is not finite ultimately. The converse is also true as in (\ref{eq14}), which is the key to the consistency result. Notice that this result is not a trivial corollary of Theorem \ref{thm:1} since we do not require a positive ratio of samples. Actually, the identifiability condition in Assumption 1 is a necessity in establishing (\ref{eq11}).

Then we can complete the proof for Theorem \ref{thm:5} by showing a stronger result.

\begin{theorem}\label{thm:ec1}
Suppose tuning hyperparameter $\gamma_{t}(c)$ is adaptive to $\{\mathcal{E}_{t}\}_{t\geq 1}$, and then the proposed TTTS-C with tuning hyperparameter $\gamma_{t}(c)$ is consistent, i.e., ${\sum_{t=1}^{T}\xi_{t}(d)}\rightarrow \infty$, $a.s.$, as $T\rightarrow \infty$, $\forall c\in\mathcal{C}$, $\forall d\in\mathcal{D}_{c}$.
\end{theorem}

\proof{Proof.}
    We argue by contradiction. To this end, suppose the set of inadequately sampled context-design pairs $\mathscr{I} = \{(c,d):c\in\mathcal{C}, d\in\mathcal{D}_{c}, \lim_{T\rightarrow \infty}\sum_{t=1}^{T}{\xi_{t}(d)}<\infty\}$ is non-empty. We first assume that $\sum_{t=1}^{\infty}\gamma_{t}(c) = \infty$. Let $\hat{\bm{\theta}}_{2,k}^{(t)}$, $\forall k\geq 1$ denote the independent copies of $\hat{\bm{\theta}}_{2}^{(t)}$ generated in Algorithm \ref{alg:1} or Algorithm \ref{alg:2}. The probability of sampling on context $c$ and design $d$ conditioned on $\mathcal{E}_{t}$ at time $t$ can be lower bounded by
    \begin{equation}\label{eq17}
        \mathbb{P}(\xi_{t}(d) \vert \mathcal{E}_{t}) \geq \mathbb{P}\left(\left(\hat{\mu}_{1}^{(t)}\right)_{d} > \max_{d^{\prime}\in\mathcal{D}_{c}\backslash\{d\}} \left(\hat{\mu}_{1}^{(t)}\right)_{d^{\prime}} \bigg\vert \mathcal{E}_{t}\right) \cdot \mathbb{P}\left(\left(\hat{\mu}_{2,1}^{(t)}\right)_{d} < \min_{d^{\prime}\in\mathcal{D}_{c}\backslash\{d\}} \left(\hat{\mu}_{2,1}^{(t)}\right)_{d^{\prime}} \bigg\vert \mathcal{E}_{t}\right) \cdot \gamma_{t}(c)\frac{1}{\vert \mathcal{D}_{c}\vert} \cdot \frac{1}{\vert\mathcal{C}\vert}.
    \end{equation}
    
    This inequality holds because the right-hand side provides a lower bound for the probability of the event where the posterior sample $\hat{\bm{\theta}}_{1}^{(t)}$ indicates that design $d$ is the best design under context $c$, and the first independent copy $\hat{\bm{\theta}}_{2,1}^{(t)}$ of the posterior sample $\hat{\bm{\theta}}_{2}^{(t)}$ indicates that design $d$ is the worst design under context $c$, so that context $c$ and design $d$ will be simulated. The factor $1/\vert \mathcal{C}\vert$ arises since $\Delta_{t}$ may also include contexts other than $c$ in the event we described above, and similarly, for the factor $1/\vert \mathcal{D}_{c}\vert$.

    By Lemma \ref{lem:ec3}, for any $(c,d)\in\mathcal{P}$, any sufficiently small number $\varepsilon>0$, there exists $\bar{T}\geq 1$ such that for any $t\geq \bar{T}$, 
    \begin{align*}
        \mathbb{P}\left(\left(\hat{\mu}_{1}^{(t)}\right)_{d} > \max_{d^{\prime}\in\mathcal{D}_{c}\backslash\{d\}} \left(\hat{\mu}_{1}^{(t)}\right)_{d^{\prime}} \bigg\vert \mathcal{E}_{t}\right) \geq \Pi_{t}\left( \bigcap_{d^{\prime}\neq d} \left\{ \mu_{d^{\prime}} < \mu^{*}_{d^{*}(c)} + \varepsilon/2 \right\}  \bigcap \left\{ \mu_{d} > \mu^{*}_{d^{*}(c)} + \varepsilon \right\} \right) > \varepsilon,
    \end{align*}
    and similarly for the second factor in (\ref{eq17}). Then we see that $$\sum_{t=\bar{T}}^{\infty}{\psi_{t}(d)} = \sum_{t=\bar{T}}^{\infty}{\mathbb{P}(\xi_{t}(d) \vert \mathcal{E}_{t})} \geq \frac{\varepsilon^{2}}{\vert\mathcal{C}\vert\vert\mathcal{D}\vert}\sum_{t=\bar{T}}^{\infty}{\gamma_{t}(c)}$$ diverges according to the assumption $\sum_{t=1}^{\infty}\gamma_{t}(c) = \infty$. Conversely, if $\sum_{t=1}^{\infty}\gamma_{t}(c) < \infty$, then $\sum_{t=1}^{\infty}1-\gamma_{t}(c) = \infty$, and a similar argument apply. By Levy's extension of the Borel-Cantelli lemma \citep{will91}, $\lim_{T\rightarrow \infty}\sum_{t=1}^{T}{\xi_{t}(d)}<\infty$ holds with zero probability. This completes the proof.
    \halmos
\endproof

\section{Characterizing the optimal sampling ratios.}
\proof{Proof of Proposition \ref{prop:1}.}
    (a) Fix the second argument $y$ for $G_{d}(x, y)$. By definition, $G_{d}(x, y) = \inf_{(\theta_{d}, \theta_{d^{*}(c)}): \mu_{d} \geq \mu_{d^{*}(c)}}{x D(\theta_{d}^{*}\Vert \theta_{d}) + y D(\theta_{d^{*}(c)}^{*}\Vert \theta_{d^{*}(c)})}$. Suppose for $\tilde{x}$ and $\breve{x}$ with $\tilde{x}<\breve{x}$, the infimum is taken at $(\tilde{\theta}_{d}, \tilde{\theta}_{d^{*}(c)})$ and $(\breve{\theta}_{d}, \breve{\theta}_{d^{*}(c)})$, respectively. Then we see that
    \begin{align*}
        G_{d}(\tilde{x}, y) &= \tilde{x} D(\theta_{d}^{*}\Vert \tilde{\theta}_{d}) + y D(\theta_{d^{*}(c)}^{*}\Vert \tilde{\theta}_{d^{*}(c)}) \\
        &\leq \tilde{x} D(\theta_{d}^{*}\Vert \breve{\theta}_{d}) + y D(\theta_{d^{*}(c)}^{*}\Vert \breve{\theta}_{d^{*}(c)}) \\
        &\leq  \breve{x} D(\theta_{d}^{*}\Vert \breve{\theta}_{d}) + y D(\theta_{d^{*}(c)}^{*}\Vert \breve{\theta}_{d^{*}(c)}) \\
        &= G_{d}(\breve{x}, y),
    \end{align*}
    where the first inequality follows from the definition of $(\tilde{\theta}_{d}, \tilde{\theta}_{d^{*}(c)})$, and the second follows from the non-negativity of KL divergence. For the second argument, a symmetric proof applies.
        
    (b) The first statement follows immediately by observing that $G_{d}(x, y)$ is the infimum of a set of linear functions in $(x, y)$. Then the local continuity at any inner point $(x, y)\in \left(\mathbb{R}^{+}\right)^{2}$ follows consequently. Now it suffices to show that the local continuity at the boundary of $\left(\mathbb{R}^{+}\right)^{2}$ holds. Without loss of generality, suppose $\{(x_{n}, y_{n})\}_{n\geq 1}$ is a sequence of inner points of $\left(\mathbb{R}^{+}\right)^{2}$, i.e., $x_{n}>0$ and $y_{n}>0$, with $x_{n} \rightarrow x_{0} \geq 0$ and $y_{n} \rightarrow 0$. By $G_{d}(x_{n}, y_{n}) \leq x_{n} D(\theta^{*}_{d} \Vert \theta^{*}_{d}) + y_{n} D(\theta^{*}_{d^{*}(c)} \Vert \theta^{*}_{d}) \leq y_{n} \sup_{\theta\in \Theta}D(\theta^{*}_{d^{*}(c)} \Vert \theta) \rightarrow 0$, we see that $G_{d}(x_{n}, y_{n}) \rightarrow G_{d}(x_{0}, 0) = 0$. To sum up, $G_{d}(\cdot, \cdot)$ is a continuous function.
        
    (c) It follows immediately from the definition
    \begin{align*}
        G_{d}(hx, hy) &= \inf_{(\theta_{d}, \theta_{d^{*}(c)}): \mu_{d} \geq \mu_{d^{*}(c)}}{hx D(\theta_{d}^{*}\Vert \theta_{d}) + hy D(\theta_{d^{*}(c)}^{*}\Vert \theta_{d^{*}(c)})} \\
        &= h \inf_{(\theta_{d}, \theta_{d^{*}(c)}): \mu_{d} \geq \mu_{d^{*}(c)}}{x D(\theta_{d}^{*}\Vert \theta_{d}) + y D(\theta_{d^{*}(c)}^{*}\Vert \theta_{d^{*}(c)})} \\
        &= h G_{d}(x, y). \halmos
    \end{align*}
    
\endproof

Then we prove Proposition \ref{prop:2}.
\proof{Proof of Proposition \ref{prop:2}.}
    The optimization problem (\ref{opt:2}) can be rewritten as
    \begin{align}
        \Gamma^{*} = \max_{\alpha,\beta\geq 0}&\quad z \label{opt:8}\\
        s.t. &\quad z \leq \alpha(c) z_{c}, & &\hspace{-3em}\forall c\in\mathcal{C}, \notag \\
        &\quad z_{c} \leq G_{d}\left(\beta(c,d^{*}(c)), \beta(c, d)\right), & &\hspace{-3em}\forall c\in\mathcal{C}, \forall d\in\mathcal{U}_{c}, \notag \\
        & \quad \sum_{c\in\mathcal{C}}{\alpha(c)} = 1, \quad  \sum_{d\in \mathcal{D}_{c}}{\beta(c,d)} = 1, & &\hspace{-3em} \forall c\in\mathcal{C}. \notag
    \end{align}
    
    Therein, decision variables $z$ and $z_{c}, ~c\in\mathcal{C}$ are introduced to facilitate analysis. The KKT conditions provide sufficient and necessary conditions \citep{rockafellar70} for (\ref{opt:8}), since the constraints are apparently strictly feasible and satisfy the Slater's condition. The KKT conditions are as follows
    \begin{gather*}
        \sum_{d^{\prime}\in\mathcal{U}_{c}} \Lambda_{2}(c, d^{\prime})\frac{\partial}{\partial x_{1}}G_{d^{\prime}}(\beta(c, d^{*}(c)), \beta(c, d^{\prime})) - \Lambda_{4}(c) = 0,\quad \forall c\in\mathcal{C}, \\
        \Lambda_{2}(c, d)\frac{\partial}{\partial x_{2}}G_{d}(\beta(c, d^{*}(c)), \beta(c, d)) - \Lambda_{4}(c) = 0,\quad \forall c\in\mathcal{C}, \forall d\in\mathcal{U}_{c}, \\
        1 - \sum_{c\in\mathcal{C}} \Lambda_{1}(c) = 0, \quad \Lambda_{1}(c)z_{c} - \Lambda_{3} = 0,\quad \Lambda_{1}(c)\alpha(c) - \sum_{d^{\prime}\in\mathcal{U}_{c}}\Lambda_{2}(c, d^{\prime}) = 0,\quad \forall c\in\mathcal{C},\\
        \Lambda_{1}(c) \left( \alpha(c)z_{c} - z \right) = 0, \quad \Lambda_{2}(c, d) \left(G_{d}(\beta(c, d^{*}(c)), \beta(c, d)) - z_{c}\right) = 0, \quad \forall c\in\mathcal{C}, \forall d\in\mathcal{U}_{c},
    \end{gather*}
    where $\Lambda_{1}(c)\geq 0, \Lambda_{2}(c, d)\geq 0, \Lambda_{3}\in\mathbb{R}$, and $\Lambda_{4}(c)\in\mathbb{R}$ are dual variables, and for all $c\in\mathcal{C}$ and $d\in\mathcal{D}_{c}$,
    \begin{equation*}
        \left(\frac{\partial}{\partial x_{1}}G_{d}(-\beta(c, d^{*}(c)), \beta(c, d)), -\frac{\partial}{\partial x_{2}}G_{d}(\beta(c, d^{*}(c)), \beta(c, d))\right)
    \end{equation*}
    denotes a subgradient of $-G_{d}(\cdot, \cdot)$ evaluated at $(\beta(c, d^{*}(c)), \beta(c, d))$.
    
    Note that any optimal solution $\alpha^{*}, \beta^{*}$ together with $z^{*}, z^{*}_{c}$ is strictly positive. According to
    $\Lambda_{1}(c)z^{*}_{c} + \Lambda_{3} = 0$, we have either $\Lambda_{1}(c) = \Lambda_{3} = 0$, $\forall c\in\mathcal{C}$, or $\Lambda_{1}(c) > 0$ and $\Lambda_{3} \neq 0$, $\forall c\in\mathcal{C}$. However, the former contradicts $1 - \sum_{c^{\prime}\in\mathcal{C}}\Lambda_{1}(c^{\prime}) =0$, hence $\Lambda_{1}(c), \Lambda_{3} \neq 0$. Finally, we show that $\Lambda_{2}(c, d)>0$ and $\Lambda_{4}(c)\neq 0$, $\forall c\in\mathcal{C}, d\in\mathcal{D}_{c}$. Again assume for the sake of contradiction that $\Lambda_{4}(c) = 0$ for some $c\in\mathcal{C}$. Then following the first two conditions and the non-negativity of $\Lambda_{2}$ and subgradients $\partial G_{d}/\partial x_{i}$, we have 
    \begin{equation*}
        \Lambda_{2}(c, d)\frac{\partial}{\partial x_{1}}G_{d}(\beta^{*}(c, d^{*}(c)), \beta^{*}(c, d)) = \Lambda_{2}(c, d)\frac{\partial}{\partial x_{2}}G_{d}(\beta^{*}(c, d^{*}(c)) , \beta^{*}(c, d)) = 0.
    \end{equation*}
    
    Note that $(0, 0)$ is not a subgradient of $-G_{d}(\cdot, \cdot)$ due to the fact that $G_{d}(\cdot, \cdot)$ is homogeneous of degree 1. Therefore, $\Lambda_{2}(c, d) = 0$, $\forall~d\in\mathcal{D}_{c}$, a contradiction to $\Lambda_{1}(c)\alpha^{*}(c) > 0$. 
    
    Now, $\Lambda_{4}(c) \neq 0$ implies $\partial G_{d}(\beta^{*}(c, d^{*}(c)) , \beta^{*}(c, d))/\partial x_{2} \neq 0$, $\forall c\in\mathcal{C}, d\in\mathcal{D}_{c}$. Combining the first two KKT conditions, we have
    \begin{equation*}
        \sum_{d\in\mathcal{U}_{c}}{\frac{\partial G_{d}(\beta^{*}(c,d^{*}(c)), \beta^{*}(c,d))/\partial x_{1}}{\partial G_{d}(\beta^{*}(c,d^{*}(c)), \beta^{*}(c,d))/\partial x_{2}}} = 1,\quad \forall c\in\mathcal{C}.
    \end{equation*}
    
    And $\Lambda_{1}(c)>0$, $\Lambda_{2}(c, d)>0$, and the last two KKT conditions yield
    \begin{equation*}
        \alpha^{*}(c)G_{d}(\beta^{*}(c,d^{*}(c)), \beta^{*}(c,d)) = z^{*}, \quad \forall c\in\mathcal{C}, d\in\mathcal{U}_{c}.
    \end{equation*}

    On the other hand, we show that any feasible solution $(\alpha^{*}, \beta^{*}, z^{*})$ satisfying the conditions in Proposition \ref{prop:2} consititutes an optimal solution to (\ref{opt:8}). Actually, it follows simply by choosing $z_{c} = z^{*} / \alpha^{*}(c)$, $\Lambda_{1}(c) = \alpha^{*}(c)$, $\Lambda_{2}(c, d) = -\Lambda_{4}(c)/(\partial G_{d}(\beta^{*}(c, d^{*}(c)) , \beta^{*}(c, d))/\partial x_{2})$, $\Lambda_{3} = z^{*}$, and $\Lambda_{4} = (\alpha^{*}(c))^{2} / \sum_{d^{\prime}\in\mathcal{U}_{c}}(\partial G_{d}(\beta^{*}(c, d^{*}(c)) , \beta^{*}(c, d^{\prime}))/\partial x_{2})$.
    \halmos
\endproof

Similar arguments work for the following two propositions.
\proof{Proof of Proposition \ref{prop:3}.} 
The optimization problem (\ref{opt:3}) can be written similarly as (\ref{opt:8}) with an additional constraint $\beta(c,d^{*}(c)) = \gamma(c)$. 
    \begin{align}
        \Gamma^{*} = \max_{\alpha,\beta\geq 0}&\quad z \label{opt:9}\\
        s.t. &\quad z \leq \alpha(c) z_{c}, & &\hspace{-3em}\forall c\in\mathcal{C}, \notag \\
        &\quad z_{c} \leq G_{d}\left(\beta(c,d^{*}(c)), \beta(c, d)\right), & &\hspace{-3em}\forall c\in\mathcal{C}, \forall d\in\mathcal{U}_{c}, \notag \\
        & \quad \sum_{c\in\mathcal{C}}{\alpha(c)} = 1, \quad  \sum_{d\in \mathcal{D}_{c}}{\beta(c,d)} = 1, & &\hspace{-3em} \forall c\in\mathcal{C}, \notag \\
        & \quad \beta(c, d^{*}(c)) = \gamma(c), & &\hspace{-3em} \forall c\in\mathcal{C}. \notag
    \end{align}
    
Then the corresponding KKT conditions coincide with those of (\ref{opt:8}), except that the first equation is replaced by feasible conditions $\beta(c,d^{*}(c)) = \gamma(c)$. For any feasible solution satisfying (\ref{eq:11}), KKT conditions are met automatically. 

On the other hand, consider an optimal solution $\alpha^{*}, \beta^{*}, z_{c}^{*}, z^{*}$ to (\ref{opt:9}). Note that $\alpha^{*}(c) = z^{-1}_{c}/\sum_{c^{\prime}\in\mathcal{C}}z^{-1}_{c^{\prime}}$ and $z^{*} = 1/\sum_{c^{\prime}\in\mathcal{C}}z^{-1}_{c^{\prime}}$.
We then show that for each $c\in\mathcal{C}$, there exists $\tilde{\beta}$ such that $G_{d}(\gamma(c), \tilde{\beta}(c,d)) = z^{*}_{c}$ for all $d\in\mathcal{U}_{c}$. This follows immediately if $G_{d}(\gamma(c), \beta^{*}(c,d)) = z^{*}_{c}$ for all $d\in\mathcal{U}_{c}$. On the contrary, if $G_{d}(\gamma(c), \beta^{*}(c,d)) > z^{*}_{c}$ for some $d$, there exists $d^{\prime}\in\mathcal{U}_{c}$ such that $G_{d^{\prime}}(\gamma(c), \beta) = z^{*}_{c}$ for all $\beta \geq \beta^{*}(c, d^{\prime})$. Otherwise, for each $d^{\prime}\in\mathcal{U}_{c}\backslash\{d\}$, there exists $\beta_{0}>\beta^{*}(c,d^{\prime})$ such that $G_{d^{\prime}}(\gamma(c), \beta_{0}) > G_{d^{\prime}}(\gamma(c), \beta^{*}(c, d^{\prime}))$. According to Proposition 2, $G_{d^{\prime}}(\cdot, \cdot)$ is concave, and thus for $\varepsilon \leq \beta_{0} - \beta^{*}(c, d^{\prime})$
\begin{equation*}
    \frac{G_{d^{\prime}}(\gamma(c), \beta^{*}(c,d^{\prime}) + \varepsilon) - G_{d^{\prime}}(\gamma(c), \beta^{*}(c,d^{\prime}))}{\varepsilon} \geq \frac{G_{d^{\prime}}(\gamma(c), \beta_{0}) - G_{d^{\prime}}(\gamma(c), \beta^{*}(c,d^{\prime}))}{\beta_{0} - \beta^{*}(c,d^{\prime})} > 0.
\end{equation*}

Since $G_{d}(\cdot, \cdot)$ is uniformly continuous, the solution $\bar{\beta}(c, d) = \beta^{*}(c, d) - (\vert \mathcal{U}_{c}\vert - 1)\varepsilon >0$ and $\bar{\beta}(c, d^{\prime}) = \beta^{*}(c, d^{\prime}) + \varepsilon > 0$ results in
\begin{equation*}
    \min_{d\in\mathcal{U}_{c}}G_{d}(\gamma(c), \bar{\beta}(c, d)) > \min_{d\in\mathcal{U}_{c}}G_{d}(\gamma(c), \beta^{*}(c, d)).
\end{equation*}

Moreover, there exist $\bar{\alpha}, \bar{z}_{c}$, and $\bar{z}$, with $\bar{z} > z^{*}$, such that $\bar{\alpha}, \bar{\beta}, \bar{z}_{c}$, and $\bar{z}$ yield a feasible solution with a larger objective, which contradicts the optimality condition. Then, we fix $d^{\prime}\in\mathcal{U}_{c}\backslash\{d\}$ with $G_{d^{\prime}}(\gamma(c), \beta) = z^{*}_{c}$ for all $\beta \geq \beta^{*}(c, d^{\prime})$. Finally, we define $\tilde{\beta}(c, d)$ implicitly by an arbitrary solution to $G_{d}(\gamma(c), \tilde{\beta}(c, d)) = z^{*}_{c}$ for $d\in\mathcal{U}_{c}\backslash\{d^{\prime}\}$. Note that it's well-defined by the intermidiate value theorem since $G_{d}(\gamma(c), 0) = 0$, $G_{d}(\gamma(c), \beta^{*}(c, d)) \geq z^{*}_{c}$, and $G_{d}(\cdot, \cdot)$ is continuous. Define $\tilde{\beta}(c, d^{\prime}) = 1- \gamma(c) - \sum_{d\in\mathcal{U}_{c}\backslash\{d^{\prime}\}}{\tilde{\beta}(c, d)}$, and then $(\alpha^{*}, \tilde{\beta}, z^{*}_{c}, z^{*})$ suffices an optimal solution to (\ref{opt:3}) that satisfies the optimality condition (\ref{eq:11}).

The second statement follows immediately by noting that when $\gamma(c) = \beta^{*}(c, d^{*}(c))$, where $(\alpha^{*}, \beta^{*})$ constitutes an optimal solution to (\ref{opt:2}), then an optimal solution of (\ref{opt:3}) is also optimal to (\ref{opt:2}). The conclusion follows immediately from Proposition \ref{prop:2}.
\halmos
\endproof

\proof{Proof of Proposition \ref{prop:5}.}
The optimization problem (\ref{opt:5}) can be rewritten as 
\begin{align}
    \Gamma^{*} = \max_{\alpha,\beta\geq 0}&\quad z \label{opt:10}\\
    s.t. &\quad z \leq \alpha(c) z_{c}, & &\hspace{-3em}\forall c\in\mathcal{C}, \notag \\
    &\quad z_{c} \leq G_{d, d^{\prime}}\left(\beta(c,d), \beta(c, d^{\prime})\right), & &\hspace{-3em}\forall c\in\mathcal{C}, \forall d\in\mathcal{P}_{c}, d\in\mathcal{U}_{c}, \notag \\
    & \quad \sum_{c\in\mathcal{C}}{\alpha(c)} = 1, \quad  \sum_{d\in \mathcal{D}_{c}}{\beta(c,d)} = 1, & &\hspace{-3em} \forall c\in\mathcal{C}. \notag
\end{align}

The first order conditions and complementary conditions with regard to the decision variable $\beta$ and the second inequality constraint are given by 
\begin{gather}
    \sum_{d^{\prime}\in\mathcal{U}_{c}} \Lambda_{2}(c, d,  d^{\prime})\frac{\partial}{\partial x_{1}}G_{d, d^{\prime}}(\beta(c, d), \beta(c, d^{\prime})) - \Lambda_{4}(c) = 0,\quad \forall c\in\mathcal{C}, \forall d\in\mathcal{P}_{c}, \label{eq58}\\
    \sum_{d\in\mathcal{P}_{c}} \Lambda_{2}(c, d, d^{\prime})\frac{\partial}{\partial x_{2}}G_{d, d^{\prime}}(\beta(c, d), \beta(c, d^{\prime})) - \Lambda_{4}(c) = 0,\quad \forall c\in\mathcal{C}, \forall d^{\prime}\in\mathcal{U}_{c}, \label{eq59}
\end{gather}
and
\begin{equation}
    \Lambda_{2}(c, d, d^{\prime}) \left(G_{d, d^{\prime}}(\beta(c, d), \beta(c, d^{\prime})) - z_{c}\right) = 0, \quad \forall c\in\mathcal{C}, \forall d\in\mathcal{P}_{c}, d^{\prime}\in\mathcal{U}_{c}, \label{eq60}
\end{equation}
where $\partial/\partial x_{i}$'s are defined as in Proposition \ref{prop:2}, and $\Lambda_{2}(c, d, d^{\prime})$ and $\Lambda_{4}(c)$ are non-negative Lagrangian multipliers. Recall that the rate functions $G_{d, d^{\prime}}(\cdot, \cdot)$ are homogeneous of degree 1, and thus for any $r>0$,
\begin{align*}
    & (r - 1)G_{d,d^{\prime}}(\beta(c, d), \beta(c, d^{\prime})) \\
    = & G_{d,d^{\prime}}(r\beta(c, d), r\beta(c, d^{\prime})) - G_{d,d^{\prime}}(\beta(c, d), \beta(c, d^{\prime})) \\
    \leq & (r - 1)\beta(c, d)\frac{\partial}{\partial x_{1}}G_{d, d^{\prime}}(\beta(c, d), \beta(c, d^{\prime})) + (r - 1)\beta(c, d^{\prime})\frac{\partial}{\partial x_{2}}G_{d, d^{\prime}}(\beta(c, d), \beta(c, d^{\prime})),
\end{align*}
which can only happen when 
\begin{equation*}
    G_{d,d^{\prime}}(\beta(c, d), \beta(c, d^{\prime})) = \beta(c, d)\frac{\partial}{\partial x_{1}}G_{d, d^{\prime}}(\beta(c, d), \beta(c, d^{\prime})) + \beta(c, d^{\prime})\frac{\partial}{\partial x_{2}}G_{d, d^{\prime}}(\beta(c, d), \beta(c, d^{\prime})).
\end{equation*}

Multiplying (\ref{eq58}) and (\ref{eq59}) with $\beta(c,d)$ and $\beta(c, d^{\prime})$, respectively, and summing up these equations, we see that
\begin{equation*}
    \sum_{d\in\mathcal{P}_{c}}\sum_{d^{\prime}\in\mathcal{U}_{c}}\Lambda_{2}(c, d, d^{\prime})G_{d,d^{\prime}}(\beta(c, d), \beta(c, d^{\prime})) = \Lambda_{4}(c)\sum_{d\in\mathcal{P}_{c}}\beta(c, d) + \Lambda_{4}(c)\sum_{d^{\prime}\in\mathcal{U}_{c}}\beta(c, d^{\prime}) = \Lambda_{4}(c). 
\end{equation*}

According to (\ref{eq60}), $$\sum_{d\in\mathcal{P}_{c}}\sum_{d^{\prime}\in\mathcal{U}_{c}}\Lambda_{2}(c, d, d^{\prime})z_{c} = \Lambda_{4}(c).$$

Note the fact that any optimal solution $(\alpha^{*}, \beta^{*}, z^{*}, z^{*}_{c})$ to (\ref{opt:10}) is positive, we see that either $\Lambda_{4}(c) > 0$ or $\Lambda_{2}(c, d, d^{\prime}) = 0$ for all $d\in\mathcal{P}_{c}$ and $d^{\prime}\in\mathcal{U}_{c}$. The latter, however, is impossible due to the following KKT conditions
\begin{equation*}
    1 - \sum_{c\in\mathcal{C}}\Lambda_{1}(c) = 0,\quad \Lambda_{1}(c)z_{c} - \Lambda_{3} = 0,\quad \Lambda_{1}(c)\alpha(c) - \sum_{d\in\mathcal{P}_{c}}\sum_{d^{\prime}\in\mathcal{U}_{c}}\Lambda_{2}(c, d, d^{\prime})=0,
\end{equation*}
where $\Lambda_{1}(c)$ and $\Lambda_{3}$ are non-negative Lagrangian multipliers.

Since $\Lambda_{4}(c) > 0$, combining (\ref{eq58}) and (\ref{eq59}), for any $d\in\mathcal{P}_{c}$ and $d^{\prime}\in\mathcal{U}_{c}$,
\begin{equation*}
    \max_{\tilde{d}\in\mathcal{P}_{c}}\Lambda_{2}(c, \tilde{d}, d^{\prime}) > 0,\quad\text{ and }\quad \max_{\tilde{d}^{\prime}\in\mathcal{U}_{c}}\Lambda_{2}(c, d, \tilde{d}^{\prime}) > 0.
\end{equation*}

Together with the complementary slackness condition (\ref{eq60}), we have
\begin{equation*}
    \min_{\tilde{d}\in\mathcal{P}_{c}}G_{\tilde{d}, d^{\prime}}(\beta(c, \tilde{d}), \beta(c, d^{\prime})) = \min_{\tilde{d}^{\prime}\in\mathcal{U}_{c}} G_{d, \tilde{d}^{\prime}}(\beta(c, d), \beta(c, \tilde{d}^{\prime})) = z_{c}.
\end{equation*}

The remaining part of the proof is a simple tautology of the proof for Proposition \ref{prop:2}, which we omit to avoid redundancy.
\halmos
\endproof

\section{Rate-optimality of TTTS-C.}\label{sec:rateoptproof}
In this section, we establish the asymptotic optimality of TTTS-C. Again, we first focus on the case where $m_{c} = 1$. We prove Lemma \ref{lem:1}, Lemma \ref{lem:2}, and Lemma \ref{lem:3} in Section \ref{sec:optratelemma}, and show Theorem \ref{thm:3} and Corollary \ref{corol:1} in Section \ref{sec:optratethm1}.

\subsection{Sampling ratios of TTTS-C.}\label{sec:optratelemma}
The three lemmas below derive the (conditional) sampling ratio put into each design under TTTS-C. We restate Lemma \ref{lem:1} for demonstration as below.
\contextbestratio*

\proof{Proof.} Equality (\ref{eq:5}) follows from the definition of $\psi_{t}$ naturally. Let $\hat{\bm{\theta}}_{2,k}^{(t)}$, $\forall~k\geq 1$ denote the independent copies of $\hat{\bm{\theta}}_{2}^{(t)}$ generated in Algorithm \ref{alg:1}, and $\hat{d}_{2,k}^{(t)}(c) = \arg\max_{d\in\mathcal{D}_{c}}(\hat{\mu}^{(t)}_{2,k})_{d}$, accordingly. We define an array of stopping times by $\kappa_{c} = \min\{k\geq 1: \hat{d}_{2,k}^{(t)}(c) \neq \hat{d}_{1}^{(t)}(c)\}$. Conditioned on $\mathcal{E}_{t-1}$ and $\hat{d}_{1}(\cdot)$, each $\kappa_{c}$ follows an independent geometric distribution with parameter $\left(1 - \pi_{(c,\hat{d}_{1}^{(t)}(c))}^{(t-1)}\right)$. By construction, for all $\Delta\in \mathscr{P}(\mathcal{C})$,
\begin{align*}
    \mathbb{P}\left(\Delta_{t} = \Delta \middle\vert \mathcal{E}_{t-1}, \hat{d}_{1}(\cdot)\right) &= \sum_{\tau = 1}^{\infty}\mathbb{P}\left(\kappa_{c^{\prime}} = \tau, \forall~c^{\prime}\in\Delta, \;\kappa_{c^{\prime}} > \tau, \forall~c^{\prime^{\prime}}\notin\Delta \middle\vert \mathcal{E}_{t-1}, \hat{d}_{1}(\cdot)\right)\\
    &= \sum_{\tau = 1}^{\infty} \prod_{c^{\prime}\in\Delta}{\left(\pi_{(c^{\prime},\hat{d}_{1}^{(t)}(c^{\prime}))}^{(t-1)}\right)^{\tau - 1}\left(1 - \pi_{(c^{\prime},\hat{d}_{1}^{(t)}(c^{\prime}))}^{(t-1)}\right)}\prod_{c^{\prime}\notin\Delta}{\left(\pi_{(c^{\prime},\hat{d}_{1}^{(t)}(c^{\prime}))}^{(\tau-1)}\right)^{\tau}} \\
    &= \frac{\prod_{c\in\Delta}{\left(1 - \pi_{(c,\hat{d}_{1}^{(t)}(c))}^{(t-1)}\right)}\prod_{c\notin\Delta}{\left(\pi_{(c,\hat{d}_{1}^{(t)}(c))}^{(t-1)}\right)}}{1 - \prod_{c\in\mathcal{C}}{\left(\pi_{(c,\hat{d}_{1}^{(t)}(c))}^{(t-1)}\right)}}.
\end{align*}

Therefore, 
\begin{align*}
    \mathbb{P}\left(C_{t} = c \middle\vert \mathcal{E}_{t-1}, \hat{d}_{1}(\cdot)\right) 
    &= \sum_{\Delta\in \mathscr{P}(\mathcal{C})}\mathbb{P}\left(\Delta_{t} = \Delta \middle\vert \mathcal{E}_{t-1}, \hat{d}_{1}(\cdot)\right)\cdot \mathbb{P}\left(C_{t} = c \middle\vert \mathcal{E}_{t-1}, \hat{d}_{1}(\cdot), \Delta_{t} = \Delta\right) \\
    &= \sum_{\Delta\in \mathscr{P}(\mathcal{C})}\mathbb{P}\left(\Delta_{t} = \Delta \middle\vert \mathcal{E}_{t-1}, \hat{d}_{1}(\cdot)\right)\cdot\frac{\bm{1}\{c\in\Delta\}}{\vert \Delta\vert} \\
    &= \sum_{\Delta\in \mathscr{P}(\mathcal{C}): c\in\Delta}\frac{\prod_{c^{\prime}\in\Delta}{\left(1 - \pi_{(c^{\prime},\hat{d}_{1}^{(t)}(c^{\prime}))}^{(t-1)}\right)}\prod_{c^{\prime}\notin\Delta}{\left(\pi_{(c^{\prime},\hat{d}_{1}^{(t)}(c^{\prime}))}^{(t-1)}\right)}}{1 - \prod_{c^{\prime}\in\mathcal{C}}{\left(\pi_{(c^{\prime},\hat{d}_{1}^{(t)}(c^{\prime}))}^{(t-1)}\right)}} \frac{1}{\vert\Delta\vert}.
\end{align*}

Then (\ref{eq:4}) follows immediately from 
\begin{equation*}
    \alpha_{t}(c) = \mathbb{P}\left(C_{t} = c \middle\vert \mathcal{E}_{t-1}\right) = \mathbb{E}\left[\mathbb{P}\left(C_{t} = c \middle\vert \mathcal{E}_{t-1}, \hat{d}_{1}(\cdot)\right) \middle\vert \mathcal{E}_{t-1}\right].
\end{equation*}

Then it suffices to show (\ref{eq:3}). Note that for each $c\in\mathcal{C}$, the estimated second best design $\hat{d}_{2,k}^{(t)}(c)$ under context $c$ follows a categorical distribution over $\mathcal{D}_{c}$, and that for $d\in\mathcal{D}_{c}$ with $d \neq \hat{d}_{1}^{(t)}(c)$,
\begin{equation*}
    \mathbb{P}\left( \hat{d}_{2,\kappa_{c}}^{(t)}(c) = d \middle\vert \mathcal{E}_{t-1}, \hat{d}_{1}(\cdot), \kappa_{c}\right) = \pi^{(t-1)}_{(c, d)} \big/ \left(1 - \pi^{(t-1)}_{(c, \hat{d}_{1}^{(t)}(c))}\right).
\end{equation*}

This implies that conditioned on $\mathcal{E}_{t-1}$ and $\hat{d}_{1}(\cdot)$, the random variable $\hat{d}_{2,\kappa_{c}}^{(t)}(c)$ is independent of $\kappa_{c}$, and thus independent of $C_{t}$. Therefore, for each $c\in\mathcal{C}$ and $d\in\mathcal{D}_{c}$ with $d \neq \hat{d}_{1}^{(t)}(c)$, we have
\begin{equation*}
    \mathbb{P}\left(\hat{d}^{(t)}_{2,\kappa_{c}}(c) = d \middle\vert \mathcal{E}_{t-1}, \hat{d}_{1}(\cdot)\right) = \frac{\pi^{(t-1)}_{(c, d)}}{\left(1 - \pi^{(t-1)}_{(c, \hat{d}_{1}^{(t)}(c))}\right)},
\end{equation*}
and
\begin{align*}
    \mathbb{P}\left(D_{t} = d \middle\vert \mathcal{E}_{t-1}, \hat{d}_{1}(\cdot)\right) = &\mathbb{P}\left(C_{t} = c, D_{t} = d \middle\vert \mathcal{E}_{t-1}, \hat{d}_{1}(\cdot)\right) \\
    = & \mathbb{P}\left(C_{t} = c \middle\vert \mathcal{E}_{t-1}, \hat{d}_{1}(\cdot)\right) \cdot \bm{1}\{\hat{d}_{1}^{(t)}(c) = d\} \cdot \gamma(c) \\
    + & \mathbb{P}\left(C_{t} = c, \hat{d}_{2,\kappa_{c}}^{(t)}(c) = d \middle\vert \mathcal{E}_{t-1}, \hat{d}_{1}(\cdot)\right) \cdot \bm{1}\{\hat{d}_{1}^{(t)}(c) \neq d\} \cdot (1-\gamma(c)) \\
    = & \left. \mathbb{P}\left(C_{t} = c \middle\vert \mathcal{E}_{t-1}, \hat{d}_{1}(\cdot)\right) \cdot \right\{\bm{1}\{\hat{d}_{1}^{(t)}(c) = d\} \cdot \gamma(c) \\
    + & \left. \mathbb{P}\left(\hat{d}_{2,\kappa_{c}}^{(t)}(c) = d \middle\vert \mathcal{E}_{t-1}, \hat{d}_{1}(\cdot)\right) \cdot \bm{1}\{\hat{d}_{1}^{(t)}(c) \neq d\} \cdot (1-\gamma(c)) \right\},
\end{align*}
where the last equality follows from the conditional independence of $\hat{d}_{2,\kappa_{c}}^{(t)}(c)$ and $C_{t}$. Then (\ref{eq:3}) follows immediately from the two equations above and
\begin{equation*}
    \beta_{t}(c, d) = \mathbb{P}\left(D_{t} = d \middle\vert \mathcal{E}_{t-1}\right) = \mathbb{E}\left[\mathbb{P}\left(D_{t} = d \middle\vert \mathcal{E}_{t-1}, \hat{d}_{1}(\cdot)\right) \middle\vert \mathcal{E}_{t-1}\right].\halmos
\end{equation*}
\endproof

Next, we present the proof of Lemma \ref{lem:2}.
\proof{Proof of Lemma \ref{lem:2}.} Fix $c\in\mathcal{C}$ and $d \neq d^{*}(c)$. For any fixed value of $d_{1}(-c)$, we have a lower bound of (\ref{eq:6})
\begin{equation}\label{eq34}
    J^{(t-1)}_{d_{1}(-c)} \geq A^{(t-1)}_{d_{1}(-c)}(1-\gamma(c))\frac{\pi^{(t-1)}_{(c, d)}\pi^{(t-1)}_{(c, d^{*}(c))}}{1 - B^{(t-1)}_{d_{1}(-c)}\cdot\pi^{(t-1)}_{(c, d^{*}(c))}}.
\end{equation}

On the other hand, (\ref{eq:7}) can be upper bounded by 
\begin{equation}\label{eq35}
    K^{(t-1)}_{d_{1}(-c)} \leq A^{(t-1)}_{d_{1}(-c)}\vert \mathcal{D}_{c}\vert \max_{d^{\prime}\in \mathcal{U}_{c}}\frac{\pi^{(t-1)}_{(c, d^{\prime})}\left(1 -\pi^{(t-1)}_{(c, d^{\prime})}\right)}{1 - B^{(t-1)}_{d_{1}(-c)}\cdot\pi^{(t-1)}_{(c, d^{\prime})}} = A^{(t-1)}_{d_{1}(-c)}\vert \mathcal{D}_{c} \vert \frac{\pi^{(t-1)}_{(c, d^{*}(c))}\left(1 -\pi^{(t-1)}_{(c, d^{*}(c))}\right)}{1 - B^{(t-1)}_{d_{1}(-c)}\cdot\pi^{(t-1)}_{(c, d^{*}(c))}}.
\end{equation}

Note that the numerical inequality $x(1-x) \geq y(1-y)$ for $1/2<x<1$ and $0<y\leq 1 - x$, and the observation that $\sum_{d^{\prime}\in \mathcal{\mathcal{D}}_{c}\backslash\{d^{*}(c)\}}\pi^{(t-1)}_{(c, d^{\prime})} = 1 - \pi^{(t-1)}_{(c, d^{*}(c))}$ implies the maximum is attained at $d^{\prime} = d^{*}(c)$. Combining (\ref{eq34}) and (\ref{eq35}), 
\begin{equation*}
    \frac{J^{(t-1)}_{d_{1}(-c)}}{K^{(t-1)}_{d_{1}(-c)}} \geq \frac{1-\gamma(c)}{\vert \mathcal{D}_{c} \vert} \frac{\pi^{(t-1)}_{(c, d)}}{1 - \pi^{(t-1)}_{(c, d^{*}(c))}},
\end{equation*}
which is irrelevant to $d_{1}(-c)$. Hence following (\ref{eq:5}),
\begin{equation*}
    \beta_{t}(c,d) = \frac{\sum_{d_{1}(-c)}{J^{(t-1)}_{d_{1}(-c)}}}{\sum_{d_{1}(-c)}{K^{(t-1)}_{d_{1}(-c)}}} \geq \frac{1-\gamma(c)}{\vert \mathcal{D}_{c} \vert} \frac{\pi^{(t-1)}_{(c, d)}}{1 - \pi^{(t-1)}_{(c, d^{*}(c))}}.
\end{equation*}

Similarly, we have
\begin{equation*}
    J^{(t-1)}_{d_{1}(-c)} \leq A^{(t-1)}_{d_{1}(-c)}\frac{\pi^{(t-1)}_{(c, d)}}{1 - B^{(t-1)}_{d_{1}(-c)}\cdot\pi^{(t-1)}_{(c, d^{*}(c))}},
\end{equation*}
and
\begin{equation*}
    K^{(t-1)}_{d_{1}(-c)} \geq A^{(t-1)}_{d_{1}(-c)}\frac{\pi^{(t-1)}_{(c, d^{*}(c))}\left(1 - \pi^{(t-1)}_{(c, d^{*}(c))}\right)}{1 - B^{(t-1)}_{d_{1}(-c)}\cdot\pi^{(t-1)}_{(c, d^{*}(c))}}.
\end{equation*}

Therefore, 
\begin{equation*}
    \beta_{t}(c,d) = \frac{\sum_{d_{1}(-c)}{J^{(t-1)}_{d_{1}(-c)}}}{\sum_{d_{1}(-c)}{K^{(t-1)}_{d_{1}(-c)}}} \leq \frac{\pi^{(t-1)}_{(c, d)}}{\pi^{(t-1)}_{(c, d^{*}(c))}\left(1 - \pi^{(t-1)}_{(c, d^{*}(c))}\right)} \leq 2\frac{\pi^{(t-1)}_{(c, d)}}{1 - \pi^{(t-1)}_{(c, d^{*}(c))}}.
\end{equation*}

For $c\in\mathcal{C}$ and $d = d^{*}(c)\in\mathcal{D}_{c}$, we first focus on a fixed value of $d_{1}(-c)$, which coincides with $d^{*}(-c)$, i.e., $d_{1}(c^{\prime})=d^{*}(c^{\prime})$ for all $c^{\prime}\in\mathcal{C}\backslash\{c\}$. By Theorem \ref{thm:1}, we see that $\pi^{(t-1)}_{(c,d^{*}(c))} \rightarrow 1$. Then for sufficiently large $t$ and any $d^{\prime}\in\mathcal{U}_{c}$, we have $\pi^{(t-1)}_{(c,d^{*}(c))} \geq \pi^{(t-1)}_{(c,d^{\prime})}$, and thus
\begin{equation}\label{eq36}
    \frac{\pi^{(t-1)}_{(c,d^{\prime})}\cdot\pi^{(t-1)}_{(c,d^{*}(c))}}{1 - B^{(t-1)}_{d_{1}(-c)}\cdot \pi^{(t-1)}_{(c,d^{\prime})}} \leq \frac{\pi^{(t-1)}_{(c,d^{\prime})}\cdot\pi^{(t-1)}_{(c,d^{*}(c))}}{1 - \pi^{(t-1)}_{(c,d^{\prime})}} \leq \frac{(1 - \pi^{(t-1)}_{(c,d^{*}(c))})\cdot\pi^{(t-1)}_{(c,d^{*}(c))}}{1 - (1 - \pi^{(t-1)}_{(c,d^{*}(c))})} = 1 - \pi^{(t-1)}_{(c,d^{*}(c))},
\end{equation}
where the last inequality follows from the monotonicity of the function $x/(1-x)$. Now, (\ref{eq:6}) can be lower bounded by
\begin{equation}\label{eq37}
    J^{(t-1)}_{d_{1}(-c)} \geq A^{(t-1)}_{d_{1}(-c)}\gamma(c) \frac{\pi^{(t-1)}_{(c,d^{*}(c))}\left(1 - \pi^{(t-1)}_{(c,d^{*}(c))}\right)}{1 - B^{(t-1)}_{d_{1}(-c)}\cdot \pi^{(t-1)}_{(c,d^{*}(c))}},
\end{equation}
and upper bounded by
\begin{equation}\label{eq38}
    \begin{aligned}
        J^{(t-1)}_{d_{1}(-c)} &\leq A^{(t-1)}_{d_{1}(-c)}\left(\gamma(c) \frac{\pi^{(t-1)}_{(c,d^{*}(c))}\left(1 - \pi^{(t-1)}_{(c,d^{*}(c))}\right)}{1 - B^{(t-1)}_{d_{1}(-c)}\cdot \pi^{(t-1)}_{(c,d^{*}(c))}} + (1 - \gamma(c))\vert \mathcal{D}_{c} \vert \max_{d^{\prime}\in\mathcal{U}_{c}}\frac{\pi^{(t-1)}_{(c,d^{\prime})}\cdot\pi^{(t-1)}_{(c,d^{*}(c))}}{1 - B^{(t-1)}_{d_{1}(-c)}\cdot \pi^{(t-1)}_{(c,d^{\prime})}}\right) \\
        &\leq A^{(t-1)}_{d_{1}(-c)}\left(\gamma(c) \frac{\pi^{(t-1)}_{(c,d^{*}(c))}\left(1 - \pi^{(t-1)}_{(c,d^{*}(c))}\right)}{1 - B^{(t-1)}_{d_{1}(-c)}\cdot \pi^{(t-1)}_{(c,d^{*}(c))}} + (1 - \gamma(c))\vert \mathcal{D}_{c} \vert \left(1 - \pi^{(t-1)}_{(c,d^{*}(c))}\right)\right),
    \end{aligned}
\end{equation}
where the last inequality follows from (\ref{eq36}). Similarly, we can bound (\ref{eq:7}), for sufficiently large $t$, by 
\begin{equation}\label{eq39}
    K^{(t-1)}_{d_{1}(-c)} \geq A^{(t-1)}_{d_{1}(-c)}\frac{\pi^{(t-1)}_{(c,d^{*}(c))}\left(1 - \pi^{(t-1)}_{(c,d^{*}(c))}\right)}{1 - B^{(t-1)}_{d_{1}(-c)}\cdot \pi^{(t-1)}_{(c,d^{*}(c))}},
\end{equation}
and
\begin{equation}\label{eq40}
    K^{(t-1)}_{d_{1}(-c)} \leq A^{(t-1)}_{d_{1}(-c)}\left(\frac{\pi^{(t-1)}_{(c,d^{*}(c))}\left(1 - \pi^{(t-1)}_{(c,d^{*}(c))}\right)}{1 - B^{(t-1)}_{d_{1}(-c)}\cdot \pi^{(t-1)}_{(c,d^{*}(c))}} + \left(1 - \pi^{(t-1)}_{(c,d^{*}(c))}\right)\right).
\end{equation}

Combining (\ref{eq37})-(\ref{eq40}), we see that
\begin{equation}\label{eq41}
    \gamma(c)/\left(1 + \delta_{t}\right) \leq \frac{J^{(t-1)}_{d_{1}(-c)}}{K^{(t-1)}_{d_{1}(-c)}} \leq \gamma(c) \cdot \left(1 + \delta_{t}\right),
\end{equation}
where 
\begin{equation*}
    \delta_{t} = \vert \mathcal{D}_{c}\vert \frac{1 - B^{(t-1)}_{d_{1}(-c)}\cdot\pi^{(t-1)}_{(c,d^{*}(c))}}{\pi^{(t-1)}_{(c,d^{*}(c))}}.
\end{equation*}

Recall that by Theorem \ref{thm:1}, with probability 1, for each $\varepsilon > 0$, there exists an $N$ (not necessarily deterministic) such that for each $t\geq N$ and $c^{\prime}\in\mathcal{C}$, we have
\begin{equation}\label{eq42}
    \pi^{(t-1)}_{(c, d^{*}(c))}\geq 1 - \varepsilon / 2\vert\mathcal{C}\vert^{\vert \mathcal{D}\vert}.
\end{equation}

Then by the numerical inequality $1/(1-x) - (1-x)\leq 3x$ for $0\leq x\leq 1/2$, we see that $\delta_{t}\leq 3\varepsilon$. Moreover, $A^{(t-1)}_{d_{1}(-c)}, B^{(t-1)}_{d_{1}(-c)}\leq \varepsilon / 2\vert\mathcal{C}\vert^{\vert \mathcal{D}\vert}$ for any $d_{1}(-c)\neq d^{*}(-c)$. By definitions (\ref{eq:6}) and (\ref{eq:7}), $J^{(t-1)}_{d_{1}(-c)}/A^{(t-1)}_{d_{1}(-c)}$ and $K^{(t-1)}_{d_{1}(-c)}/A^{(t-1)}_{d_{1}(-c)}$ are increasing in $B^{(t-1)}_{d_{1}(-c)}$. Therefore, we conclude that for all $d_{1}(-c)\neq d^{*}(-c)$,
\begin{equation}\label{eq43}
    \frac{J^{(t-1)}_{d_{1}(-c)}}{J^{(t-1)}_{d^{*}(-c)}} \leq \frac{A^{(t-1)}_{d_{1}(-c)}}{A^{(t-1)}_{d^{*}(-c)}} \leq \frac{\varepsilon}{\vert\mathcal{C}\vert^{\vert \mathcal{D}\vert - 1}}, \quad \text{ and } \quad \frac{K^{(t-1)}_{d_{1}(-c)}}{K^{(t-1)}_{d^{*}(-c)}} \leq \frac{A^{(t-1)}_{d_{1}(-c)}}{A^{(t-1)}_{d^{*}(-c)}} \leq \frac{\varepsilon}{\vert\mathcal{C}\vert^{\vert \mathcal{D}\vert - 1}} .
\end{equation}

Combining (\ref{eq41})-(\ref{eq43}), we see that
\begin{equation*}
    \beta_{t}(c, d^{*}(c)) = \frac{J^{(t-1)}_{d^{*}(-c)} + \sum_{d_{1}(-c)\neq d^{*}(-c)}J^{(t-1)}_{d_{1}(-c)}}{K^{(t-1)}_{d^{*}(-c)} + \sum_{d_{1}(-c)\neq d^{*}(-c)}K^{(t-1)}_{d_{1}(-c)}} \leq \gamma(c)\cdot (1+\delta_{t})(1 + \varepsilon) \leq \gamma(c)(1 + 5\varepsilon).
\end{equation*}

This implies that with probability 1, $\limsup_{t\rightarrow\infty}\beta_{t}(c, d^{*}(c)) \leq \gamma(c)$. Similarly, we see that
\begin{equation*}
    \beta_{t}(c, d^{*}(c)) \geq \gamma(c)/(1 + 5\varepsilon),
\end{equation*}
and thus $\liminf_{t\rightarrow\infty}\beta_{t}(c, d^{*}(c)) \geq \gamma(c)$. This completes the proof. \halmos
\endproof

The next proof establishes Lemma \ref{lem:3} for the aggregate sampling ratio for each context.
\proof{Proof of Lemma \ref{lem:3}.}
Fix a context $c\in\mathcal{C}$. By Lemma \ref{lem:1}, the sampling effort for context $c$ can be lower bounded by reducing the summation in (\ref{eq:4}) to the dominating term with $d_{1}(\cdot) = d^{*}(c)$ and $\Delta = \{c\}$, i.e., with probability 1, for sufficiently large $t$,
\begin{equation}\label{eq49}
    \alpha_{t}(c) \geq \frac{\prod_{c^{\prime}\in\mathcal{C}}\left(\pi^{(t-1)}_{c^{\prime}, d^{*}(c^{\prime})}\right)}{1 - \prod_{c^{\prime}\in\mathcal{C}}\left(\pi^{(t-1)}_{c^{\prime}, d^{*}(c^{\prime})}\right)}\left(1 - \pi^{(t-1)}_{c, d^{*}(c)}\right)\prod_{c^{\prime}\in\mathcal{C}\backslash\{c\}}\left(\pi^{(t-1)}_{c^{\prime}, d^{*}(c^{\prime})}\right) \geq \frac{1}{2}\frac{1 - \pi^{(t-1)}_{c, d^{*}(c)}}{1 - \prod_{c^{\prime}\in\mathcal{C}}\left(\pi^{(t-1)}_{c^{\prime}, d^{*}(c^{\prime})}\right)},
\end{equation}
where the last inequality follows from Theorem \ref{thm:5}. The second inequality in (\ref{eq47}) follows immediately from the second term in (\ref{eq43}) by choosing $\varepsilon$ to be small enough. 

It remains to show (\ref{eq48}). By taking $\tilde{\bm{\Theta}} = \bigcup_{d^{\prime}\in\mathcal{D}_{c}}\{\mu_{d^{\prime}} > \mu_{d^{*}(c)}\}$ in Theorem \ref{thm:1}, combining (\ref{eq49}), we have
\begin{align*}
    \alpha_{t}(c) &\geq \frac{1}{2}\exp\left\{-(t-1)\left(\bar{\alpha}_{t-1}(c)\min_{d\in\mathcal{U}_{c}}G_{d}(\bar{\beta}_{t-1}(c, d^{*}(c)), \bar{\beta}_{t-1}(c, d)) + \varepsilon_{t-1}\right)\right\} \\
    &\geq \frac{1}{2}\exp\left\{-2\Gamma^{*}\cdot (t-1)\bar{\alpha}_{t-1}(c)\right\}, 
\end{align*}
where $\varepsilon_{t}$ is a vanishing sequence as $t\rightarrow 0$. Note the inequality that
\begin{align}
    T\bar{\alpha}_{T}(c) &= \sum_{t=1}^{T}{\alpha_{t}(c)} = (T-1)\bar{\alpha}_{T-1}(c) + \alpha_{T}(c) \notag\\ 
    &\geq (T-1)\bar{\alpha}_{T-1}(c) + \frac{1}{2}\exp\left\{-2\Gamma^{*}\cdot (T-1)\bar{\alpha}_{T - 1}(c)\right\}.\label{eq46}
\end{align}

Note that $T\bar{\alpha}_{T}(c)$ is non-decreasing with respect to $T$. Therefore, we have either $T\bar{\alpha}_{T}(c) \rightarrow \infty$ or $\lim_{T\rightarrow\infty}T\bar{\alpha}_{T}(c) =: L < \infty$. The latter immediately leads to a contradiction by $\frac{1}{2}\exp\{-2L\Gamma^{*}\} \leq T\bar{\alpha}_{T}(c) - (T-1)\bar{\alpha}_{T-1}(c) \rightarrow 0$. Now, the derivative of the right-hand side of (\ref{eq46}) with respect to $(T-1)\bar{\alpha}_{T-1}(c)$, i.e., $1 - \Gamma^{*}\exp\left\{-2\Gamma^{*}\cdot(T-1)\bar{\alpha}_{T - 1}(c)\right\}$, is non-negative for $T$ large enough since $(T-1)\bar{\alpha}_{T - 1}(c) \rightarrow \infty$. Therefore, if $T\bar{\alpha}_{T}(c) \geq \frac{1}{2\Gamma^{*}}\ln\left(C + \Gamma^{*}\cdot(T - 1)\right)$, then by (\ref{eq46}),
\begin{equation*}
    (T + 1)\bar{\alpha}_{T + 1}(c) \geq \frac{1}{2\Gamma^{*}}\ln\left(C + \Gamma^{*}\cdot(T - 1)\right) + \frac{1}{2}\cdot \frac{1}{C + \Gamma^{*}\cdot(T - 1)} \geq \frac{1}{2\Gamma^{*}} \ln\left(C + \Gamma^{*}\cdot T \right),
\end{equation*}
where the last inequality follows from the numerical inequality $x\geq \ln(1 + x)$ with $x = \Gamma^{*} / \left(C + \Gamma^{*}\cdot (T - 1)\right)$. Now it suffices to construct the random variable $C$ and apply the mathematical induction. Let $\Tilde{T}$ be the smallest number such that all of the aforementioned asymptotic statements hold. Then we complete the proof by setting $C = \exp\{2\Gamma^{*}\cdot \tilde{T}\bar{\alpha}_{\Tilde{T}}(c)\} - \Gamma^{*}\cdot (\tilde{T} - 1)$.
\halmos
\endproof

\subsection{Proof for $m_{c} = 1$.}\label{sec:optratethm1}

We first show that TTTS-C for selecting contextual best designs asymptotically satisfies the balance condition (\ref{eq:11}).

\rateoptone*
Immediately we have the following corollary.
\rateopt*

\proof{Proof.}
    Consider any subsequences $\{\bar{\alpha}_{T_{n}}\}_{n\geq 1}$ and $\{\bar{\beta}_{T_{n}}\}_{n\geq 1}$ of $\{\bar{\alpha}_{T}\}_{n\geq 1}$ and $\{\bar{\beta}_{T}\}_{n\geq 1}$, respectively. By Theorem \ref{thm:3}, we have
    \begin{equation}\label{eq28}
        \lim_{n\rightarrow \infty}\bar{\alpha}_{T_{n}}(c)G_{d}(\bar{\beta}_{T_{n}}(c,d^{*}(c)), \bar{\beta}_{T_{n}}(c,d)) - z = 0, \quad \forall c\in\mathcal{C}, d\in\mathcal{D}_{c}.
    \end{equation}
    
    Since $\bar{\alpha}_{T}$ and $\bar{\beta}_{T}$ are bounded, there exist cluster points $\alpha_{0}$ and $\beta_{0}$, and further subsequences $\{\bar{\alpha}_{T_{n_{k}}}\}_{k\geq 1}$ and $\{\bar{\beta}_{T_{n_{k}}}\}_{k\geq 1}$ that converge to the cluster points, respectively. By Lemma \ref{lem:2}, we have $\beta_{0}(c, d^{*}(c)) = \gamma(c), ~\forall c\in\mathcal{C}$. Therefore, $(\alpha_{0}, \beta_{0})$ is feasible for the optimization problem (\ref{opt:3}). Moreover, $(\alpha_{0}, \beta_{0})$ is optimal to (\ref{opt:3}), which follows immediately by noting that $G_{d}(\cdot, \cdot)$ is continuous for all $d\in\mathcal{D}$ and (\ref{eq28}). As a consequence,
    \begin{equation*}
    \begin{aligned}
        \lim_{k\rightarrow \infty} -\frac{1}{T_{n_{k}}}\ln \Pi_{T_{n_{k}}}\left( \bigcup_{c\in\mathcal{C}}\bigcup_{d\in\mathcal{D}\backslash\{d^{*}(c)\}}\bm{\Theta}_{(c, d)} \right) = & \lim_{k\rightarrow \infty}\min_{c\in\mathcal{C}}\min_{d\in\mathcal{D}_{c}\backslash\{d^{*}\}}\bar{\alpha}_{T_{n_{k}}}(c)G_{d}(\bar{\beta}_{T_{n_{k}}}(c,d^{*}(c)), \bar{\beta}_{T_{n_{k}}}(c,d)) \\
        = & \Gamma_{\bm{\gamma}}^{*}.
    \end{aligned}
    \end{equation*}
    
    That is, each subsequence has a convergent subsequence that converges to $\Gamma^{*}_{\bm{\gamma}}$, which leads to the first statement. The second statement is a direct consequence of Corollary \ref{corol:1} and the definition of $\Gamma_{\bm{\gamma}}^{*}$.\halmos
\endproof

Then we return to show the main result.
\proof{Proof of Theorem \ref{thm:3}.} 
This proof consists of three steps.

\noindent \textbf{Step 1: Consistency.} By Theorem \ref{thm:5} and Lemma \ref{lem:ec3}, for each context $c$ and each design $d\in\mathcal{U}_{c}$, and for $\varepsilon = \min_{c\in\mathcal{C}}\min_{d, d^{\prime}\in\mathcal{D}_{c}: d^{\prime}\neq d^{\prime\prime}}\vert \mu_{d^{\prime}} - \mu_{d^{\prime\prime}}\vert / 2$, as $t\rightarrow \infty$,
\begin{equation*}
    \pi^{(t)}_{(c, d)} = \Pi_{t}(\bm{\Theta}_{(c, d)}) \leq \Pi_{t}(\left\{\bm{\theta}\in\bm{\Theta} \middle| \big\vert \mu_{d} - \mu_{d}^{*} \big\vert\geq \varepsilon\right\}) + \Pi_{t}(\left\{\bm{\theta}\in\bm{\Theta} \middle| \big\vert \mu_{d^{*}(c)} - \mu_{d^{*}(c)}^{*} \big\vert\geq \varepsilon\right\}) \rightarrow 0.
\end{equation*}

Therefore, the posterior probability of a sub-optimal design being the context-dependent best is diminishing. As a consequence, Lemma \ref{lem:2} is valid.

\noindent \textbf{Step 2: Local balance of sampling ratios in a fixed context $c$.}

\noindent \textbf{Step 2.1.} Define $z_{c} := \Gamma^{*}_{\gamma(c)}$. We claim that, with probability 1,
\begin{equation}\label{eq44}
    \limsup_{T\rightarrow\infty}G_{d}(\gamma(c), \bar{\beta}_{T}(c, d)) - z_{c} \leq 0,\quad \forall d\in\mathcal{U}_{c}.
\end{equation}

We first show that the limit inferior of $\left(G_{d}(\gamma(c), \bar{\beta}_{T}(c, d)) - z_{c} \right)$ in (\ref{eq44}) is non-positive. Suppose otherwise that the limit inferior is $\delta > 0$. Then by Lemma \ref{lem:2}, with probability 1, for sufficiently large $T$,
\begin{equation}\label{eq45}
    \beta_{T}(c, d) \leq 2\frac{\pi^{(T-1)}_{(c, d)}}{1-\pi^{(T-1)}_{(c, d^{*}(c))}}\leq 2\frac{\pi^{(T-1)}_{(c, d)}}{\max_{d^{\prime}\in\mathcal{U}_{c}}\pi^{(T-1)}_{(c, d^{\prime})}}.
\end{equation}

According to the second statement of Theorem \ref{thm:1}, there exists a vanishing sequence of $\varepsilon_{T}$ such that $\varepsilon_{T}\rightarrow 0$, and with probability 1, for all $d^{\prime}\in\mathcal{D}_{c}$ and sufficiently large $T$,
\begin{gather}
\begin{aligned}
    & \exp\{-(T-1)\bar{\alpha}_{T-1}(c)\left(G_{d^{\prime}}(\gamma(c), \bar{\beta}_{T-1}(c, d^{\prime})) + \varepsilon_{T-1}\right)\} \leq \pi^{(T-1)}_{(c, d^{\prime})} \label{eq50}\\ 
    \leq & \exp\{-(T - 1)\bar{\alpha}_{T-1}(c)\left(G_{d^{\prime}}(\gamma(c), \bar{\beta}_{T-1}(c, d^{\prime})) - \varepsilon_{T-1}\right)\}. \notag
    \end{aligned}
\end{gather}

With a slight abuse of notation, we will refer to $\{\varepsilon_{T}\}_{T\geq 1}$ as several different vanishing sequences introduced below. Since the maximum of a finite number of vanishing sequences also vanishes, this notion can be formalized by redefining $\{\varepsilon_{T}\}_{T\geq 1}$ as maximum of these sequences, while we omit these tedious statements.

Let $\Gamma^{*}_{\gamma_{T}(c)}$ be the optimal value defined in optimization problem (\ref{opt:4}) with $\gamma(c)$ replaced by $\gamma_{T}(c) := \bar{\beta}_{T}(c, d^{*}(c))$. Then by definition, we have
$$\min_{d^{\prime}\in\mathcal{U}_{c}}G_{d}(\gamma_{T}(c), \bar{\beta}_{T}(c, d^{\prime})) \leq \Gamma^{*}_{\gamma_{T}(c)}.$$

According to Lemma \ref{lem:2}, we see that $\gamma_{T}(c) \rightarrow \gamma(c)$, and $\Gamma^{*}_{\gamma_{T}(c)} \rightarrow \Gamma^{*}_{\gamma(c)}$. Note that $G_{d}(\cdot, \cdot)$ is a concave function on a compact set, and thus is uniformly continuous. As a result, there exists a vanishing sequence $\{\tilde{\varepsilon}_{T}\}_{T\geq 1}$ such that 
\begin{equation}\label{eq51}
    \min_{d^{\prime}\in\mathcal{U}_{c}}G_{d}(\gamma(c), \bar{\beta}_{T}(c, d^{\prime})) \leq \Gamma^{*}_{\gamma(c)} + \tilde{\varepsilon}_{T} = z_{c} + \tilde{\varepsilon}_{T}.
\end{equation}

Combining (\ref{eq45}) with (\ref{eq50}), we see that for $T$ sufficiently large,
\begin{equation}\label{eq52}
    \begin{aligned}
        \beta_{T}(c, d) &\leq \exp\bigg\{-(T - 1)\bar{\alpha}_{T-1}(c)\bigg(G_{d}(\gamma(c), \bar{\beta}_{T-1}(c, d)) - \min_{d^{\prime}\in\mathcal{U}_{c}}G_{d^{\prime}}(\gamma(c), \bar{\beta}_{T-1}(c, d^{\prime})) - 2\varepsilon_{T-1}\bigg)\bigg\}\\
        &\leq \exp\{-(T - 1)\bar{\alpha}_{T-1}(c)\left(G_{d}(\gamma(c), \bar{\beta}_{T-1}(c, d)) - \Gamma^{*}_{\gamma(c)} - \tilde{\varepsilon}_{T} -2\varepsilon_{T-1}\right)\} \\
        &\leq \exp\{-(T-1)\bar{\alpha}_{T-1}(c)\cdot\delta / 2\},
    \end{aligned}
\end{equation}
where the second inequality follows from (\ref{eq51}). This results in negligible conditional sampling effort 
\begin{align*}
    \bar{\beta}_{T}(c, d) &= \frac{\sum_{t=1}^{T}{\alpha_{t}(c)\beta_{t}(c, d)}}{\sum_{t=1}^{T}{\alpha_{t}(c)}} \\
    &= \frac{\sum_{t=1}^{\tilde{T} - 1}{\alpha_{t}(c)\beta_{t}(c, d)}}{\sum_{t=1}^{T}{\alpha_{t}(c)}} + \frac{\sum_{t=\tilde{T}}^{T}{\alpha_{t}(c)\beta_{t}(c, d)}}{\sum_{t=1}^{T}{\alpha_{t}(c)}} \\
    &\leq \frac{\sum_{t=1}^{\tilde{T} - 1}{\alpha_{t}(c)\beta_{t}(c, d)}}{\sum_{t=1}^{T}{\alpha_{t}(c)}} + \frac{\sum_{t=\tilde{T}}^{T}{\alpha_{t}(c)\exp\{-(t-1)\bar{\alpha}_{t-1}(c)\cdot\delta / 2\}}}{\sum_{t=1}^{T}{\alpha_{t}(c)}},
\end{align*}
where $\tilde{T}$ is an arbitrary positive integer such that the aforementioned asymptotic bounds hold for $t\geq \tilde{T}$. The first term in the equality vanishes eventually, while the second term can be bounded using the Abel transformation
\begin{align*}
     & \frac{\sum_{t=\tilde{T}}^{T}{\alpha_{t}(c)\exp\{-(T-1)\bar{\alpha}_{T-1}(c)\cdot\delta / 2\}}}{\sum_{t=1}^{T}{\alpha_{t}(c)}} \\
    = &\frac{{\alpha_{T}(c)\exp\{-(T-1)\bar{\alpha}_{T-1}(c)\cdot\delta / 2\}}}{\sum_{t=1}^{T}{\alpha_{t}(c)}} \\
     &+ \frac{\sum_{t=\tilde{T}}^{T - 1}{\left(\sum_{\tau=\tilde{T}}^{t}\alpha_{\tau}(c)\right)\left(\exp\{-(t-1)\bar{\alpha}_{t-1}(c)\cdot\delta / 2\} - \exp\{-t\bar{\alpha}_{t}(c)\cdot\delta / 2\}\right)}}{\sum_{t=1}^{T}{\alpha_{t}(c)}} \\
    \leq &\exp\{-(T-1)\bar{\alpha}_{T-1}(c)\cdot\delta / 2\} + \sum_{t=\tilde{T}}^{T - 1}{\bigg(\exp\{-(t-1)\bar{\alpha}_{t-1}(c)\cdot\delta / 2\} - \exp\{-t\bar{\alpha}_{t}(c)\cdot\delta / 2\}\bigg)} \\
    = &\exp\{-(\tilde{T}-1)\bar{\alpha}_{\tilde{T}-1}(c)\cdot\delta / 2\}.
\end{align*}

Therefore, we see that $\limsup_{T\rightarrow\infty}\bar{\beta}_{T}(c, d) \leq \exp\{-(\tilde{T}-1)\bar{\alpha}_{\tilde{T}-1}(c)\cdot\delta / 2\}$ since $T\bar{\alpha}_{T}(c) \rightarrow \infty$ by Theorem \ref{thm:3}. Note that $\tilde{T}$ can be chosen arbitrarily large. Sending $\tilde{T}\rightarrow \infty$, we see that $\lim_{T\rightarrow\infty}\bar{\beta}_{T}(c, d) = 0$. This results in 
\begin{equation*}
    \lim_{T\rightarrow\infty}G_{d}(\bar{\beta}_{T}(c, d^{*}(c)), \bar{\beta}_{T}(c, d)) = G_{d}(\gamma(c), 0) = 0,
\end{equation*}
which contradicts the supposition that the limit inferior of $\left(G_{d}(\gamma(c), \bar{\beta}_{T}(c, d)) - z_{c} \right)$ in (\ref{eq44}) is positive. 

Then we can show that the limit superior of $\left(G_{d}(\gamma(c), \bar{\beta}_{T}(c, d)) - z_{c} \right)$ in (\ref{eq44}) is non-positive. Again we argue by contradiction. Suppose otherwise that the limit superior is $\delta > 0$. This can only occur when there is an increasing sequence of time steps $T_{1}<T_{2}<\ldots<T_{n}<\dots$, such that for any integer $k\geq 1$, $G_{d}(\gamma(c), \bar{\beta}_{T_{2k - 1}}(c, d)) - \Gamma^{*}_{\gamma(c)} \leq \delta/2$, $G_{d}(\gamma(c), \bar{\beta}_{T_{2k}}(c, d)) - \Gamma^{*}_{\gamma(c)} \geq \delta$, and $\delta/2 < G_{d}(\gamma(c), \bar{\beta}_{T}(c, d)) - \Gamma^{*}_{\gamma(c)} < \delta$ for any $T_{2k - 1} < T < T_{2k}$. Note that by definition, 
\begin{equation}\label{eq55}
    \left\vert \bar{\beta}_{T+1} - \bar{\beta}_{T}\right\vert = \left\vert \frac{\sum_{t=1}^{T+1}{\alpha_{t}(c)\beta_{t}(c, d)}}{\sum_{t=1}^{T+1}{\alpha_{t}(c)}} - \frac{\sum_{t=1}^{T}{\alpha_{t}(c)\beta_{t}(c, d)}}{\sum_{t=1}^{T}{\alpha_{t}(c)}}\right\vert \leq \frac{1}{\sum_{t=1}^{T}{\alpha_{t}(c)}} \rightarrow 0,
\end{equation}
hence $\vert G_{d}(\gamma(c), \bar{\beta}_{T+1}(c, d)) - G_{d}(\gamma(c), \bar{\beta}_{T}(c, d))\vert \rightarrow 0$. As a result, for sufficiently large $k$, we have $G_{d}(\gamma(c), \bar{\beta}_{T_{2k - 1}}(c, d)) - \Gamma^{*}_{\gamma(c)} > \delta / 4$. Since $G_{d}(\cdot, \cdot)$ is uniformly continuous, there exists $\tilde{\delta} > 0$ such that $\vert x - y \vert < \tilde{\delta}$ implies $\vert G_{d}(\gamma(c), x) - G_{d}(\gamma(c), y) \vert < \delta/2$. This in turn implies that 
\begin{equation*}
    \bar{\beta}_{T_{2k}}(c, d) - \bar{\beta}_{T_{2k - 1}}(c, d) \geq \tilde{\delta}.
\end{equation*}

Rearranging the terms, we see that
\begin{equation*}
    \sum_{t=T_{2k - 1} + 1}^{T_{2k}}\Big(\alpha_{t}(c)\beta_{t}(c, d)\Big) \geq \sum_{t=T_{2k - 1} + 1}^{T_{2k}}\left(\alpha_{t}(c) (\bar{\beta}_{T}(c, d) + \tilde{\delta})\right).
\end{equation*}

However, similar to (\ref{eq52}), for $T_{2k - 1}\leq T\leq T_{2k}$ and $k$ sufficiently large,
\begin{equation*}
    \beta_{T}(c, d) \leq \exp\{-(T-1)\bar{\alpha}_{T-1}(c)\cdot \delta / 8\} \rightarrow 0,
\end{equation*}
leading to a contraction.

\noindent \textbf{Step 2.2.} We have now established (\ref{eq44}). In the following, we will show that with probability 1,
\begin{equation}\label{eq53}
    \liminf_{T\rightarrow\infty}G_{d}(\gamma(c), \bar{\beta}_{T}(c, d)) - \Gamma^{*}_{\gamma(c)} \geq 0,\quad \forall d\in\mathcal{U}_{c}.
\end{equation}

We start by showing that
\begin{equation}
    \label{eq54}\limsup_{T\rightarrow\infty}\min_{d^{\prime}\in\mathcal{U}_{c}}G_{d^{\prime}}(\gamma(c), \bar{\beta}_{T}(c, d^{\prime})) - \Gamma^{*}_{\gamma(c)} \geq 0.
\end{equation}

Note that (\ref{eq44}) still holds if $z_{c}$ is replaced with $\tilde{z}_{c}$, where
$$\tilde{z}_{c} := \limsup_{T\rightarrow\infty}\min_{d^{\prime}\in\mathcal{U}_{c}}G_{d^{\prime}}(\gamma(c), \bar{\beta}_{T}(c, d^{\prime}));$$ 
in other words,
\begin{equation}
    \limsup_{T\rightarrow\infty}G_{d}(\gamma(c), \bar{\beta}_{T}(c, d)) - \tilde{z}_{c} \leq 0,\quad \forall d\in\mathcal{D}_{c},
\end{equation}
and the proof is exactly the same as that for (\ref{eq44}) in Step 2.1, except for the inequality (\ref{eq52}). Thus, we omit repetition to avoid redundancy. Let $\beta^{*}_{\bm{\gamma}}$ be an optimal solution to (\ref{opt:3}) that satisfies optimal condition (\ref{eq:10}). If $\tilde{z}_{c} \leq \Gamma^{*}_{\gamma(c)} - \delta$ for some $\delta > 0$, then
$$\limsup_{T\rightarrow\infty}G_{d}(\gamma(c), \bar{\beta}_{T}(c, d)) \leq \Gamma^{*}_{\gamma(c)} - \delta = G_{d}(\gamma(c), \beta^{*}_{\bm{\gamma}}(c, d)) - \delta,\quad \forall d\in\mathcal{D}_{c}.$$

Since $G_{d}(\cdot, \cdot)$ is monotonically increasing and uniformly continuous, there exists a positive number $\varepsilon>0$ such that $\bar{\beta}_{T}(c, d) < \beta^{*}_{\bm{\gamma}}(c, d) - \varepsilon$ for each $d\in\mathcal{U}_{c}$. However,
$$\begin{aligned}
1 - \gamma(c) - (\vert \mathcal{D}_{c}\vert - 1)\varepsilon & > \sum_{d\in\mathcal{U}_{c}}\bar{\beta}_{T}(c, d) \\
& = 1- \bar{\beta}_{T}(c, d^{*}(c)) \rightarrow 1 - \gamma(c),
\end{aligned}$$
leading to a contradiction. Therefore, we have $\tilde{z}_{c} \geq \Gamma^{*}_{\gamma(c)}$.

To complete the proof, we leverage (\ref{eq54}) to show a stronger condition than (\ref{eq53}) holds, namely
\begin{equation*}
    \liminf_{T\rightarrow\infty}\min_{d^{\prime}\in\mathcal{U}_{c}}G_{d^{\prime}}(\gamma(c), \bar{\beta}_{T}(c, d^{\prime})) - \Gamma^{*}_{\gamma(c)} \geq 0.
\end{equation*}

Again, let us argue by contradiction, and assume that the limit inferior is less than $-\delta$ for some $\delta > 0$. Then by (\ref{eq54}), there exist an increasing sequence of time steps $T_{1}<T_{2}<\ldots<T_{n}<\dots$, such that for any integer $k\geq 1$, $\min_{d^{\prime}\in\mathcal{U}_{c}}G_{d^{\prime}}(\gamma(c), \bar{\beta}_{T_{2k - 1}}(c, d^{\prime})) - \Gamma^{*}_{\gamma(c)} \geq -\delta/2$, and $\min_{d^{\prime}\in\mathcal{U}_{c}}G_{d^{\prime}}(\gamma(c), \bar{\beta}_{T_{2k}}(c, d^{\prime})) - \Gamma^{*}_{\gamma(c)} \leq -\delta$. Furthermore, $-\delta < \min_{d^{\prime}\in\mathcal{U}_{c}}G_{d^{\prime}}(\gamma(c), \bar{\beta}_{T}(c, d^{\prime})) - \Gamma^{*}_{\gamma(c)} < -\delta/2$ for any $T_{2k - 1} < T < T_{2k}$. Let $d_{k}$ denote $\argmin_{d^{\prime}\in\mathcal{U}_{c}} G_{d^{\prime}}(\gamma(c), \bar{\beta}_{T_{2k}}(c, d^{\prime}))$. Then we see that
$$ G_{d_{k}}(\gamma_{c}, \bar{\beta}_{T_{2k - 1}}(c, d_{k})) - G_{d_{k}}(\gamma_{c}, \bar{\beta}_{T_{2k}}(c, d_{k})) > \delta / 2. $$

Recall that $G_{d}(\cdot, \cdot)$ is universally uniformly continuous and monotonically increasing in both arguments. There exists a positive number $\varepsilon>0$ such that for all $k\geq 1$, $\bar{\beta}_{T_{2k - 1}}(c, d_{k}) - \bar{\beta}_{T_{2k}}(c, d_{k}) > \varepsilon.$ By Lemma \ref{lem:2}, for sufficiently large $T$ and $ T^{\prime}$, 
\begin{align*}
     &\left\vert \sum_{d^{\prime}\in\mathcal{U}_{c}}\bar{\beta}_{T}(c, d^{\prime}) - \sum_{d^{\prime}\in\mathcal{U}_{c}}\bar{\beta}_{T^{\prime}}(c, d^{\prime})\right\vert \\
    \leq &\left\vert \sum_{d^{\prime}\in\mathcal{U}_{c}}\bar{\beta}_{T}(c, d^{\prime}) - (1-\gamma(c))\right\vert + \left\vert \sum_{d^{\prime}\in\mathcal{U}_{c}}\bar{\beta}_{T^{\prime}}(c, d^{\prime}) - (1-\gamma(c))\right\vert \\
    \leq &\varepsilon / 2.
\end{align*}

Hence there exists $d^{\prime}_{k}\neq d_{k}$ such that $\bar{\beta}_{T_{2k - 1}}(c, d^{\prime}_{k}) - \bar{\beta}_{T_{2k - 1}}(c, d^{\prime}_{k}) < -\varepsilon/(2\vert\mathcal{D}_{c}\vert)$. By (\ref{eq55}), for sufficiently large $k$, there exists a time step $\tilde{T}_{k}$ with $T_{2k-1}\leq \tilde{T}_{k} < T_{2k}$, such that $\forall~ \tilde{T}_{k} \leq T < T_{2k}$, 
$$G_{d^{\prime}_{k}}(\gamma_{c}, \bar{\beta}_{T}(c, d^{\prime}_{k})) - \Gamma^{*}_{\gamma(c)} \geq -\frac{2}{3}\delta, $$
and
$$G_{d_{k}}(\gamma_{c}, \bar{\beta}_{T}(c, d_{k})) - \Gamma^{*}_{\gamma(c)} \leq -\frac{5}{6}\delta. $$

This leads to $\beta_{T}(c, d^{\prime}_{k}) \leq \exp\{-(T-1)\bar{\alpha}_{T-1}\cdot\delta / 12\}$ for sufficiently large $k$, and $\tilde{T}_{k} < T < T_{2k}$. However, it contradicts $\bar{\beta}_{T_{2k - 1}}(c, d^{\prime}_{k}) - \bar{\beta}_{T_{2k}}(c, d^{\prime}_{k}) < -\varepsilon/(2\vert\mathcal{D}_{c}\vert)$, which leads to
\begin{equation*}
    \sum_{t=\tilde{T}_{k} + 1}^{T_{2k}}\Big(\alpha_{t}(c)\beta_{t}(c, d^{\prime}_{k})\Big) \geq \sum_{t=\tilde{T}_{k} + 1}^{T_{2k}}\left(\alpha_{t}(c) \left(\bar{\beta}_{T}(c, d^{\prime}_{k}) + \frac{\varepsilon}{2\vert\mathcal{D}_{c}\vert}\right)\right).
\end{equation*}

To partially conclude, we have
\begin{equation*}
    \lim_{T\rightarrow\infty}G_{d}(\gamma(c), \bar{\beta}_{T}(c, d)) = z_{c} = \Gamma^{*}_{\gamma(c)},\quad \forall d\in\mathcal{U}_{c}.
\end{equation*}

That is, the conditional sampling effort leads to the optimal value of rate function almost surely in each context.

\noindent \textbf{Step 3: Global balance of sampling ratios to contexts.} 
We complete the proof by showing that $\lim_{T\rightarrow \infty}\bar{\alpha}_{T}(c) = \alpha^{*}_{\bm{\gamma}}(c)$ for any fixed context $c\in\mathcal{C}$. According to Corollary \ref{corol:1}, 
\begin{align*}
    &1 - \prod_{c^{\prime}\in\mathcal{C}}\left(\pi^{(T-1)}_{c^{\prime}, d^{*}(c^{\prime})}\right) 
    = \Pi_{T-1}\left(\bigcup_{c^{\prime}\in\mathcal{C}}\bigcup_{d^{\prime}\in\mathcal{D}_{c^{\prime}}\backslash\{d^{*}(c^{\prime})\}}\bm{\Theta}_{(c^{\prime},d^{\prime})}\right) \\
    \leq & \exp\left\{-(T-1)\left(\min_{c^{\prime}\in\mathcal{C}}\bar{\alpha}_{T-1}(c^{\prime})\min_{d^{\prime}\in\mathcal{D}_{c^{\prime}}\backslash\{d^{*}(c^{\prime})\}}G_{d^{\prime}}(\bar{\beta}_{T-1}(c^{\prime}, d^{*}(c^{\prime})), \bar{\beta}_{T-1}(c^{\prime}, d^{\prime})) - \frac{\varepsilon_{T-1}}{2}\right)\right\} \\
    \leq & \exp\left\{-(T-1)\left(\min_{c^{\prime}\in\mathcal{C}}\bar{\alpha}_{T-1}(c^{\prime})\Gamma^{*}_{\gamma(c^{\prime})} - \varepsilon_{T-1}\right)\right\},
\end{align*}
where $\{\varepsilon_{T}\}_{T\geq 1}$ is a vanishing sequence of positive numbers, and the last inequality follows from the universally uniform continuity of rate functions and Step 2. For the other direction, we also have
\begin{equation}\label{eq56}
    1 - \prod_{c^{\prime}\in\mathcal{C}}\left(\pi^{(T-1)}_{c^{\prime}, d^{*}(c^{\prime})}\right) \geq \exp\left\{-(T-1)\left(\min_{c^{\prime}\in\mathcal{C}}\bar{\alpha}_{T-1}(c^{\prime})\Gamma^{*}_{\gamma(c^{\prime})} + \varepsilon_{T-1}\right)\right\}.
\end{equation}

Similarly, by taking $\Tilde{\bm{\Theta}} = \bm{\Theta}^{(c,d^{*}(c))}$ in Theorem 1, 
\begin{equation}\label{eq57}
    \exp\left\{-(T-1)\left(\bar{\alpha}_{T-1}(c)\Gamma^{*}_{\gamma(c)} + \varepsilon_{T-1}\right)\right\} \leq 1 - \pi^{(T-1)}_{c, d^{*}(c)} \leq \exp\left\{-(T-1)\left(\bar{\alpha}_{t-1}(c)\Gamma^{*}_{\gamma(c)} - \varepsilon_{T-1}\right)\right\}.
\end{equation}

We first show that $\liminf_{T\rightarrow \infty} \bar{\alpha}_{T}(c) \leq \alpha^{*}_{\bm{\gamma}}(c)$. Assume, for the sake of contradiction, that there exist some $\delta>0$, $\bar{\alpha}_{T}(c) > \alpha^{*}_{\bm{\gamma}}(c)(1 + \delta)$ for any sufficiently large $T$. That being the case, we have
$$\bar{\alpha}_{T-1}(c)\Gamma^{*}_{\gamma(c)} - \min_{c^{\prime}\in\mathcal{C}}\bar{\alpha}_{T-1}(c^{\prime})\Gamma^{*}_{\gamma(c^{\prime})} > (1 + \delta)\alpha^{*}_{\bm{\gamma}}(c)\Gamma^{*}_{\gamma(c)} - \Gamma^{*}_{\bm{\gamma}} = \delta\cdot\Gamma^{*}_{\bm{\gamma}}.$$

Combining this with inequalities (\ref{eq56}) and (\ref{eq57}), we obtain $\alpha_{T}(c) \leq \exp\{-(T-1)(\delta\Gamma^{*}_{\bm{\gamma}} - 2\varepsilon_{T-1})\} \leq \exp\{-(T-1)\cdot \delta\Gamma^{*}_{\bm{\gamma}} / 2\}$, which implies that $\lim_{T\rightarrow \infty}\bar{\alpha}_{T}(c) = 0$, leading to a contradiction to our assumption that $\bar{\alpha}_{T}(c) > \alpha^{*}_{\bm{\gamma}}(c)(1 + \delta)$.

We further show that $\limsup_{T\rightarrow \infty} \bar{\alpha}_{T}(c) \leq \alpha^{*}_{\bm{\gamma}}(c)$. Otherwise, there exist some $\delta>0$ and an increasing sequence of time steps $T_{1}<T_{2}<\ldots<T_{n}<\dots$, such that for any integer $k\geq 1$, $\bar{\alpha}_{T_{2k-1}}(c) \leq \alpha^{*}_{\bm{\gamma}}(c) (1 + \delta/2)$, $\bar{\alpha}_{T_{2k}}(c) \geq \alpha^{*}_{\bm{\gamma}}(c) (1 + \delta)$, and for all $T_{2k-1}<T<T_{2k}$, $\alpha^{*}_{\bm{\gamma}}(c) (1 + \delta/2)< \bar{\alpha}_{T_{2k}}(c) < \alpha^{*}_{\bm{\gamma}}(c) (1 + \delta)$. Since $\vert \bar{\alpha}_{T+1}(c) - \bar{\alpha}_{T}(c) \vert \leq 1/(T+1)\rightarrow 0$, for any sufficiently large $k$, $\bar{\alpha}_{T_{2k-1}}(c) \geq \alpha^{*}_{\bm{\gamma}}(c) (1 + \delta/4)$. As a result, $\alpha_{T}(c) \leq \exp\{-(T-1)\cdot \delta\Gamma^{*}_{\bm{\gamma}}/8\}$ for $T_{2k-1}<T<T_{2k}$. According to (\ref{eq56}) and (\ref{eq57}),
\begin{align*}
    \delta\cdot \Gamma^{*}_{\bm{\gamma}}/2 &< \bar{\alpha}_{T_{2k}}(c) - \bar{\alpha}_{T_{2k-1}}(c) = \frac{1}{T_{2k}}\sum_{t=1}^{T_{2k}}{\alpha_{t}(c)} - \frac{1}{T_{2k-1}}\sum_{t=1}^{T_{2k-1}}{\alpha_{t}(c)} \\
    &=\frac{T_{2k} - T_{2k-1}}{T_{2k}}\left(\frac{1}{T_{2k}- T_{2k-1}}\sum_{t=T_{2k-1}+1}^{T_{2k}}{\alpha_{t}(c)} - \frac{1}{T_{2k-1}}\sum_{t=1}^{T_{2k-1}}{\alpha_{t}(c)}\right) \\
    &\leq \frac{T_{2k} - T_{2k-1}}{T_{2k}} \exp\{-(T-1)\cdot \delta\Gamma^{*}_{\bm{\gamma}}/8\} \rightarrow 0,
\end{align*}
leading to a contradiction to our assumption that $\delta>0$.

We come to the conclusion that $\limsup_{T\rightarrow \infty}\bar{\alpha}_{T}(c) \leq \alpha^{*}(c)$. Note that by definition, $\sum_{c\in\mathcal{C}}\bar{\alpha}_{T}(c) = 1,$ there must be $\lim_{T\rightarrow \infty}\bar{\alpha}_{T}(c) = \alpha^{*}(c)$. \halmos
\endproof

\begin{remark}
The proof can, to a much extent, be simplified if we assume uniqueness of the optimal allocation and strict monotinicity of rate functions. In such cases, Step 2.1 implies that $\limsup_{T\rightarrow\infty}\bar{\beta}_{T}(c, d) \leq \beta^{*}(c, d)$, and immediately we have $\lim_{T\rightarrow\infty}\bar{\beta}_{T}(c, d) = \beta^{*}(c, d)$ since $\sum_{d\in\mathcal{D}_{c}}\bar{\beta}_{T}(c, d) = \sum_{d\in\mathcal{D}_{c}}\beta^{*}(c, d) = 1$ for all $T\geq 1$ and $c\in\mathcal{C}$. Thus, Step 2.2 can be omitted, and this reduces to a similar argument to \citet{russo2020simple}.
\end{remark}

Then, we have a direct result for Algorithm \ref{alg:4} with a tuning hyperparameter.
\begin{corollary}[Asymptotic Optimality]\label{corol:ec1}
    Under TTTS-C with tuning hyperparameter in Algorithm \ref{alg:2} and Algorithm \ref{alg:4}, we have
    \begin{equation*}
        \lim_{T\rightarrow \infty} -\frac{1}{T}\ln \Pi_{T}\left( \bigcup_{c\in\mathcal{C}}\bigcup_{d\in\mathcal{D}\backslash\{d^{*}(c)\}}\bm{\Theta}_{(c, d)} \right) = \Gamma^{*}.
    \end{equation*}
\end{corollary}
\proof{Proof.} Let $\bm{\gamma}_{t}$ be the hyperparameter used at time $t$. According to Theorem \ref{thm:ec1}, each design $d\in\mathcal{D}$ receives infinite simulation samples ultimately with probability 1. Therefore, $\bm{\gamma}_{t} \rightarrow \bm{\gamma}^{*}$, and thus $\lim_{t\rightarrow\infty}\Gamma^{*}_{\bm{\gamma}_{t}} = \Gamma^{*}_{\bm{\gamma}^{*}} = \Gamma^{*}$. The latter limit holds since the constrained optimal value $\Gamma^{*}_{\bm{\gamma}}$ is continuous in $\bm{\gamma}$, according to the envelop theorem \citep[see][Corollary 3]{milgrom02}. Note that Lemma \ref{lem:2} and Lemma \ref{lem:3} will still be valid if $\gamma(c)$ is replaced with $\gamma^{*}(c)$. Now, what remains is to simply repeat the proof of Theorem \ref{thm:3}, which completes the proof.
\halmos
\endproof

\subsection{Extension to top-$m_{c}$ selection.}
Now, we can extend the analysis for the large deviations rate for the generic top-$m_{c}$ selection problem. To this end, the sampling policy can be expressed explicitly and their bounds are derived as in Lemmas \ref{lem:2} and \ref{lem:3}.
\begin{lemma}\label{lem:ec4}
    For any $c\in\mathcal{C}$ and $d\in\mathcal{D}_{c}$, the conditional sampling ratio of TTTS-C with hyperparameter $\bm{\gamma}$ satisfies:
    \begin{enumerate}
        \item for $d\in\mathcal{P}_{c}$, we have
        $$\lim_{t\rightarrow \infty}\sum_{d\in\mathcal{P}_{c}}\beta_{t}(c, d) = \gamma(c),$$
        and with probability 1, for sufficiently large $t$,
        $$ \frac{\gamma(c)}{\vert\mathcal{D}_{c}\vert^{2}} \frac{\sum_{\tilde{P}_{c}:d\notin \tilde{P}_{c}}\pi^{(t-1)}_{(c, \tilde{P}_{c})}}{1-\pi^{(t-1)}_{(c, \mathcal{P}_{c})}} \leq \beta_{t}(c, d) \leq 2 \frac{\sum_{\tilde{P}_{c}:d\notin \tilde{P}_{c}}\pi^{(t-1)}_{(c, \tilde{P}_{c})}}{1-\pi^{(t-1)}_{(c, \mathcal{P}_{c})}},$$
        \item for $d^{\prime}\in\mathcal{U}_{c}$, we have
        $$\lim_{t\rightarrow \infty}\sum_{d^{\prime}\in\mathcal{U}_{c}}\beta_{t}(c, d^{\prime}) = 1 - \gamma(c),$$
        and with probability 1, for sufficiently large $t$,
        $$ \frac{1-\gamma(c)}{\vert\mathcal{D}_{c}\vert^{2}} \frac{\sum_{\tilde{P}_{c}:d^{\prime}\in \tilde{P}_{c}}\pi^{(t-1)}_{(c, \tilde{P}_{c})}}{1-\pi^{(t-1)}_{(c, \mathcal{P}_{c})}} \leq \beta_{t}(c, d) \leq 2 \frac{\sum_{\tilde{P}_{c}:d^{\prime}\in \tilde{P}_{c}}\pi^{(t-1)}_{(c, \tilde{P}_{c})}}{1-\pi^{(t-1)}_{(c, \mathcal{P}_{c})}}.$$
    \end{enumerate}
    Moreover, with probability 1, for sufficiently large $t$,
    $$ \frac{1}{2}\frac{1 - \pi^{(t-1)}_{(c, \mathcal{P}_{c})}}{1 - \prod_{c^{\prime}\in\mathcal{C}}\left(\pi^{(t-1)}_{(c^{\prime}, \mathcal{P}_{c^{\prime}})}\right)} \leq \alpha_{t}(c) \leq 2\frac{1 - \pi^{(t-1)}_{(c, \mathcal{P}_{c})}}{1 - \prod_{c^{\prime}\in\mathcal{C}}\left(\pi^{(t-1)}_{(c^{\prime}, \mathcal{P}_{c^{\prime}})}\right)}.$$
\end{lemma}
\proof{Proof.} Let $\pi_{(c, P)}^{(t-1)} := \Pi_{t-1}(\bm{\Theta}^{(c, P)})$ denote the posterior probability of $P = \mathcal{P}_{c}$, i.e., $P$ consisting of the top-$m_{c}$ designs, for some subset $P\subseteq \mathcal{D}_{c}$ with $\vert P \vert = m_{c}$. Then similar to Lemma \ref{lem:1}, we have the following expression for the sampling policy under TTTS-C for selecting top-$m_{c}$ designs. For any fixed $c\in\mathcal{C}$ and $d\in\mathcal{D}_{c}$,
\begin{equation}\label{eq:16}
\begin{aligned}
    &\psi_{t}(d) = \sum_{P: d\in P_{c}} \prod_{c^{\prime}\in\mathcal{C}}{\left(\pi_{(c^{\prime},P_{c^{\prime}})}^{(t-1)}\right)} \sum_{\Delta\ni c} \frac{\prod_{c^{\prime}\in\Delta}{\left(1 - \pi_{(c^{\prime},P_{c^{\prime}})}^{(t-1)}\right)}\prod_{c^{\prime}\notin\Delta}{\left(\pi_{(c^{\prime},P_{c^{\prime}})}^{(t-1)}\right)}}{1 - \prod_{c^{\prime}\in\mathcal{C}}{\left(\pi_{(c^{\prime},P_{c^{\prime}})}^{(t-1)}\right)}} \cdot\frac{\sum_{\tilde{P}_{c}: d\notin \tilde{P}_{c}}{\pi^{(t-1)}_{(c, \tilde{P}_{c})}}/\left\vert P_{c} - \tilde{P}_{c} \right\vert}{\left(1 - \pi^{(t-1)}_{(c, P_{c})}\right)} \frac{\gamma(c)}{\vert \Delta\vert} \\
    &+ \sum_{P: d\notin P_{c}} \prod_{c^{\prime}\in\mathcal{C}}{\left(\pi_{(c^{\prime},P_{c^{\prime}})}^{(t-1)}\right)}\sum_{\Delta\ni c}\frac{\prod_{c^{\prime}\in\Delta}{\left(1 - \pi_{(c^{\prime},P_{c^{\prime}})}^{(t-1)}\right)}\prod_{c^{\prime}\notin\Delta}{\left(\pi_{(c^{\prime},P_{c^{\prime}})}^{(t-1)}\right)}}{1 - \prod_{c^{\prime}\in\mathcal{C}}{\left(\pi_{(c^{\prime},P_{c^{\prime}})}^{(t-1)}\right)}} \cdot\frac{\sum_{\tilde{P}_{c}: d\in \tilde{P}_{c}}{\pi^{(t-1)}_{(c, \tilde{P}_{c})}}/\left\vert \tilde{P}_{c} - P_{c} \right\vert}{\left(1 - \pi^{(t-1)}_{(c, P_{c})}\right)} \frac{1 - \gamma(c)}{\vert\Delta\vert} ,
\end{aligned}
\end{equation}
and 
\begin{equation}\label{eq:17}
    \alpha_{t}(c) = \sum_{P}\frac{\prod_{c^{\prime}\in\mathcal{C}}{\left(\pi_{(c^{\prime},P_{c^{\prime}})}^{(t-1)}\right)}} {1 - \prod_{c^{\prime}\in\mathcal{C}}{\left(\pi_{(c^{\prime},P_{c^{\prime}})}^{(t-1)}\right)}} \sum_{\Delta\ni c}\prod_{c^{\prime}\in\Delta}{\left(1 - \pi_{(c^{\prime},P_{c^{\prime}})}^{(t-1)}\right)}\prod_{c^{\prime}\notin\Delta}{\left(\pi_{(c^{\prime},P_{c^{\prime}})}^{(t-1)}\right)} \frac{1}{\vert \Delta\vert} ,
\end{equation}
where the summation $\sum_{P}$ is taken over all combinations of subsets $P_{c^{\prime}}\subseteq \mathcal{D}_{c^{\prime}}$ with $\vert P_{c^{\prime}}\vert = m_{c^{\prime}}$ for context $c^{\prime}$ ranging over $\mathcal{C}$, while the summation $\sum_{\tilde{P}_{c}}$ is taken over $\tilde{P}_{c}\subseteq \mathcal{D}_{c}$ with $\vert \tilde{P}_{c}\vert = m_{c}$ for the fixed context $c$. 

Note that for $d\in\mathcal{P}_{c}$, the second summation in Equation (\ref{eq:16}) converges to 0 due to Theorem \ref{thm:ec1}. Therefore, we have
\begin{equation*}
    \begin{aligned}
         &\lim_{t\rightarrow\infty}\sum_{d\in\mathcal{P}_{c}}\psi_{t}(d) \\
        = &\lim_{t\rightarrow\infty}\sum_{d\in\mathcal{P}_{c}}\sum_{P: d\in P_{c}} \prod_{c^{\prime}\in\mathcal{C}}{\left(\pi_{(c^{\prime},P_{c^{\prime}})}^{(t-1)}\right)} \sum_{\Delta\ni c} \frac{\prod_{c^{\prime}\in\Delta}{\left(1 - \pi_{(c^{\prime},P_{c^{\prime}})}^{(t-1)}\right)}\prod_{c^{\prime}\notin\Delta}{\left(\pi_{(c^{\prime},P_{c^{\prime}})}^{(t-1)}\right)}}{1 - \prod_{c^{\prime}\in\mathcal{C}}{\left(\pi_{(c^{\prime},P_{c^{\prime}})}^{(t-1)}\right)}} \\
        &\quad\quad\quad\quad\quad\quad\quad\quad\quad\quad\quad\quad\quad\quad \cdot\frac{\sum_{\tilde{P}_{c}: d\notin \tilde{P}_{c}}{\pi^{(t-1)}_{(c, \tilde{P}_{c})}}/\left\vert P_{c} - \tilde{P}_{c} \right\vert}{\left(1 - \pi^{(t-1)}_{(c, P_{c})}\right)} \frac{\gamma(c)}{\vert \Delta\vert} \\
        = &\lim_{t\rightarrow\infty}\sum_{P} \prod_{c^{\prime}\in\mathcal{C}}{\left(\pi_{(c^{\prime},P_{c^{\prime}})}^{(t-1)}\right)} \sum_{\Delta\ni c} \frac{\prod_{c^{\prime}\in\Delta}{\left(1 - \pi_{(c^{\prime},P_{c^{\prime}})}^{(t-1)}\right)}\prod_{c^{\prime}\notin\Delta}{\left(\pi_{(c^{\prime},P_{c^{\prime}})}^{(t-1)}\right)}}{1 - \prod_{c^{\prime}\in\mathcal{C}}{\left(\pi_{(c^{\prime},P_{c^{\prime}})}^{(t-1)}\right)}} \\
        &\quad\quad\quad\quad\quad\quad\quad\quad\quad\quad\quad\quad \cdot \sum_{\tilde{P}_{c}: \tilde{P}_{c}\neq P_{c}}\frac{{\pi^{(t-1)}_{(c, \tilde{P}_{c})}}}{\left(1 - \pi^{(t-1)}_{(c, P_{c})}\right)} \frac{\left\vert (P_{c} - \tilde{P}_{c}) \bigcap \mathcal{P}_{c} \right\vert}{\left\vert P_{c} - \tilde{P}_{c} \right\vert} \frac{\gamma(c)}{\vert \Delta\vert}. \\
    \end{aligned}
\end{equation*}

An analogous argument as in the proof for Lemma \ref{lem:1} yields that the summand with $P_{c} = \mathcal{P}_{c}$ dominates the summation, i.e.,
\begin{equation*}
\begin{aligned}
     &\lim_{t\rightarrow\infty}\sum_{d\in\mathcal{P}_{c}}\beta_{t}(c, d) = \lim_{t\rightarrow\infty}\sum_{d\in\mathcal{P}_{c}}\psi_{t}(d) / \alpha_{t}(c) \\
    = & \lim_{t\rightarrow\infty}\frac{ \prod_{c^{\prime}\in\mathcal{C}}{\left(\pi_{(c^{\prime},\mathcal{P}_{c^{\prime}})}^{(t-1)}\right)} \sum_{\Delta\ni c} \frac{\prod_{c^{\prime}\in\Delta}{\left(1 - \pi_{(c^{\prime},\mathcal{P}_{c^{\prime}})}^{(t-1)}\right)}\prod_{c^{\prime}\notin\Delta}{\left(\pi_{(c^{\prime},\mathcal{P}_{c^{\prime}})}^{(t-1)}\right)}}{1 - \prod_{c^{\prime}\in\mathcal{C}}{\left(\pi_{(c^{\prime},\mathcal{P}_{c^{\prime}})}^{(t-1)}\right)}}\cdot \sum_{\tilde{P}_{c}: \tilde{P}_{c}\neq \mathcal{P}_{c}}\frac{{\pi^{(t-1)}_{(c, \tilde{P}_{c})}}}{\left(1 - \pi^{(t-1)}_{(c, \mathcal{P}_{c})}\right)} \frac{\left\vert (\mathcal{P}_{c} - \tilde{P}_{c}) \bigcap \mathcal{P}_{c} \right\vert}{\left\vert \mathcal{P}_{c} - \tilde{P}_{c} \right\vert} \frac{\gamma(c)}{\vert \Delta\vert} }{ \prod_{c^{\prime}\in\mathcal{C}}{\left(\pi_{(c^{\prime},\mathcal{P}_{c^{\prime}})}^{(t-1)}\right)} \sum_{\Delta\ni c} \frac{\prod_{c^{\prime}\in\Delta}{\left(1 - \pi_{(c^{\prime},\mathcal{P}_{c^{\prime}})}^{(t-1)}\right)}\prod_{c^{\prime}\notin\Delta}{\left(\pi_{(c^{\prime},\mathcal{P}_{c^{\prime}})}^{(t-1)}\right)}}{1 - \prod_{c^{\prime}\in\mathcal{C}}{\left(\pi_{(c^{\prime},\mathcal{P}_{c^{\prime}})}^{(t-1)}\right)}}\cdot \frac{1}{\vert \Delta\vert} } \\
    = &\gamma(c),
\end{aligned}
\end{equation*}
where the last equation follows from $(\mathcal{P}_{c} - \tilde{P}_{c}) \bigcap \mathcal{P}_{c} = \mathcal{P}_{c} - \tilde{P}_{c}$ and $\sum_{\tilde{P}_{c}: \tilde{P}_{c}\neq \mathcal{P}_{c}}{{\pi^{(t-1)}_{(c, \tilde{P}_{c})}}}/{\left(1 - \pi^{(t-1)}_{(c, \mathcal{P}_{c})}\right)} = 1$. Similarly, $\sum_{d^{\prime}\in\mathcal{U}_{c}}\beta_{t}(c, d^{\prime}) \rightarrow 1 - \gamma(c)$.

As for the inequalities, the proof resembles those for Lemma \ref{lem:2} and Lemma \ref{lem:3}, which we omit for simplicity. \halmos
\endproof

\proof{Proof for Theorem \ref{thm:4}.} The proof also breaks down into three steps. The first step is to establish the consistency of TTTS-C, which is a direct consequence of Theorem \ref{thm:ec1}. Therefore, for all $c\in\mathcal{C}$, we have $\pi_{(c,\mathcal{P}_{c})}^{(t-1)} \rightarrow 1$. The second step and the third step, respectively, are to show that for all $c\in\mathcal{C}$, $d\in\mathcal{P}_{c}$, $d^{\prime}\in\mathcal{U}_{c}$,
$$\lim_{T\rightarrow\infty}\min_{\tilde{d}\in\mathcal{P}_{c}}G_{\tilde{d},d^{\prime}}(\bar{\beta}_{T}(c, \tilde{d}), \bar{\beta}_{T}(c, d^{\prime})) - z_{T,c} = \lim_{T\rightarrow\infty}\min_{\tilde{d}^{\prime}\in\mathcal{U}_{c}}G_{d,\tilde{d}^{\prime}}(\bar{\beta}_{T}(c, d), \bar{\beta}_{T}(c, \tilde{d}^{\prime})) - z_{T,c} = 0,$$
where $z_{T,c} := \min_{\tilde{d}\in\mathcal{P}_{c}}\min_{\tilde{d}^{\prime}\in\mathcal{U}_{c}}G_{\tilde{d}, \tilde{d}^{\prime}}(\bar{\beta}_{T}(c, \tilde{d}), \bar{\beta}_{T}(c, \tilde{d}^{\prime}))$, and that for all $c\in\mathcal{C}$, 
$$\lim_{T\rightarrow\infty}\bar{\alpha}_{T}(c)z_{T,c} - z_{T} = 0,$$
where $z_{T} := \min_{c^{\prime}\in\mathcal{C}}\bar{\alpha}_{T}(c)z_{T,c}$. The argument repeats that of Theorem \ref{thm:3} with Lemma \ref{lem:ec4} in place of Lemma \ref{lem:2} and Lemma \ref{lem:3}. This completes the proof.
\halmos
\endproof

\section{Further discussions on top-$m_{c}$ selection.}
In this section, we further discuss the optimal sampling ratios for top-$m_{c}$ selection problems. Recall that the optimal sampling ratio for each context is inversely proportional to the optimal rate attained by allocating samples within each context. Therefore, we only have to focus on the context-free case, where $\vert \mathcal{C}\vert = 1$. Following this section, we leave out the subscript $c$ for simplicity. 

\subsection{Uniqueness of optimal allocation.}\label{sec:unique}
We first consider the uniqueness of optimal allocation problem associated with (context-free) top-$m_{c}$ selection
\begin{align}
    &\max_{\bm{\psi}} & & \hspace{-12em} G(\bm{\psi}) \label{opt:11}\\
    &\text{ s.t.} & & \hspace{-12em} \sum_{d\in \mathcal{P}}{\psi(d)} + \sum_{d^{\prime}\in \mathcal{U}}{\psi(d)} = 1, \notag \\
    & & & \hspace{-12em} \bm{\psi} \geq 0, \notag
\end{align}
where \begin{equation*}
    G(\bm{\psi}):= \min_{(d,d^{\prime})\in \mathcal{P}\times\mathcal{U}} G_{d,d^{\prime}}(\psi(d), \psi(d^{\prime})).
\end{equation*}

\begin{proposition}\label{prop:7}
Suppose $G_{d,d^{\prime}}(\cdot, \cdot)$'s are strictly concave as bivariate functions, then the solution to the optimization problem (\ref{opt:11}) is unique.
\end{proposition}

Recall that the optimal sampling ratios of (\ref{opt:5}) for each context is uniquely determined by being inversely proportional to $\Gamma^{*}_{\gamma^{*}(c)}$, and thus we only have to show the optimal conditional sampling ratio for each context under a context is unique. Then Proposition \ref{prop:4} is a direct consequence of Proposition \ref{prop:7}.

\begin{remark}
We make some remarks on the strict concavity assumption. It rules out cases where the sampling effort required to distinguish two designs can be substituted. For example, consider a problem with two designs, denoted as $\mathcal{P}=\{d\}$ and $\mathcal{U} = \{d^{\prime}\}$, with $m = 1$. Suppose the rate function $G_{d,d^{\prime}}(\psi(d), \psi(d^{\prime}))$ equals $\Gamma^{*}\times (\psi(d) + \psi(d^{\prime}))$ locally for $\vert \psi(d) + \psi(d^{\prime}) - 1 \vert \leq 0.1$, and $\vert \psi(d) - \psi(d^{\prime})\vert \leq 0.1$. In this case, the optimal allocation includes a continuum of solutions with $\psi(d) + \psi(d^{\prime}) = 1$, violating the strict concavity. Around $(\psi(d), \psi(d^{\prime})) = (1/2, 1/2)$, compensating for the effort allocated to $d$ at the expanse of the effort for $d^{\prime}$ does not affect the large deviations rate, leading to multiple optimal solutions. On the other hand, the strict concavity implies strict monotonicity, as $G_{d,d^{\prime}}(\cdot, \cdot)$ is monotonically increasing in both arguments, which facilitates the proof. We note that strict concavity is not uncommon under general model assumptions. For example, Gaussian distributions, Bernoulli distributions, Exponential distributions, and other commonly used distributions lead to strictly concave rate functions. Similarly, large deviations rate functions for mean performances of samples under a frequentist perspective have also been shown to be strictly concave \citep{glynn2004large}.
\end{remark}

The proof leverages the concave property of the objective $G(\bm{\psi})$, although it may not be strictly concave as a multivariate function. The idea is to iteratively eliminate all potentially variable coordinates of the optima by utilizing the strict concavity of rate functions. 

\proof{Proof of Proposition \ref{prop:7}.}
We argue by contradiction. Assume that $\bm{\psi}^{(0)}$ and $\bm{\psi}^{(1)}$ are two distinct optimal solutions to problem (\ref{opt:11}). Then $\Gamma^{*}=G(\bm{\psi}^{(0)})=G(\bm{\psi}^{(1)})$ by definition. With a slight abuse of notations, define by \begin{equation*}
    \bm{\psi}^{(r)} := r \bm{\psi}^{(1)} + (1-r) \bm{\psi}^{(0)},\quad r \in [0,1]
\end{equation*} 
the convex combinations of $\bm{\psi}^{(0)}$ and $\bm{\psi}^{(1)}$. 

Observe that $\bm{\psi}^{(r)}$ maximizes the objective for $r \in [0,1]$. It simply follows from
\begin{align*}
    \Gamma^{*} \geq G(\bm{\psi}^{(r)}) \geq r G(\bm{\psi}^{(1)}) + (1-r) G(\bm{\psi}^{(0)}) = \Gamma^{*},
\end{align*}
where the second inequality holds due to the fact that $G(\cdot)$, as a function minimizing over concave functions $G_{d,d^{\prime}}(\cdot,\cdot)$, is concave in $\bm{\psi}$ as well. 

Then we can inductively construct nest intervals $[r_{L,n},r_{R,n}]$ along with a sequence of partitions of $\mathcal{P}\times\mathcal{U}$, i.e., $\{(A_{n},B_{n}):n\in\mathbb{N}\}$ such that $A_{n}\bigcup B_{n}$ is a disjoint union equal to $\mathcal{P}\times\mathcal{U}$. By construction, we will show that (a) $A_{n}$ has cardinality $n$; (b) $A_{n}$ consists of $(d,d^{\prime})$-pairs such that the coordinates corresponding to $d$ (and $d^{\prime}$) in $\bm{\psi}^{(0)}$ and $\bm{\psi}^{(1)}$ coincide; and (c) for any $n\geq 1$ and $r\in [r_{L,n},r_{R,n}]$, there holds $$\Gamma^{*}=\min_{(d,d^{\prime})\in A_{n}}G_{d,d^{\prime}}(\psi^{(r)}(d), \psi^{(r)}(d^{\prime})).$$ 

$A_{n}$ can been deemed as a desired set of indices whose corresponding coordinates in optimal solutions must coincide. On the other hand, $B_{n}$ is a suspicious set whose elements correspond to coordinates in $\bm{\psi}^{(r)}$ that may vary. The induction procedure goes on until we can guarantee that $\bm{\psi}^{(0)}=\bm{\psi}^{(1)}$.

To be specific, let $r_{L,0}=0$, $r_{R,0}=1$, $P_{0}=\O$, and $Q_{0} = \mathcal{P}\times\mathcal{U}$. We then provide the inductive step in constructing these quantities. Assume that the aforementioned quantities are well defined up to index $(n-1)$. We claim that at least one of the following case happens:

\begin{enumerate}[\text{Case} 1.]
    \item $\psi^{(0)}(d)=\psi^{(1)}(d)$, and $\psi^{(0)}(d^{\prime})=\psi^{(1)}(d^{\prime})$, $~\forall (d,d^{\prime}) \in B_{n-1}$;
    \item $\min_{(d,d^{\prime})\in B_{n-1}}G_{d,d^{\prime}}(\psi^{(r)}(d),\psi^{(r)}(d^{\prime})) = \Gamma^{*}$, $~\forall r \in (r_{L,n-1},r_{R,n-1})$.
\end{enumerate}

For $n=1$, the claim is trivial since Case 2 is true. For $n\geq 2$, we argue by contradiction. Assume otherwise that neither of this two cases happens, that is, there exists an $(\tilde{d}, \tilde{d}^{\prime})$-pair in $B_{n-1}$ such that at least one of $\psi^{(0)}(\tilde{d})\neq\psi^{(1)}(\tilde{d})$ and $\psi^{(0)}(\tilde{d}^{\prime})\neq\psi^{(1)}(\tilde{d}^{\prime})$ is true, and for some $r^{0}\in (r_{L,n-1},r_{R,n-1})$, $$\min_{(d,d^{\prime})\in B_{n-1}}G_{d,d^{\prime}}(\psi^{(r^{0})}(d),\psi^{(r^{0})}(d^{\prime})) > G(\bm{\psi}^{(r^{0})})=\Gamma^{*}.$$ 

Without loss of generality, we assume $\psi^{(0)}(\tilde{d}) < \psi^{(1)}(\tilde{d})$, and therefore $\psi^{(r)}(\tilde{d})$ is monotonically increasing in $r$. 

Since $G_{d,d^{\prime}}(\cdot,\cdot)$'s and $G(\cdot)$ are continuous functions, there exists $\varepsilon>0$ such that for all $\Tilde{\bm{\psi}}$ with $\Vert \Tilde{\bm{\psi}} - \bm{\psi}^{(r^{0})}\Vert_{\infty} \leq \varepsilon$, 
\begin{align}
    & \sum_{(d,d^{\prime})\in \mathcal{P}\times\mathcal{U}}{\left\vert G_{d,d^{\prime}}(\Tilde{\psi}(d),\Tilde{\psi}(d^{\prime})) - G_{d,d^{\prime}}(\psi^{(r^{0})}(d), \psi^{(r^{0})}(d^{\prime})) \right\vert + \left\vert G(\Tilde{\bm{\psi}}) - G(\bm{\psi}^{(r^{0})})\right\vert } \notag \\
    < & \min_{(d,d^{\prime})\in B_{n-1}}G_{d,d^{\prime}}(\psi^{(r^{0})}(d),\psi^{(r^{0})}(d^{\prime})) - \Gamma^{*}.\label{eq19}
\end{align}

Let $R_{n-1}$ be the set of indices in elements of $A_{n-1}$, that is,
$$R_{n-1} := \left\{ \bar{d} : (\bar{d}, d^{\prime})\in A_{n-1} \text{ for some } d^{\prime}\in\mathcal{U} \text{ or } (d, \bar{d})\in A_{n-1} \text{ for some } d\in\mathcal{P}\right\}.$$

Note that by (b), $d\notin R_{n-1}$, i.e., there exists no $\bar{d}$ such that $(d, \bar{d})\in A_{n-1}$. Then one can choose $0<\delta<\varepsilon$ to be small enough so that $\Tilde{\bm{\psi}}$ is a feasible solution to (\ref{opt:11}) given by
\begin{equation}
    \Tilde{\psi}(\bar{d}) = \begin{cases}
    \psi^{(r^{0})}(d) - \delta, & \text{if }\bar{d} = d, \\
    \psi^{(r^{0})}(\bar{d}) + \delta/\vert R_{n-1} \vert, & \text{if } \bar{d}\in R_{n-1}, \\
    \psi^{(r^{0})}(\bar{d}), & \text{otherwise.}
    \end{cases} \label{eq20}
\end{equation}

This solution yields a larger objective
\begin{equation}
    G(\Tilde{\bm{\psi}}) = \min_{(d,d^{\prime})\in A_{n-1}}{G_{d,d^{\prime}}(\Tilde{\psi}(d), \Tilde{\psi}(d^{\prime}))} > \min_{(d,d^{\prime})\in A_{n-1}}{G_{d,d^{\prime}}(\psi^{(r^{0})}(d), \psi^{(r^{0})}(d^{\prime}))} = \Gamma^{*}, \label{eq21}
\end{equation}
which contradicts the definition of $\Gamma^{*}$. The first equation follows from triangular inequality and (\ref{eq19}), and the second inequality follows from the fact that $G_{d,d^{\prime}}(\cdot,\cdot)$ is strictly increasing in both arguments.

Now, we have shown that at least one of cases must be true. In Case 1, we terminate the construction procedure. In Case 2, we define for each $(d,d^{\prime})$-pair in $B_{n-1}$ a subset of $[r_{L,n-1},r_{R,n-1}]$ as $$W_{d,d^{\prime}}^{n}:=\left\{r\in[r_{L,n-1},r_{R,n-1}]: G_{d,d^{\prime}}(\psi^{(r)}(d),\psi^{(r)})(d^{\prime}) = \min_{(i^{\prime},j^{\prime})\in B_{n-1}}{G_{i^{\prime}j^{\prime}}(\psi_{i^{\prime}}^{(r)},\psi_{j^{\prime}}^{(r)})}=\Gamma^{*}\right\}.$$

Obviously $\bigcup_{(d,d^{\prime})\in B_{n-1}}{W_{d,d^{\prime}}^{n}} = [r_{L,n-1},r_{R,n-1}]$, and for each $(d,d^{\prime})\in B_{n-1}$, $W_{d,d^{\prime}}^{n}$ is a closed interval since $G_{d,d^{\prime}}(\psi_{i}^{(r)}, \psi_{j}^{(r)})$ is continuous and concave in $r$. Then the Baire category theorem \citep[see][Theorem 48.2]{munkres2000topology} implies that there exists at least one pair $(d_{n},d^{\prime}_{n})\in B_{n-1}$ such that $W_{d_{n},d^{\prime}_{n}}^{n}$ has nonempty interior. Now, we can define \begin{align*}
    &r_{L,n} := \inf W_{d_{n},d^{\prime}_{n}}^{n},\quad r_{R,n} := \sup W_{d_{n},d^{\prime}_{n}}^{n},\\
    &A_{n} = A_{n-1} \bigcup \{(d_{n},d^{\prime}_{n})\}, \quad B_{n} = B_{n-1}\backslash \{(d_{n},d^{\prime}_{n})\}.
\end{align*}

Note that (a) and (c) hold naturally by construction. To justify (b), it suffices to show that $\psi^{(0)}(d_{n}) = \psi^{(1)}(d^{\prime}_{n})$. By the definition of $W_{d_{n},d^{\prime}_{n}}^{n}$, we see that $G_{d_{n},d^{\prime}_{n}}(\psi^{(r)}(d_{n}),\psi^{(r)}(d^{\prime}_{n})) = \Gamma^{*}$ for $r\in [r_{L,n},r_{R,n}]$. Note that if $\psi^{(r^{L,n})}(d_{n}) \neq \psi^{(r^{R,n})}(d_{n})$ or $\psi^{(r^{L,n})}(d^{\prime}_{n}) \neq \psi^{(r^{R,n})}(d^{\prime}_{n})$, then  
\begin{align*}
    \Gamma^{*} &= G_{d_{n},d^{\prime}_{n}}\left(\psi^{\left(\frac{r_{L,n}+r_{R,n}}{2}\right)}(d_{n}), \psi^{\left(\frac{r_{L,n}+r_{R,n}}{2}\right)}(d^{\prime}_{n})\right) \\
    &> \frac{1}{2}G_{d_{n},d^{\prime}_{n}}(\psi^{(r_{L,n})}(d_{n}), \psi^{(r_{L,n})}(d^{\prime}_{n})) + \frac{1}{2}G_{d_{n},d^{\prime}_{n}}(\psi^{(r_{R,n})}(d_{n}), \psi^{(r_{R,n})}(d^{\prime}_{n})) = \Gamma^{*},
\end{align*}
since $G_{d_{n}, d^{\prime}_{n}}(\cdot, \cdot)$ is strictly concave, leading to a contradiction. Therefore,
\begin{equation*}
    \psi^{(r^{L,n})}(d_{n}) = \psi^{(r^{R,n})}(d_{n}), \quad \psi^{(r^{L,n})}(d^{\prime}_{n}) = \psi^{(r^{R,n})}(d^{\prime}_{n}).
\end{equation*}

Moreover, we have $\psi^{(0)}(d_{n}) = \psi^{(1)}(d_{n})$ and $\psi^{(0)}(d^{\prime}_{n}) = \psi^{(1)}(d^{\prime}_{n})$, which completes the induction.

To conclude, we end up with a partition $(P_{n_{0}},Q_{n_{0}})$ of $\mathcal{P}\times\mathcal{U}$ together with an interval $[r_{L,n_{0}},r_{R,n_{0}}]$ for some $n_{0}\geq 1$. Based on the stopping criteria of the induction step, we have $\psi^{(0)}(d)=\psi^{(1)}(d)$ and $\psi^{(0)}(d^{\prime})=\psi^{(1)}(d^{\prime})$ for all $(d, d^{\prime})\in B_{n_{0}}$. Furthermore, by construction, this equality is also valid for all $(d, d^{\prime})\in A_{n_{0}}$, and consequently for all $(d,d^{\prime})\in\mathcal{P}\times\mathcal{U}$. This completes the proof.
\halmos
\endproof

\subsection{Necessary conditions for Gaussian sampling distributions.}\label{sec:gausscondition}
For the specific case of Gaussian designs with known heterogeneous variance, we can further characterize the optimal allocation. In this case, the rate function equals
$$G_{d, d^{\prime}}(\psi(d), \psi(d^{\prime})) = \frac{(\mu_{d}^{*} - \mu_{d^{\prime}}^{*})^{2}}{(\sigma_{d}^{*})^{2}/\psi(d) + (\sigma_{d^{\prime}}^{*})^{2}/\psi(d^{\prime})}.$$ 

Recall that Proposition \ref{prop:5} yields a necessary condition that $\exists z\in\mathbb{R}^{+}$, for all $d\in\mathcal{P},d^{\prime}\in\mathcal{U}$,
\begin{equation}\label{eq:15}
    \min_{\tilde{d}\in\mathcal{P}} \frac{(\mu_{\tilde{d}}^{*} - \mu_{d^{\prime}}^{*})^{2}}{(\sigma_{\tilde{d}}^{*})^{2}/\psi(\tilde{d}) + (\sigma_{d^{\prime}}^{*})^{2}/\psi(d^{\prime})} = \min_{\tilde{d}^{\prime}\in\mathcal{U}} \frac{(\mu_{d}^{*} - \mu_{\tilde{d}^{\prime}}^{*})^{2}}{(\sigma_{d}^{*})^{2}/\psi(d) + (\sigma_{\tilde{d}^{\prime}}^{*})^{2}/\psi(\tilde{d}^{\prime})} = z.
\end{equation}

Typically, there exists a continuum of solutions to these equations, especially when $\vert \mathcal{P}_{c}\vert \gg 1$ and $\vert \mathcal{U}_{c}\vert \gg 1$, since there are significantly more variables than equations. The solutions to the balance condition can be categorized into a combinatorial number of situations depending on whether the bivariate rate functions attain the minimum. Let $\vartheta(d, d^{\prime})\in\{0, 1\}$ be a binary variable, and let $\bm{\vartheta}$ be the vector of $\vartheta(d, d^{\prime})$, where $d\in\mathcal{P}, d^{\prime}\in\mathcal{U}$. We say that a solution to the necessary condition is of type $\bm{\vartheta}$ when $G_{d, d^{\prime}}(\psi(d), \psi(d^{\prime})) = \min_{\tilde{d}\in\mathcal{P}}\min_{\tilde{d}^{\prime}\in\mathcal{U}} G_{\tilde{d}, \tilde{d}^{\prime}}(\psi(\tilde{d}), \psi(\tilde{d}^{\prime}))$ if and only if $\vartheta(d, d^{\prime}) = 1$.

To fully characterize the optimality condition of (\ref{opt:11}), we define an equivalence relation $\overset{\bm{\vartheta}}{\sim}$ on $\mathcal{D}$ for each type $\bm{\vartheta}$.
\begin{definition}\label{def:1}
    For $\tilde{d}, \bar{d} \ in \mathcal{D}$, we call $\tilde{d} \overset{\bm{\vartheta}}{\sim} \bar{d}$ if there exists a chain of design pairs $\{(d_{i}, d^{\prime}_{i}) \in \mathcal{P} \times \mathcal{U}: i = 1, \dots, n\}$ such that
    \begin{enumerate}
        \item $\vartheta(d_{i}, d^{\prime}_{i}) = 1$ for $i = 1, \dots, n$,
        \item either $d_{i} = d_{i+1}$ or $d^{\prime}_{i} = d^{\prime}_{i+1}$ for $i = 1, \dots, n - 1$,
        \item $\tilde{d} \in \{d_{1}, d^{\prime}_{1}\}$ and $\bar{d} \in \{d_{n}, d^{\prime}_{n}\}$.
    \end{enumerate}
\end{definition}

This equivalence relationship defines $L$ equivalence classes of designs, i.e., $\mathcal{D} = \mathcal{D}^{1} \bigcup \cdots \bigcup \mathcal{D}^{L}$, such that two designs $\tilde{d}, \bar{d}$ belong to the same equivalent class $\mathcal{D}^{l}$ if and only if $\tilde{d} \overset{\bm{\vartheta}}{\sim} \bar{d}$. Denote $\mathcal{P}^{l} = \mathcal{D}^{l} \bigcap \mathcal{P}$ and $\mathcal{U}^{l} = \mathcal{D}^{l} \bigcap \mathcal{U}$, $l = 1,\dots,L$. Then a necessary optimality condition is as follows.

\begin{proposition}[Modification of Theorem 3, \citet{zhang2021asymptotically}]\label{prop:6}
    The feasible solution $\psi \geq 0$ to (\ref{opt:11}) is optimal only if there exist $z\in\mathcal{R}^{+}$ and type $\bm{\vartheta}$ with $\max_{\tilde{d}\in\mathcal{P}}\vartheta(\tilde{d}, d^{\prime}) = \max_{\tilde{d}^{\prime}\in\mathcal{U}}\vartheta(d, \tilde{d}^{\prime}) = 1$, such that for each equivalent class $\mathcal{D}^{l}$, $l=1,\dots, L$, induced by $\overset{\bm{\vartheta}}{\sim}$, the following conditions hold:
    \begin{equation*}
        \sum_{d\in\mathcal{P}^{l}} \left(\frac{\psi(d)}{\sigma_{d}^{*}}\right)^{2} = \sum_{d^{\prime}\in\mathcal{U}^{l}} \left(\frac{\psi(d^{\prime})}{\sigma_{d^{\prime}}^{*}}\right)^{2},
    \end{equation*}
    and
    \begin{equation*}
        \frac{(\mu_{d}^{*} - \mu_{d^{\prime}}^{*})^{2}}{(\sigma_{d}^{*})^{2}/\psi(d) + (\sigma_{d^{\prime}}^{*})^{2}/\psi(d^{\prime})} \geq z,
    \end{equation*}
    with equality holds if and only if $\vartheta(d, d^{\prime}) = 1$.
\end{proposition}

\proof{Proof.} The inequality and its equality condition follows from Proposition \ref{prop:5}. On the other hand, recall that the complementary slackness condition (\ref{eq60}) in the KKT conditions yield
$$\Lambda_{2}(d, d^{\prime})\left(\frac{(\mu_{d}^{*} - \mu_{d^{\prime}}^{*})^{2}}{(\sigma_{d}^{*})^{2}/\psi(d) + (\sigma_{d^{\prime}}^{*})^{2}/\psi(d^{\prime})} - z\right) = 0, \quad \forall d\in\mathcal{P}, d^{\prime}\in\mathcal{U}.$$

Hence if $\vartheta(d, d^{\prime}) = 0$, then $\Lambda_{2}(d, d^{\prime}) = 0$. Since for $\ell \neq \ell^{\prime}$ and $d\in\mathcal{P}^{\ell}$, $d^{\prime}\in\mathcal{U}^{\ell^{\prime}}$, the singleton $\{(d, d^{\prime})\}$ per se satisfies conditions 2 and 3 for the chain in Definition \ref{def:1}, we see that $\vartheta(d, d^{\prime}) = 0$. Otherwise $d\overset{\bm{\vartheta}}{\sim}d^{\prime}$, leading to a contradiction. This in turn implies that $\Lambda(d, d^{\prime}) = 0$.

Following the aforementioned facts, the first order conditions (\ref{eq58}) and (\ref{eq59}) can be rewritten as, for $l=1,\dots,L$, 
\begin{gather}
    \sum_{d^{\prime}\in\mathcal{U}^{l}} \Lambda_{2}(d, d^{\prime})\frac{\partial}{\partial x_{1}}G_{d, d^{\prime}}(\psi(d), \psi(d^{\prime})) - \Lambda_{4} = 0,\quad \forall d\in\mathcal{P}^{l}, \label{eq61}\\
    \sum_{d\in\mathcal{P}^{l}} \Lambda_{2}(d, d^{\prime})\frac{\partial}{\partial x_{2}}G_{d, d^{\prime}}(\psi(d), \psi(d^{\prime})) - \Lambda_{4} = 0,\quad \forall d^{\prime}\in\mathcal{U}^{l}, \label{eq62}
\end{gather}
where $\partial/\partial x_{i}$ denotes the derivative with regard to the $i$-th argument. Then the following calculation is a simple tautology of that in \citet{zhang2021asymptotically}, which completes the proof.
\halmos
\endproof

\begin{remark}
    The conditions stated in Proposition \ref{prop:6} do not necessarily guarantee a unique solution, and thus they may not constitute a sufficient condition for optimality. The issue aries from the fact that the specific type $\vartheta$ of the optimal allocation is unknown beforehand. However, based on our understanding, what can be inferred is that the necessary conditions allow for at most one solution for each type $\bm{\vartheta}$. Note that if two solutions, dented as $\tilde{\psi}$ and $\bar{\psi}$, are of the same type $\bm{\vartheta}$, then the inequality $\tilde{\psi}(d_{0}) < \bar{\psi}(d_{0})$ for a certain $d_{0}\in\mathcal{P}^{l}$, along with the balance condition and the strict monotonicity of rate functions, yields
    $$\tilde{\psi}(d) < \bar{\psi}(d),\quad \text{ and }\quad \tilde{\psi}(d^{\prime}) > \bar{\psi}(d^{\prime}),\quad \forall d\in\mathcal{P}^{l}, d^{\prime}\in\mathcal{U}^{l},$$
    leading to
    $$\sum_{d\in\mathcal{P}^{l}} \left(\frac{\bar{\psi}(d)}{\sigma_{d}^{*}}\right)^{2} > \sum_{d\in\mathcal{P}^{l}} \left(\frac{\tilde{\psi}(d)}{\sigma_{d}^{*}}\right)^{2} = \sum_{d^{\prime}\in\mathcal{U}^{l}} \left(\frac{\tilde{\psi}(d^{\prime})}{\sigma_{d^{\prime}}^{*}}\right)^{2} > \sum_{d^{\prime}\in\mathcal{U}^{l}} \left(\frac{\bar{\psi}(d^{\prime})}{\sigma_{d^{\prime}}^{*}}\right)^{2}, $$
    a contradiction to Proposition \ref{prop:6}. We conclude that the necessary optimality conditions offer at most a combinatorial number of candidates for the optimal allocation.
\end{remark}

These necessary conditions do not trivially extend to any sampling distributions. The derivation for the optimal conditions, actually, involves the particular structure of rate functions of Gaussian designs with known variance, i.e.,
$$\frac{\partial}{\partial x_{1}}G_{d, d^{\prime}}(\psi(d), \psi(d^{\prime})) \bigg/ \frac{\partial}{\partial x_{2}}G_{d, d^{\prime}}(\psi(d), \psi(d^{\prime})) = f_{d}(\psi(d)) / f_{d^{\prime}}(\psi(d^{\prime})),$$
where $f_{d}(x) = x^{2} / (\sigma_{d}^{*})^{2}$. This ``separation" property typically does not hold for the rate functions of general distribution families, such as Gaussian distributions with unknown variance, Bernoulli distributions, and Poisson distributions. Fo example, consider the specific case of Gaussian designs with unknown variance. In this case, the Kullback-Leibler (KL) divergence between two Gaussian random variables with parameters $\theta_{1} = (\mu_{1}, \sigma^{2}_{1})$ and $\theta_{2} = (\mu_{2}, \sigma^{2}_{2})$ is given by
$$D(\theta_{1}\Vert \theta_{2}) = \ln\left(\frac{\sigma_{2}}{\sigma_{1}}\right) + \frac{\sigma_{1}^{2} + (\mu_{1} - \mu_{2})^{2}}{2\sigma^{2}_{2}} - \frac{1}{2},$$
where $\mu_{d}$ is the mean parameter and $\eta_{d} = \sigma^{2}_{d}$ is the unknown variance parameter. Hence, the rate function can be expressed as
$$\begin{aligned}
    G_{d, d^{\prime}}(\psi(d), \psi(d^{\prime})) &= \inf_{\mu_{d} \leq \mu_{d^{\prime}}} \inf_{\sigma_{d}^{2}, \sigma_{d^{\prime}}^{2}} \psi(d)\left(\ln\left(\frac{\sigma_{d}}{\sigma_{d}^{*}}\right) + \frac{(\sigma^{*}_{d})^{2} + (\mu^{*}_{d} - \mu_{d})^{2}}{2\sigma^{2}_{d}} - \frac{1}{2}\right) \\
    & \quad\quad\quad\quad\quad\quad + \psi(d^{\prime})\left(\ln\left(\frac{\sigma_{d^{\prime}}}{\sigma_{d^{\prime}}^{*}}\right) + \frac{(\sigma^{*}_{d^{\prime}})^{2} + (\mu^{*}_{d^{\prime}} - \mu_{d^{\prime}})^{2}}{2\sigma^{2}_{d^{\prime}}} - \frac{1}{2}\right) \\
    &= \inf_{\mu_{d} \leq \mu_{d^{\prime}}} \frac{1}{2}\psi(d)\ln\left(\frac{(\sigma^{*}_{d})^{2} + (\mu^{*}_{d} - \mu_{d})^{2}}{(\sigma_{d}^{*})^{2}}\right) + \frac{1}{2}\psi(d^{\prime})\ln\left(\frac{(\sigma^{*}_{d^{\prime}})^{2} + (\mu^{*}_{d^{\prime}} - \mu_{d^{\prime}})^{2}}{(\sigma_{d^{\prime}}^{*})^{2}}\right) \\
    &= \inf_{\mu\in\mathbb{R}} \frac{1}{2}\psi(d)\ln\left(\frac{(\sigma^{*}_{d})^{2} + (\mu^{*}_{d} - \mu)^{2}}{(\sigma_{d}^{*})^{2}}\right) + \frac{1}{2}\psi(d^{\prime})\ln\left(\frac{(\sigma^{*}_{d^{\prime}})^{2} + (\mu^{*}_{d^{\prime}} - \mu)^{2}}{(\sigma_{d^{\prime}}^{*})^{2}}\right).
\end{aligned}$$

Let $\mu^{0}_{d, d^{\prime}}$ denote a point that takes the infimum above, then by the envelop theorem \citep[see][Corollary 3]{milgrom02},
\begin{gather*}
     \frac{\partial}{\partial x_{1}} G_{d, d^{\prime}}(\psi(d), \psi(d^{\prime})) = \frac{1}{2}\ln\left(\frac{(\sigma^{*}_{d})^{2} + (\mu^{*}_{d} - \mu^{0}_{d, d^{\prime}})^{2}}{(\sigma_{d}^{*})^{2}}\right), \\
     \frac{\partial}{\partial x_{2}} G_{d, d^{\prime}}(\psi(d), \psi(d^{\prime})) = \frac{1}{2}\ln\left(\frac{(\sigma^{*}_{d^{\prime}})^{2} + (\mu^{*}_{d^{\prime}} - \mu^{0}_{d, d^{\prime}})^{2}}{(\sigma_{d^{\prime}}^{*})^{2}}\right).
\end{gather*}

This fails to conform with the ``separate" form, thus rather complicating the necessary optimal conditions.
\end{APPENDICES}

\end{document}